\documentclass[aps, nofootinbib, notitlepage, prx, preprint]{revtex4-2}

\usepackage{amsfonts, amsmath, amssymb, amsthm, bbold}
\usepackage{acronym, array, bm, bbm, ccaption, color, dcolumn, epsfig, graphicx, multirow, nomencl, tabularx, upgreek,url}
\usepackage{outline}
\usepackage{ulem}
\usepackage[pdftex]{hyperref}
\hypersetup{colorlinks=true, pdfauthor=Kenneth M. Rudinger et al., pdftitle=Two-Qubit Gate Set Tomography with Fewer Circuits}
\usepackage[utf8]{inputenc}
\usepackage[ruled,linesnumbered]{algorithm2e}

\usepackage{bbm}
\usepackage{subcaption}

\usepackage[T1]{fontenc}

\usepackage{xspace}
\newcommand{\rrangle}{\rangle\kern-1.2ex~\rangle\xspace}
\newcommand{\llangle}{\langle\kern-1.2ex~\langle\xspace}

\usepackage{mathtools}

\DeclarePairedDelimiter\ket{\lvert}{\rangle}
\DeclarePairedDelimiterX\braket[2]{\langle}{\rangle}{#1 \delimsize\vert #2}
\DeclarePairedDelimiterX\expval[3]{\langle}{\rangle}{#1 \delimsize\vert #2 \delimsize\vert #3}

\DeclarePairedDelimiter\bbra{\llangle}{\rvert}
\DeclarePairedDelimiter\kket{\lvert}{\rrangle}
\DeclarePairedDelimiterX\bbraket[2]{\llangle}{\rrangle}{#1 \delimsize\vert #2}
\DeclarePairedDelimiterX\eexpval[3]{\llangle}{\rrangle}{#1 \delimsize\vert #2 \delimsize\vert #3}

\DeclarePairedDelimiter\floor{\lfloor}{\rfloor}

\DeclarePairedDelimiter\set{\{}{\}}
\newcommand{\germs}{\mathbb{g}}
\newcommand{\pfids}{\mathbb{f}}
\newcommand{\mfids}{\mathbb{h}}

\definecolor{blue(ryb)}{rgb}{0.01, 0.28, 1.0}
\definecolor{lava}{rgb}{0.81, 0.06, 0.13}
\definecolor{purplish}{rgb}{0.80, 0.00, 0.80}
\definecolor{gosharks}{rgb}{0, 0.42, 0.45}
\definecolor{darkorange}{rgb}{1.0, 0.55, 0.0}
\definecolor{cadmiumgreen}{rgb}{0.0, 0.42, 0.24}
\definecolor{pumpkin}{rgb}{1.0, 0.46, 0.09}

\SetKwComment{Comment}{$\triangleright\,\,$}{}
\DeclareMathOperator*{\argmax}{arg\,max}
\DeclareMathOperator*{\argmin}{arg\,min}

\newcommand{\edesign}[2]{#1 germs/#2 fiducials}

\begin{document}
\title{Two-Qubit Gate Set Tomography with Fewer Circuits}

\author{Kenneth M. Rudinger}\thanks{kmrudin@sandia.gov}
\author{Corey I. Ostrove}
\author{Stefan K. Seritan}
\author{Matthew D. Grace}
\author{Erik Nielsen}\thanks{Now at IonQ, Inc., College Park, MD 20740}
\author{Robin J. Blume-Kohout}
\author{Kevin C. Young}
\affiliation{Quantum Performance Laboratory, Sandia National Laboratories, Albuquerque, NM 87185 and Livermore, CA 94550}

\date{\today}

\begin{abstract}
Gate set tomography (GST) is a self-consistent and highly accurate method for the tomographic reconstruction of a quantum information processor's quantum logic operations, including gates, state preparations, and measurements.  However, GST's experimental cost grows exponentially with qubit number.  For characterizing even just two qubits, a standard GST experiment may have tens of thousands of circuits, making it prohibitively expensive for platforms.  We show that, because GST experiments are massively overcomplete, many circuits can be discarded.  This dramatically reduces GST's experimental cost while still maintaining GST's Heisenberg-like scaling in accuracy.  We show how to exploit the structure of GST circuits to determine which ones are superfluous.  We confirm the efficacy of the resulting experiment designs both through numerical simulations and via the Fisher information for said designs.  We also explore the impact of these techniques on the prospects of three-qubit GST.
\end{abstract}

\pacs{}

\maketitle
\tableofcontents
\section{Introduction}

The implementation of high-performing quantum logic gates is a critical piece of achieving full-scale quantum computation.  Reaching such high performance demands powerful gate characterization tools.  While there are a variety of gate characterization techniques \cite{Artiles2005-ts,Altepeter2003-fb,Banaszek2013-zx,Bendersky2008-xa,Bialczak2010-et,Blume-Kohout2010-hb,Blume-Kohout2010-vv,Childs2001-en,Christandl2012-am,Chuang1997-vf,DAriano2001-fx,Fiurasek2001-mp,Gale1968-mk,Granade2016-qy,Haah2017-jw,Hradil1997-fv,Kim2014-ix,Kimmel2014-nx,Lobino2008-ud,James2001-hz,OBrien2004-tr,Poyatos1997-mz,Riebe2006-tc,Smolin2012-yv,Vogel1989-vt,Weinstein2004-vn,Nielsen2020_GST,MerkelPRA13,GST2013,Greenbaum15,Blume-Kohout2017-kn,GST2015,Dehollain2016-zt,matteo2020operational,Hong2020-vc,Joshi2020-wo,Proctor2019-oi,Song2019-fg,Ware2018-cq,Zhang2020-ux,Mavadia2018-al,Rol2017-wn,White2019-ls,mkadzik2022precision,hashim2022benchmarking,rudinger2022characterizing,rudinger2021experimental,EmersonScience2007, Knill2008, wallman_randomized_2014, Magesan2011-ra, Magesan2012-bo, Gaebler2012-vq, Magesan2012-sg, Gambetta2012-yu, Corcoles2013-zs, Barends2014-ap, Wallman2015-pa, Carignan-Dugas2015-pz, Wallman2015-uq, Chasseur2015-zz, Sheldon2016-sj, Wallman2016-kx, Proctor2017-ru, Harper2017-oa, Wallman2018-wy, Huang2019-zj, McKay2019-kf, Proctor2019-ma, Erhard2019-ig, Ekert2002-ma, Levi2007-rb, Toth2010-xi, Flammia2011-rw, Da_Silva2011-jv, Moussa2012-kd, Reich2013-oj, Kimmel2015-tj, Rudinger2017-vy, Aaronson2018-tj, Mayer2018-zt, Helsen2019-fi, Huang2020-ll}, 
the focus of this work is gate set tomography (GST).  GST is a self-consistent tomographic reconstruction technique \cite{MerkelPRA13, GST2013,Greenbaum15,Blume-Kohout2017-kn,Nielsen2020_GST}  that has been used to characterize and improve the performance of a wide variety of qubit platforms \cite{Nielsen2020_GST,MerkelPRA13,GST2013,Greenbaum15,GST2015,Blume-Kohout2017-kn,Dehollain2016-zt,matteo2020operational,Hong2020-vc,Joshi2020-wo,Proctor2019-oi,Song2019-fg,Ware2018-cq,Zhang2020-ux,Mavadia2018-al,Rol2017-wn,White2019-ls,mkadzik2022precision,hashim2022benchmarking,rudinger2022characterizing,rudinger2021experimental}.  Through the use of carefully chosen high-depth circuits, GST can achieve Heisenberg scaling in accuracy, enabling extremely accurate gate characterization \cite{Blume-Kohout2017-kn}.  However, a drawback to GST is its overall experimental cost, making it challenging for some experimental groups to implement.

Our central result is that, via careful experiment design, the total number of circuits in a gate set tomography experiment can be \textit{dramatically} reduced (Fig. 1(a)), while preserving GST's Heisenberg-like scaling in accuracy (Figs. 1(b), 1(c)).  This breakthrough is predicated on two observations.  First, the list of circuits for a ``na\"{i}ve'' GST experiment is massively over-complete.  Second, many of the circuits contained therein yield relatively little additional information about gate set parameters we wish to learn.  Therefore, we should be able to discard relatively uninformative circuits from a GST experiment design without sacrificing much by way of estimate accuracy.

We use two techniques for choosing which circuits to discard from a GST experiment design: \textit{germ reduction} and \textit{fiducial pair reduction}, illustrated schematically in Fig. 1(a).  These two techniques can be combined, further reducing the number of circuits required to perform a GST experiment.  The validity of these approaches is demonstrated both analytically, by computing the Fisher information for the resulting experiment designs, and numerically, by directly simulating the resulting GST experiments and examining the accuracies of the corresponding gate set estimates.

We present results for reducing both single-qubit and two-qubit GST experiment design sizes. We find that for GST experiment designs with fixed maximum depth, non-trivial savings can be achieved for both single-qubit and two-qubit GST, while maintaining Heisenberg-like (or near-Heisenberg-like) scaling in accuracy.  For a given level of estimate accuracy, different experiment designs will require different maximum circuit depths.  In those cases, we find that the single- and two-qubit GST circuit costs can be reduced by as much as a factors of $\sim7.7$ and $\sim8.8$, respectively. It may also be the case that the limiting resource is actually wall-clock time, rather than the maximum circuit depth or total number of circuits. For an analysis of the relationship between the reduced experiment designs and wall-clock time see Appendix \ref{app:wall_clock}.

This paper is organized as follows.  In Sec.~\ref{sec:standard_gst_edesign}, we review how a standard GST experiment design is constructed, i.e., how standard GST circuits are chosen.  We then introduce a framework for \textit{circuit reduction}, i.e., reducing the size of a GST experiment design.  In Sec.~\ref{sec:results} we present our numerical results, showing how smaller GST experiment designs can offer significant experimental savings.  In Sec~\ref{sec:fisher_information}, we explain the results of Sec.~\ref{sec:results} from a more theoretical perspective by computing the Fisher information for the various experiment designs considered.  In Sec.~\ref{sec:methods} we present our methods in detail, discussing the more technical aspects of circuit reduction and corresponding details of standard GST experiment design.  We make our concluding remarks in Sec.~\ref{sec:conclusion}.  Included appendices provide additional technical details as necessary.

\begin{figure}[!htpb]
    \centering
    \includegraphics[width=\textwidth]{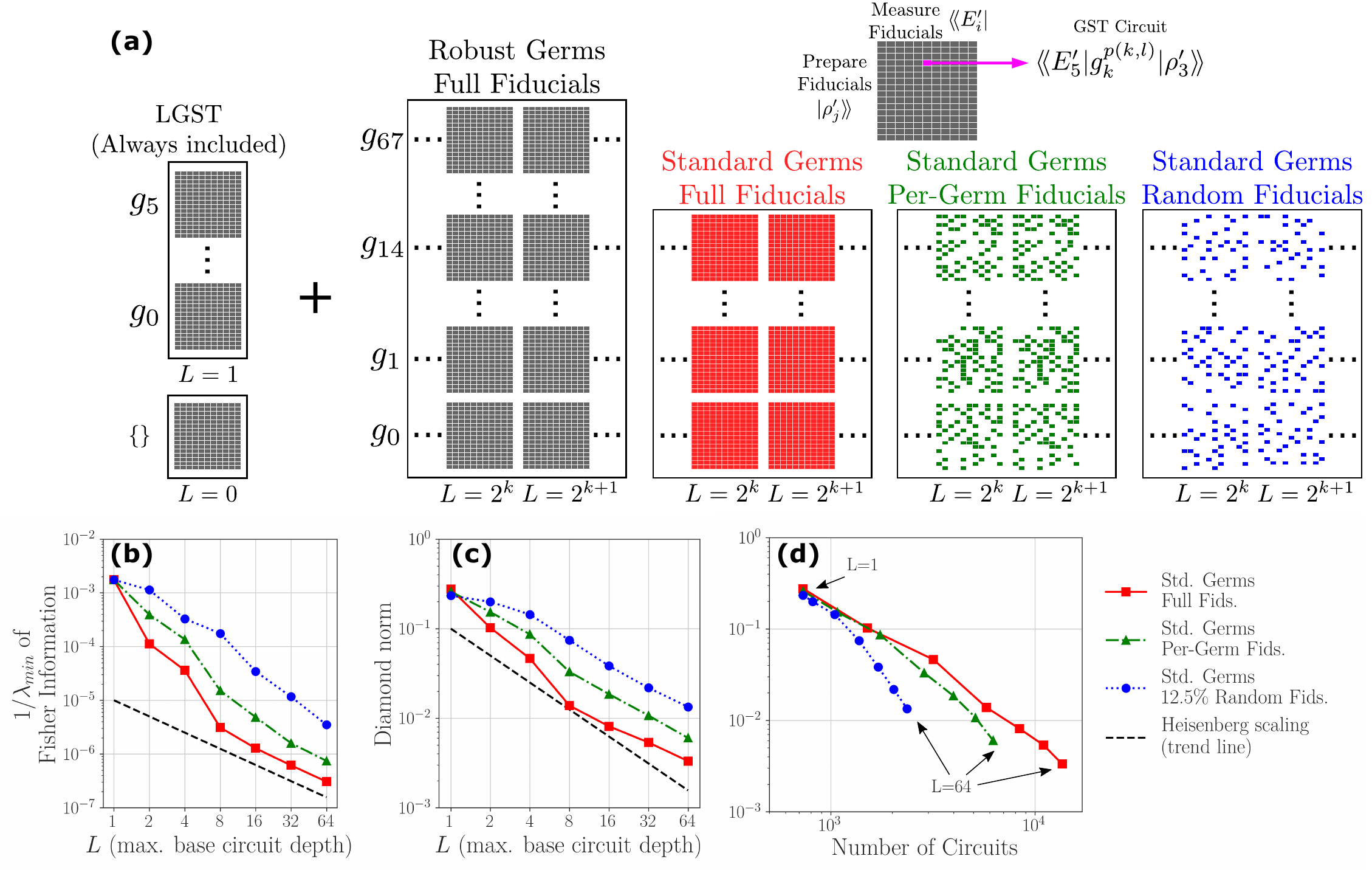}
    \caption{Overview of circuit reduction techniques and resulting accuracy. \textbf{(a)} schematically depicts a GST experiment design. Each rectangular ``plaquette'' of pixels corresponds to a particular germ (indexed by row) and power (indexed by column).  Each pixel in a plaquette represents a single GST circuit and corresponds to a fiducial pair choice.  The relaxation of the amplificationally complete (AC) condition allows truncation from the robust to standard germs, and per-germ and per-germ-power random fiducial pair reduction allow for sparser plaquettes. \textbf{(b)} and \textbf{(c)} show reduced two-qubit GST experiments maintain their asymptotic Heisenberg-like scaling of the estimation error with respect to germ power as measured by the (Cram\'er-Rao bound inspired) inverse minimum eigenvalue of the Fisher information or the diamond distance, respectively. \textbf{(d)} also shows the diamond distance as a function of the total number of circuits, a more experimentally relevant quantity.}
    \label{fig:GR-FPR-overview}
\end{figure}

\section{Standard GST experiment design}
\label{sec:standard_gst_edesign}

Before discussing how to reduce gate set tomography's experimental cost, we first describe how the circuits comprising a gate set tomography experiment are constructed.  A more thorough treatment of this subject may be found in \cite{Nielsen2020_GST}.

Gate set tomography is interested in estimation of the underlying parameters of an entire gate set $\mathcal{G}$, which consists of:
\begin{itemize}
    \item Gates, $\set{G_k}$ for $k = 1...N_G$
    \item A state preparation, $\kket{\rho}$
    \item A measurement operation $E$ with $m$ outcomes, $\set{\bbra{E_j}}$ for $j = 1...N_E^m$. \footnote{We note that GST can be used with gate sets that contain multiple native state preparations and/or measurement operations, but we may restrict our considerations to a single input state and measurement operation without loss of generality.}
\end{itemize}

\noindent The use of superket $\kket{\cdot}$ and superbra $\bbra{\cdot}$ for the state preparation $\kket{\rho}$ and the measurement effects $\bbra{E_i}$ denotes that these are represented by vectors in Hilbert-Schmidt space and by elements of the corresponding dual space, respectively; the gates $\set{G_i}$ are represented by superoperators which act by matrix multiplication on $\kket{\rho}$.  Therefore, for any circuit $\mathcal{C}$ beginning with state preparation  $\kket{\rho}$, followed by a sequence of gates $G_{k_1} \ldots G_{k_n}$\footnote{We slightly abuse notation here---a string of gate labels can be referring to either a series of gate instructions in a circuit (and occur left to right) \textit{or} the multiplication of the corresponding Pauli transfer matrices, in which case the order of gate operations is read from right to left.  We use both notations in this paper, but it should be clear from context which case is being employed.}, and ending with measurement $E$, the probability of observing outcome $j$ is given by $p(j|\mathcal{C})$:

\begin{equation}
\label{eqn:prob}
    p(j|\mathcal{C}) = \bbra{E_j}G_{k_n}\ldots G_{k_1}\kket{\rho}.
\end{equation}

\noindent Therefore, knowledge of the gate set $\mathcal{G}$ is sufficient to predict the outcome probabilities of any desired circuit.  Conversely, we may estimate $p(j|\mathcal{C})$ simply by performing the circuit $\mathcal{C}$ and recording the frequencies of the observed outcomes.  Through careful choice of a collection of circuits and appropriately processing the corresponding observed outcome frequency data, we can in turn estimate $\mathcal{G}$, i.e., we can tomographically reconstruct the gate set.

\subsection{Fiducial circuits}
In order to fully characterize the gates in $\mathcal{G}$, we need access to informationally complete (IC) sets of state preparations and measurements, i.e., those whose elements span the entirety of Hilbert-Schmidt space. This is often not possible natively on today's experimental devices, which may have only one native state preparation and measurement basis. We therefore generate the requisite set of IC states and measurements via the use of short sequences of gates; we call these short sequences ``fiducial circuits'', or simply ``fiducials''.
These are separated into \textit{preparation fiducial circuits}, or prep fiducials for short, denoted by $f_k \in \pfids$, and \textit{measurement fiducial circuits}, or measurement fiducials for short, denoted by $h_k\in\mfids$.
Taken together, the elements of $\pfids \times \mfids$ are referred to as \textit{fiducial pairs}.  Critically, we require that all gates used in the fiducials appear in the gate set $\mathcal{G}$, to ensure self-consistency.\footnote{We also note that while do not not assume strong a priori knowledge of $\mathcal{G}$, we must have some very rough sense of the action of the operations used in the fiducials in order to guarantee that the fiducials are indeed informationally complete.}

Using these state preparation and measurement fiducials, the \textit{effective} elements of the IC set of state preparations $\set{\kket{\rho_j^\prime}}$ and measurements $\set{\bbra{E_i^\prime}}$ are defined as
\begin{align}
    \kket{\rho_j^\prime} &= F_j \kket{\rho},\label{eqn:rho_j}\\
    \bbra{E_i^\prime} &= \bbra{E_{t(i)}} H_{h(i)},\label{eqn:E_i}
\end{align}
where $t(i) = \floor*{\frac{i}{m}}$ and $h(i)= i \hspace{-.5em} \mod m$ are functions mapping the single index $i$ to appropriate indices into the sets of measurement effects and measurement fiducials.

\subsection{Long-sequence GST}
\label{sec:LSGST}

A key innovation of modern GST, also called long-sequence GST (LSGST), 
is the ability to \textit{amplify} gate set parameters \cite{Blume-Kohout2017-kn}.  This is a marked improvement over its predecessor, called linear inversion GST (LGST), in which gates are self-consistently characterized, but with limited precision \cite{GST2013}.  Unlike LGST, LSGST uses circuits each of whose outcome probabilities have heightened sensitivities to some of the underlying parameters of the gate set.  This is akin to robust phase estimation \cite{Kimmel2015-tj,Rudinger2017-vy}, in which sensitivity to a single parameter in a gate set (the amount by which a gate over- or under-rotates) is amplified by repeating a gate many times, allowing for a buildup of that over-rotation.  However, unlike robust phase estimation, LSGST can amplify \textit{all} of the (non-gauge and non-SPAM) parameters of a gate set.  This is done through the inclusion of circuits that consist of repeating sub-sequences of gates called ``germs."  A set of germs will be denoted $\germs$.  In designing a GST experiment, a set of germs is chosen to be \textit{amplificationally complete} (AC), meaning that circuits in aggregate amplify all (non-gauge, non-SPAM) parameters of the gate set.  

LSGST circuits are constructed in which each of the germs is repeated increasingly many times, which serves to amplify the observable parameters of the gate set.  Thus a full LSGST experiment consists of the LGST  circuits plus tomography-like circuits of each repeated germ, e.g., circuits of the form $F_j g_k^p H_i$ for all elements of the germ and fiducial sets.  $p$ denotes the number of times a germ is repeated in a circuit, and we refer to it as the ``germ power''. In typical GST experiments we include for each germ circuits corresponding to multiple values of the germ-power, most commonly with values of $p$ that are logarithmically spaced. Rather than directly specifying the germ-powers of our circuits, we often choose instead to pick a sequence of maximum circuit depths $L_l$ and correspondingly set the germ-powers for our circuits as the largest integer $p(k,l)$ for which $\text{len}(g_k^{p(k,l)}) \leq L_l$ (and $\text{len}(\cdot)$ returns the depth of a circuit). The latter is what we use in this work, with the sequence of maximum circuits depths $L_l$ increasing by powers of two, i.e., $L_l\in[1,2,4\ldots L_\text{max}]$\footnote{When unambiguous we may drop the subscript on $L_l$.}.

Given a set of LSGST circuits, we then experimentally measure the frequencies of circuits of the following form:

\begin{equation}
    f_{ijkl} \approx \eexpval{E_i^\prime}{g_k^{p(k,l)}}{\rho_j^\prime}
    \label{eqn:frequencies}
\end{equation}

\noindent and use these frequencies to iteratively estimate gate set parameters using maximum likelihood estimation (MLE) \cite{Nielsen2020_GST}. As we only have access to these frequencies, self-consistent protocols like GST can only describe the gate set up to a similarity transform. Thus GST estimates contain a gauge degree of freedom, which  is described briefly in Appendix~\ref{app:gauge_invariance}.  Accounting for this gauge degree of freedom is necessary for the use of gauge-variant quantities such as diamond distance in Section~\ref{sec:results} and for computing the Fisher information in Section~\ref{sec:fisher_information}.  

While (long-sequence\footnote{We note that, as this paper only concerns itself with improving long-sequence GST, ``GST'' refers to its long-sequence variant for the remainder of this paper, unless explicitly stated otherwise.}) GST can achieve Heisenberg scaling in accuracy, the number of circuits it uses can be quite large, especially when compared to other characterization protocols.  For example, one of the first single-qubit GST demonstrations used over two thousand unique circuits \cite{Blume-Kohout2017-kn}, while a comparable two-qubit GST experiment required tens of thousands of circuits \cite{robin2016bbn}.  Such experiments can take a non-trivial amount of time, on the order of several hours to days to complete, during which the system can drift out of calibration \cite{Proctor2019-oi}.  While there have been experimental demonstrations of both single-qubit GST \cite{Blume-Kohout2017-kn, GST2013, zhang2022hidden, kim2022approaching, geller2021conditionally, mooney2021generation, Ware2018-cq, white2021many, zhang2021improving, moueddene2021context, chen2021quantum, luhman2021control, Joshi2020-wo, white2020demonstration, Proctor2019-oi, geller2020rigorous, wang2020high, Rudinger-PRX2019, crain2019high, hu2018experimental, hu2018experimental, Rol2017-wn, o2017density, RudingerRPE_PRL2017, GST2016} and two-qubit GST \cite{xue2022quantum, mkadzik2022precision, dahlhauser2022benchmarking, White2019-ls, dahlhauser2021characterization, hong2020demonstration, hughes2020benchmarking, govia2020bootstrapping, moueddene2020realistic, Song2019-fg} (as well as joint demonstrations of both \cite{hashim2023benchmarking, wang2021noise, cincio2020machine, xue2021computing, rudinger2021experimental, zhang2020error}), two-qubit GST remains experimentally expensive and out of reach for many platforms.  We do briefly note that to date there have been a handful of published experimental results that have utilized streamlined GST variants (e.g.,  \cite{Hong2020-vc,mkadzik2022precision}), and the underlying theory for such streamlining was touched upon in \cite{Nielsen2020_GST}.  This work, however, provides the first comprehensive description of these methods, as well as a quantification of the experimental savings they offer.  We discuss how to achieve these savings below. 

\subsection{Circuit reduction}

We've discussed the desired properties of GST experiment designs, but we've not yet discussed \textit{how} to build sets of circuits with such properties.  This task decomposes into two parallel tasks of \textit{germ selection} and \textit{fiducial selection}.  These tasks are described at length in \cite{Nielsen2020_GST}, and a recapitulation of them is given in Sec.~\ref{sec:circuit_selection}.

Of course, a primary goal of this paper is not to show how to select circuits for GST experiment designs, but how to \textit{reduce} the size of such experiment designs.  We present methods for independently shrinking the number of required germs and fiducial pairs, which are called \textit{germ reduction} and \textit{fiducial pair reduction}, respectively.  In this work, we consider multiple germ and fiducial pair reduction techniques; in this subsection we give brief functional descriptions for these various protocols.  We examine how they perform in simulation in Sec.~\ref{sec:results}, and how they perform in theory in Sec.~\ref{sec:fisher_information}.  A more thorough description of these techniques, as well as the underlying theory behind them, is given in Sec.~\ref{sec:circuit_reduction_techniques}. Explicit construction of the circuit reduction algorithms is given in Appendix~\ref{app:algorithms}.

\subsubsection{Germ reduction}
In order to describe our germ reduction methods, we first recall that we want any germ set in question to be amplificationally complete (AC) for the gate set it is probing, i.e., it should amplify the experiment's sensitivity to all (non-gauge non-SPAM) parameters of that gate set.  Therefore, the particular gate set provided to the germ selection algorithm can impact which germs are ultimately selected.  While one intuitively might choose to input the ideal (i.e., noiseless) gate set, it turns out that the ideal gate set usually contains degeneracies in its spectrum.  This means that a single germ in the germ set can amplify more gate set parameters (ones that are degenerate) than it otherwise could.  When the degeneracy is broken (e.g., when noisy gates are probed) that germ can no longer amplify all previously-degenerate parameters; some parameters are no longer amplified.

The solution to this problem historically \cite{Nielsen2020_GST} has been to provide the germ selection algorithm with not an ideal gate set, but rather one or more unitarily perturbed (from ideal) gate sets.  The resulting germ set is now robust against the effects of ``accidental'' degeneracies in the ideal gate set; we therefore call it a \textit{robust germ set}.

However, robust germ sets can be quite large, contributing to the overall cost of a GST experiment design.  In this work we therefore re-examine the decision to select germs for unitarily perturbed gate sets.  We refer to a germ set generated from only an ideal gate set as a \textit{standard germ set}.

Finally, we consider a germ set whose germs are just each of the bare gates in the gate set.  That is, for a gate set containing the gates $\set{G_k}$, the corresponding germ set is just $\germs=\set{G_k}$.  We call this the \textit{bare germ set}.  We explicitly do not expect the bare germ set to be AC; we use it in Sec~\ref{sec:one_and_two_q_fisher_info_spectra} to illustrate the behavior of deficient GST experiment designs.

\subsubsection{Fiducial pair reduction}

Another way to shrink the size of a GST experiment design is to modify how the fiducial preparation and measurement subcircuits are employed.  However, unlike in the case of germ reduction, we do not directly modify the preparation fiducial set $\pfids$ or the measurement fiducial set $\mfids$.  We instead note that every GST circuit begins with a preparation fiducial $F_j\in\pfids$ and ends with a measurement fiducial $H_i\in\mfids$, so each GST circuit has a \textit{fiducial pair} $(F_j,H_i)$ associated with it.  The ``standard'' GST experiment design contains all fiducial pairs for all germs at all powers.  That is, it contains all circuits of the form
\begin{equation}
    F_j g_k^{p(k,l)} H_i
\end{equation}
for $F_j\in\pfids$, $H_i\in\mfids$, $g_k\in\germs$, and $l\in[1..\log_2 L_\text{max}]$.  Therefore, for any given germ $g_i$ and value of $l$, the standard GST experiment design includes a number of circuits equal to $|\pfids|\cdot|\mfids|$.  We can reduce this experimental cost dramatically by removing pairs of fiducials for each $(g_k,l)$ pair.  We consider three schemes for \textit{fiducial pair reduction}, or FPR, detailed below.  All three fiducial pair reduction schemes are illustrated schematically in Fig.~\ref{fig:GR-FPR-overview}(a).

The first fiducial pair ``reduction'' scheme is not really a reduction scheme at all; it is simply the one used in standard GST experiment design.  All fiducial pairs are employed for all (germ, depth) pairs.  We call these the \textit{full fiducials}.  

The second fiducial pair reduction scheme identifies for each germ $g_i$ a set of fiducial pairs (i.e., a subset of $\pfids\times\mfids$) necessary to probe all of that germ's experimentally accessible parameters.  We call these the \textit{per-germ fiducials}.  Details on how per-germ fiducials are selected is provided in Sec.~\ref{sec:circuit_reduction_techniques}.

Finally, we consider a randomized approach to fiducial pair reduction.  Given a desired reduction factor $\alpha\in(0,1)$, for each (germ, depth) pair, we independently choose at random and without replacement, $\alpha\cdot|\pfids|\cdot|\mfids|$ fiducial pairs from $\pfids\times\mfids$.  We call these the \textit{random fiducials}.


\section{Results}
\label{sec:results}

In the following sections we present results demonstrating the precision scaling for GST estimates fit based on data from simulated noise models. To generate these numerical results, we first construct two classes of noise models. The first is a coherent-only noise model where the noise acting on each gate is purely unitary. In the error-generator framework \cite{rbk2022taxonomy} each instantiation of a noisy model will have weights for the elementary Hamiltonian generators sampled independently at random from a normal distribution with mean zero and standard deviation $\sigma$. The second class of noise models, coherent+depolarization, includes both unitary errors and uniform depolarizing errors. Here the unitary error is sampled at random exactly as described for the coherent-only model, but we additionally have a uniform depolarization errors following each gate with a fixed depolarization rate of $\eta$. 

For our one-qubit simulations we will apply these noise models to a gate set consisting of three gates labeled $G_i$, $G_{x_{\pi/2}}$ and $G_{y_{\pi/2}}$ (corresponding to idle, $R_X(\pi/2)$, and $R_Y( \pi/2)$ operations, respectively), along with a native state preparation in the $\ket{0}$ state and native measurement in the computational basis (we'll denote this as the XYI gate set for shorthand). The XYI gate set has $43$ parameters in total, of which $31$ are non-gauge parameters. For the two-qubit simulations the gate set consists of the gate $G_{x_{\pi/2}}$ and $G_{y_{\pi/2}}$ on each of the individual qubits along with the two-qubit Controlled-Z gate, with native state preparation in the $\ket{00}$ state and native measurement in the computational basis (we'll denote this as the XYCPHASE gate set for shorthand). The XYCPHASE gate set has $1263$ parameters in total, of which $1023$ are non-gauge parameters. 

\subsection{Numerical Simulations}
\label{sec:1Q_2Q_ddist_results}

For one-qubit GST we generate for each of the two classes of noisy models described above $10$ random samples, using values of $\sigma=.01$ and $\eta=.001$. Each of the randomly sampled noise models is used to generate $10$ simulated datasets each, for a total of $200$ datasets overall. Each dataset is generated with respect to the set of circuits for the largest of the experiment designs under test, which for one-qubit is the \edesign{robust}{full} experiment design, with $1000$ shots per circuit. When fitting smaller subdesigns we simply truncate the dataset to only include the relevant circuits and fit our model with respect to the truncated dataset. 

In order to evaluate the quality and precision scaling behavior of the gate set models fit by GST the diamond distance with respect to the true data-generating model is evaluated as a function of the maximum circuit depth $L$ (see Section \ref{sec:LSGST}). The iterative nature of the MLE implemented in standard GST allows for easily tracking the precision of a gate set estimates throughout the fitting process.

\begin{figure}[htbp]
    \centering
    \begin{minipage}{.32\textwidth}
    \includegraphics[height=2.15in]{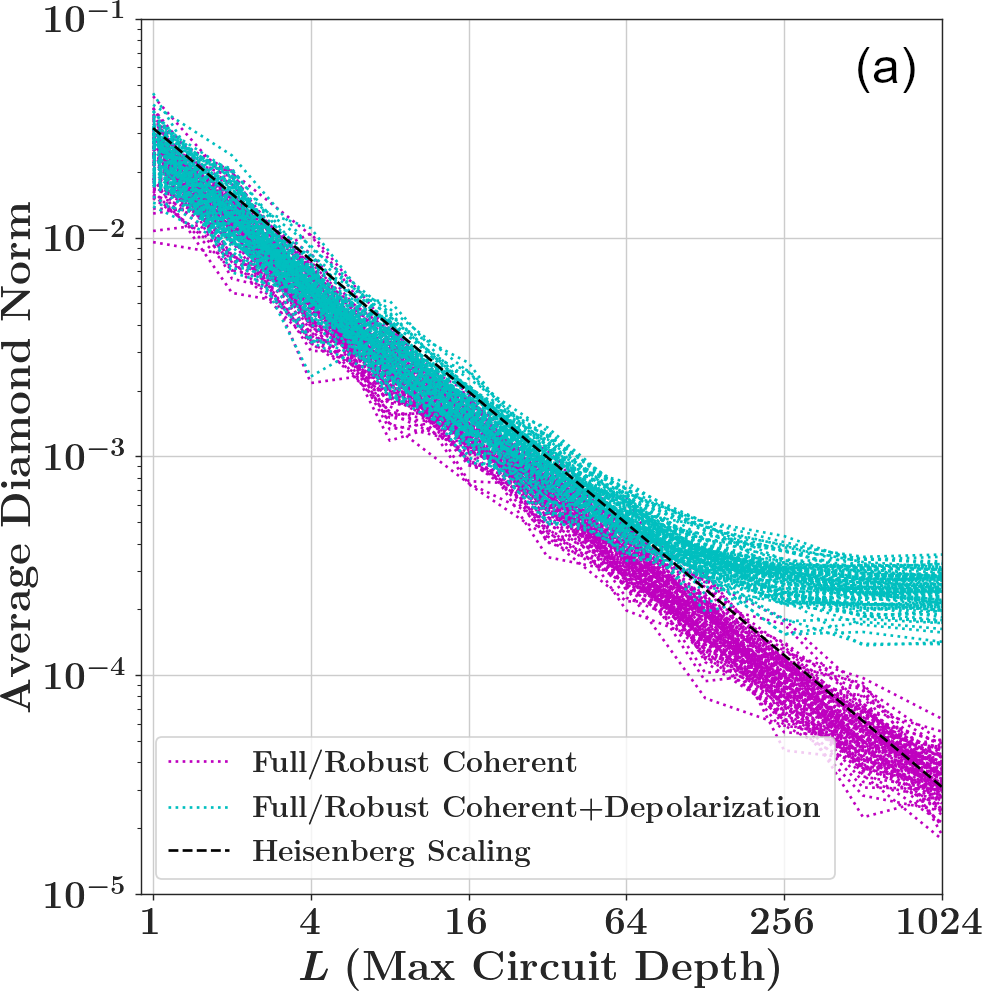}
    \end{minipage}
    \begin{minipage}{.32\textwidth}
    \includegraphics[height=2.15in]{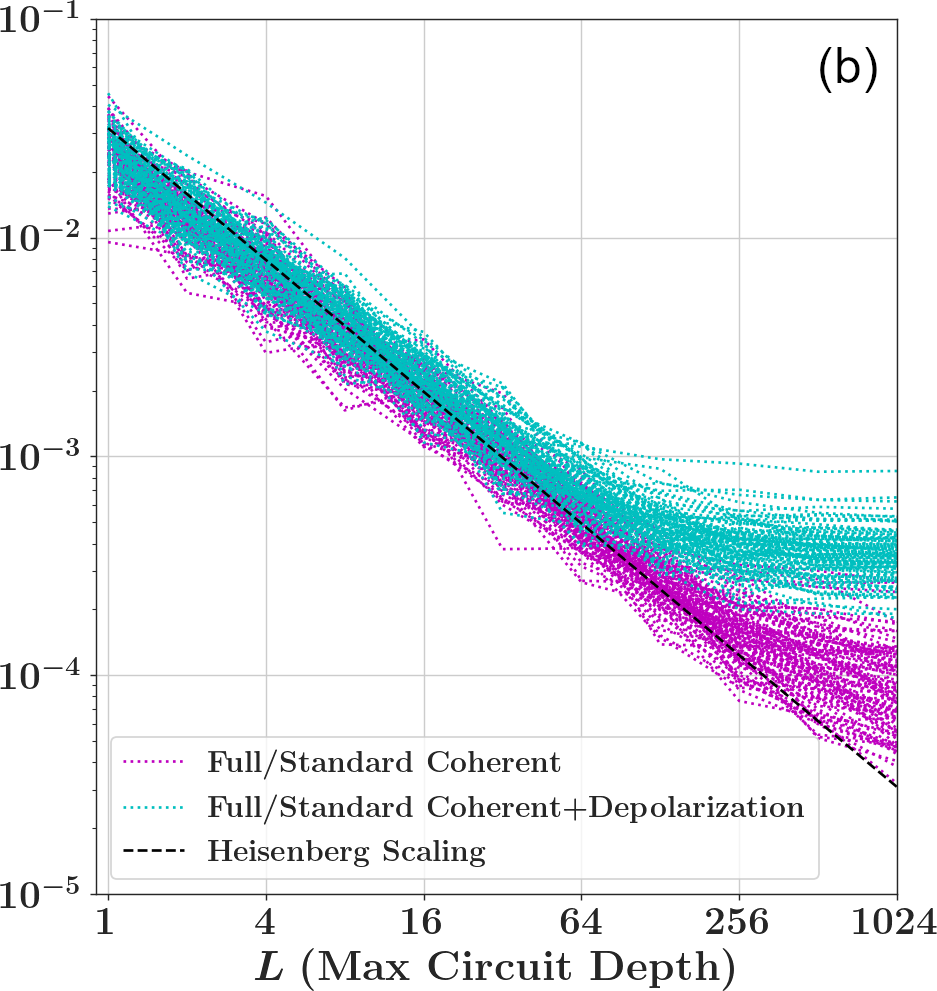}
    \end{minipage}
    \begin{minipage}{.32\textwidth}
    \includegraphics[height=2.15in]{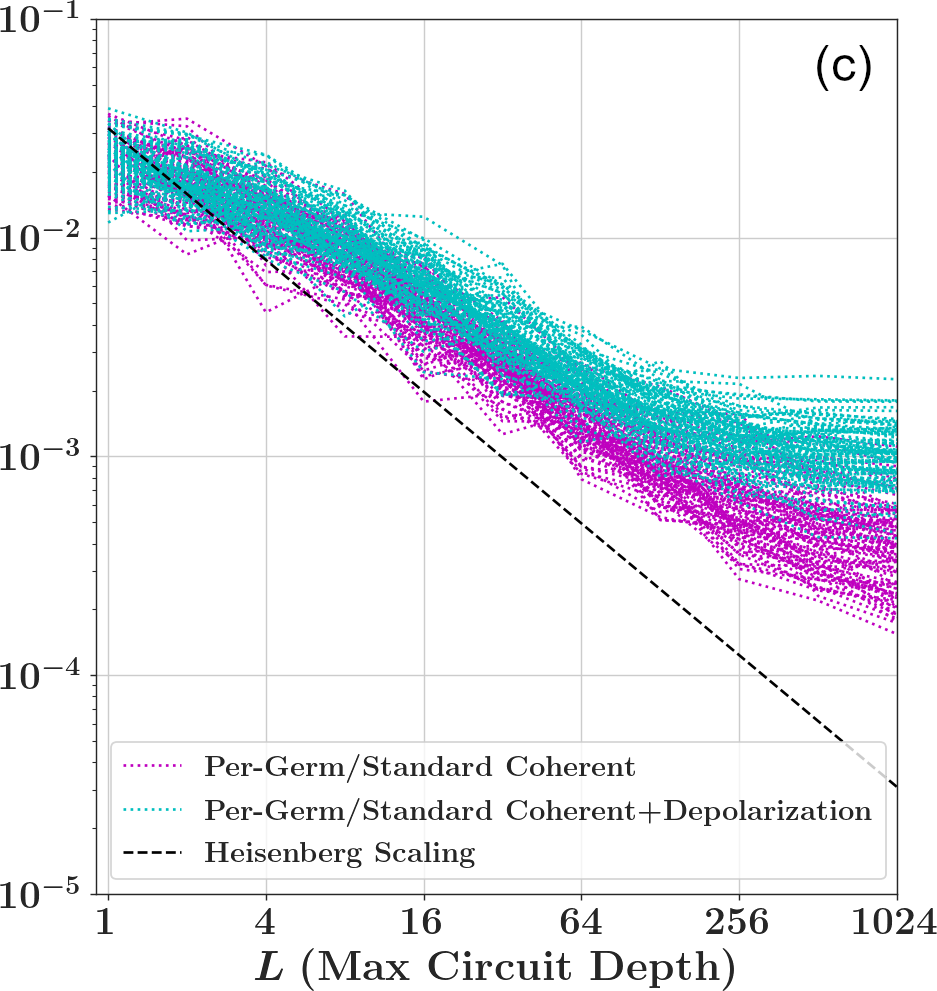}
    \end{minipage}
    \caption{Average gate set diamond distance between data-generating models and GST estimates thereof versus maximum circuit depth for two classes of single-qubit noise models. The coherent-only noise class includes only unitary errors, while the coherent+stochastic class includes both unitary errors and general stochastic dephasing. For both classes, 10 sets of parameters are sampled at random and, for each of these 10 noise model instantiations, there are 10 independent datasets are simulated. (a) \edesign{Robust}{full} (b) \edesign{standard}{full} (c) \edesign{standard}{per-germ}. Note that for coherent-only error instances, as the experiment design shrinks (moving successively from panel (a) to (c)), estimate accuracy deviates from Heisenberg scaling.  When depolarizing error is added, Heisenberg scaling is decoherence-limited (in (a) and (b)).  In (c), the experiment design is sufficiently small that Heisenberg scaling is never fully achieved; however, accuracy still improves with maximum circuit depth and may still be a good choice given experimental constraints.}
    \label{fig:ddist_vs_L_full_lite_pergerm}
\end{figure}

In Figure \ref{fig:ddist_vs_L_full_lite_pergerm}(a)-(c) we plot traces of the average gate set diamond distance\footnote{The diamond distance is a gauge-variant quantity, and so some care must be taken in its use as a metric for precision. See Appendix \ref{app:gauge_invariance} for more details.} versus $L$ for the \edesign{robust}{full}, \edesign{standard}{full} and \edesign{standard}{per-germ} experiment designs respectively. In Figure \ref{fig:ddist_vs_L_full_lite_pergerm} we've separated out the traces corresponding to the coherent-only class of noise models and the coherent+depolarization class of noise models.

\begin{figure}
    \centering
    \includegraphics[width=.7\textwidth]{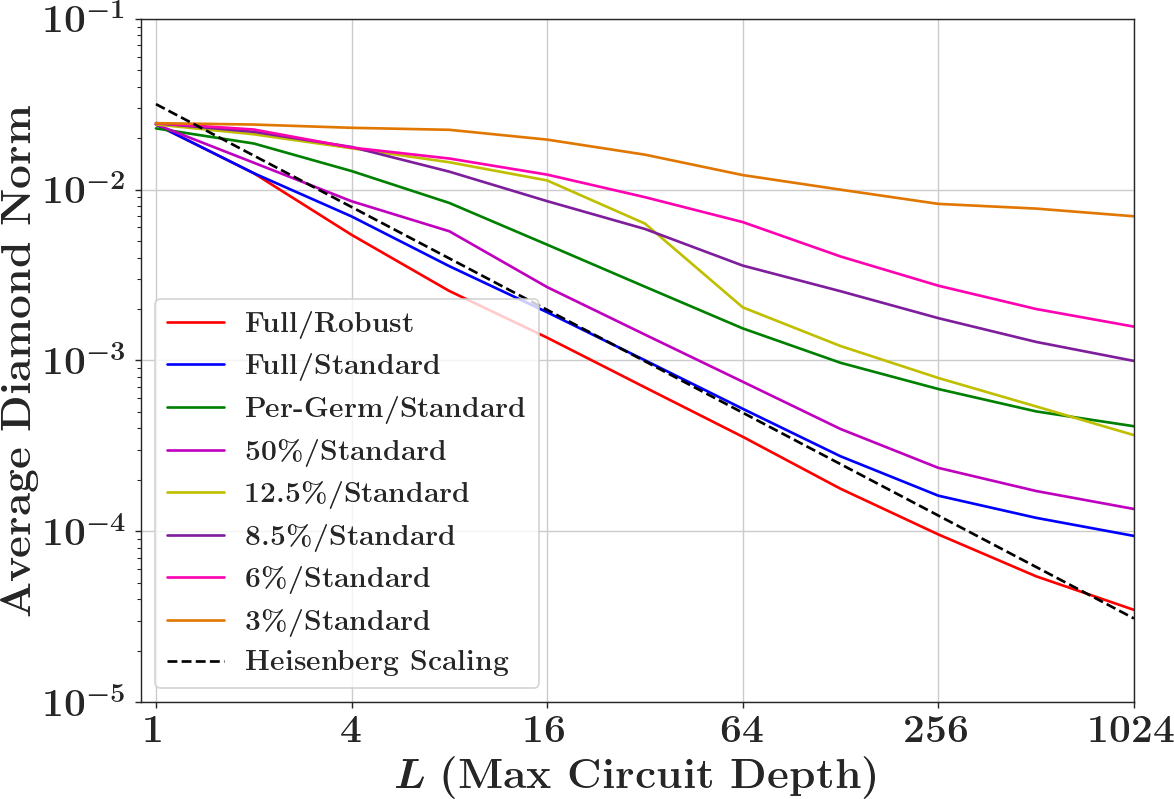}
    \caption{Average gate set diamond distances versus $L$ averaged over all single-qubit coherent-only noise model and dataset instances.}
    \label{fig:avg_ddist_vs_L_all_edesigns}
\end{figure}

Making direct visual comparisons of the diamond distance scaling behavior using the plots as in Figure \ref{fig:ddist_vs_L_full_lite_pergerm} is challenging with more than very few experiment designs, so we instead first perform an average over each of the models and datasets at each $L$. In Figure \ref{fig:avg_ddist_vs_L_all_edesigns} we plot these averages for all of the different experiment designs under consideration. As we expect to see that in the presence of stochastic error all of the experiment designs will have their sensitivity drop off once $L\sim \frac{1}{\eta}$, where $\eta$ is the stochastic error rate of out noise model, we include in Figure \ref{fig:avg_ddist_vs_L_all_edesigns} only the results for the coherent-only noise models. 

\begin{figure}
    \centering
    \includegraphics[width=.75\textwidth]{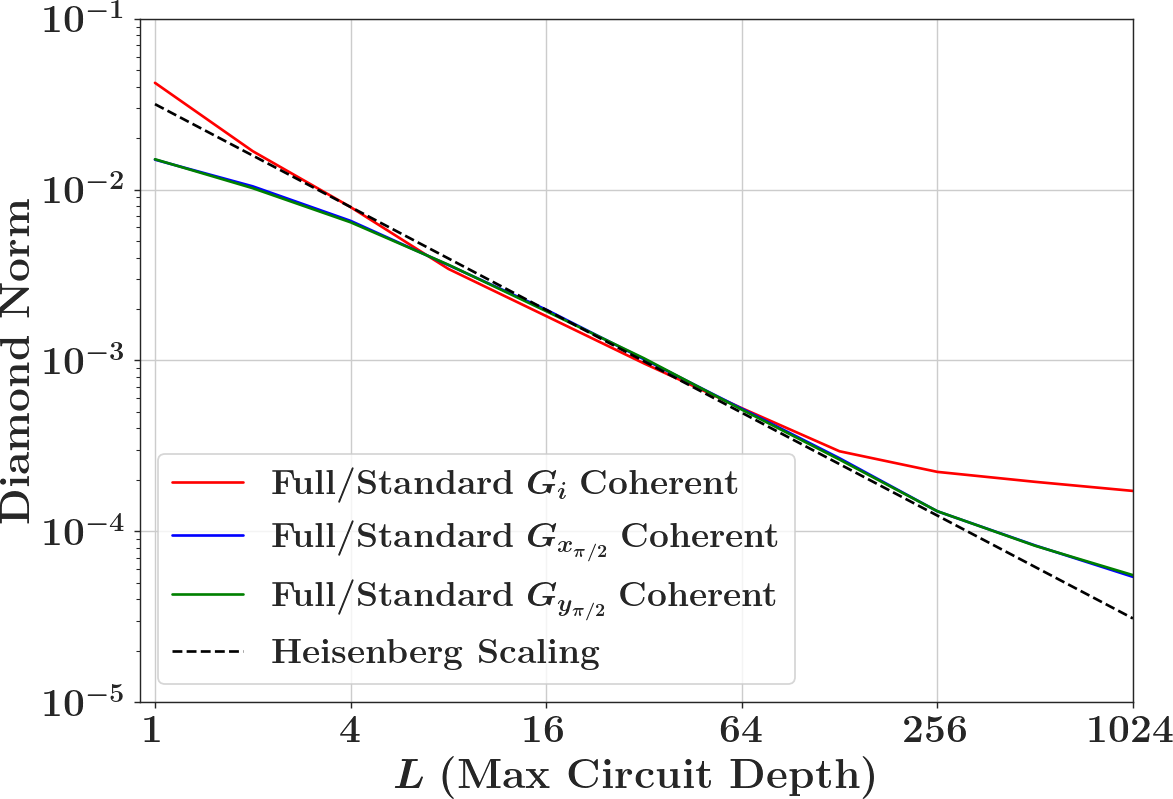}
    \caption{Per-gate diamond distance versus $L$ averaged over all of model and dataset instances for the \edesign{standard}{full} experiment design. At large $L$ our experiment design loses sensitivity to higher order perturbations to the idle gate resulting in a plateauing of the average diamond distance with respect to the true data-generating model for this gate.  (Note that the traces for $G_{x_\pi/2}$ and $G_{y_\pi/2}$ are nearly identical.}
    \label{fig:per_gate_avg_ddist_full_lite}
\end{figure}

An important, perhaps subtle, feature which can be gleaned from Figure \ref{fig:avg_ddist_vs_L_all_edesigns} is that there is a reduction in the sensitivity of our experiment designs at large $L$ when shifting from the robust germ set to the standard germ set. This is particularly evident in the results for the \edesign{standard}{full} experiment design starting at $L=256$. This reduction in sensitivity is primarily attributable to a reduction in sensitivity in the idle gate, as seen in Figure \ref{fig:per_gate_avg_ddist_full_lite} where the three contributions to the \edesign{standard}{full} diamond distance from each gate are separated. In Figure \ref{fig:per_gate_avg_ddist_full_lite} we see Heisenberg-like precision scaling for all three gates up to a depth of roughly $L=128$, after which the idle gate experiences a significant drop off in its precision. 

This plateauing at larger $L$ of the sensitivity of the \edesign{standard}{full} design (and more generally for experiment designs using the standard germ set) for the coherent-only noise model is a well understood consequence of relaxing the amplificational completeness constraint and will be discussed in Section \ref{sec:germ_reduction}. However, the fact that the sensitivity to errors in the idle gate plateaus at long $L$ values is fine so long as one is only interested in learning the dominant errors. Given the average magnitude of the coherent errors in our randomly sampled noise models, values of $L$ between $128$ and $256$ are exactly where we would expect the subdominant tilt component of the errors to start echoing out, thereafter becoming swamped by the dominant over-rotation component. 

\begin{figure}[htbp]
    \centering
    \includegraphics[width=.75\textwidth]{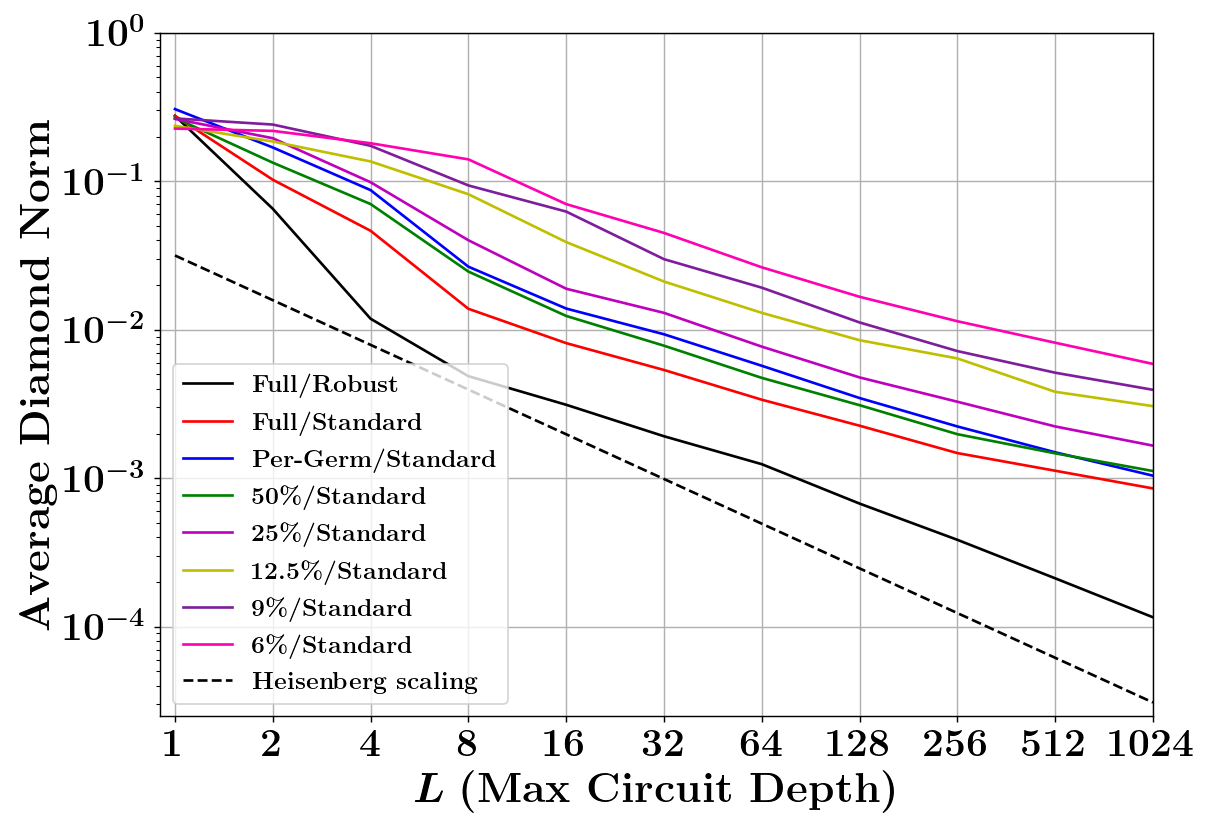}
    \caption{Average gate set diamond distances vs. max $L$ for the two-qubit XYCPHASE gate set under a coherent-only error model using various reduced experiment design. The scaling of the diamond distance for the model with additional depolarization is comparable.}
    \label{fig:two_qubit_ddist_scaling_all_edesigns}
\end{figure}

Due to the substantial increase in computational cost associated with performing the GST fits in the two-qubit case we only sample a single random model and data set for each of the two classes of noise models. In Figure \ref{fig:two_qubit_ddist_scaling_all_edesigns} we plot the average gate set diamond distance versus $L$ for the \edesign{standard}{full} and \edesign{standard}{per-germ}, as well as for \edesign{standard}{random} experiment designs for a number of different retention fractions. For the \edesign{standard}{full} and \edesign{standard}{per-germ} designs the scaling of the average gateset diamond distance with $L$ indicates Heisenberg-like precision scaling. At sufficiently large values of the retention fraction we likewise see Heisenberg-like scaling for the \edesign{standard}{random} designs but, as seen for one-qubit, for very small values of the retention fraction we eventually fail to achieve this error scaling.

\begin{figure}[htbp]
    \centering
    \begin{minipage}{.45\textwidth}
        \includegraphics[width=\textwidth]{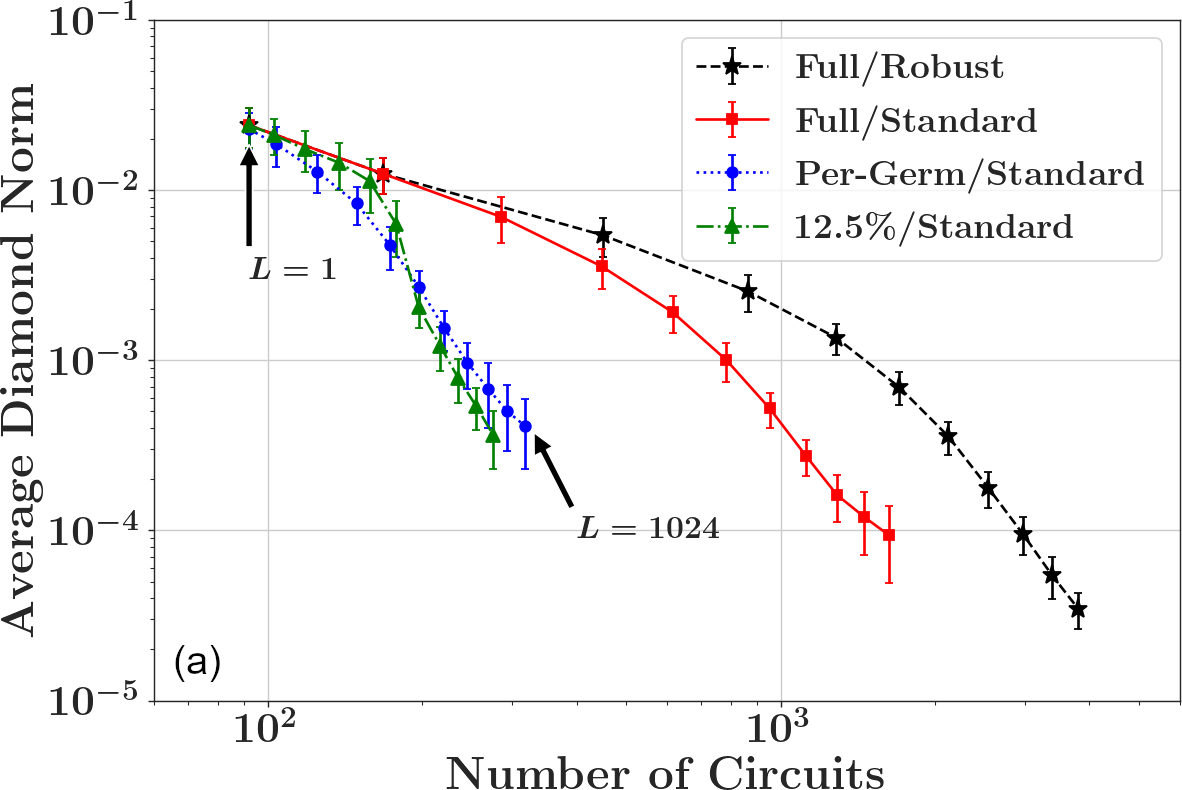}
    \end{minipage}
    \hspace{.5em}
    \begin{minipage}{.45\textwidth}
        \includegraphics[width=\textwidth]{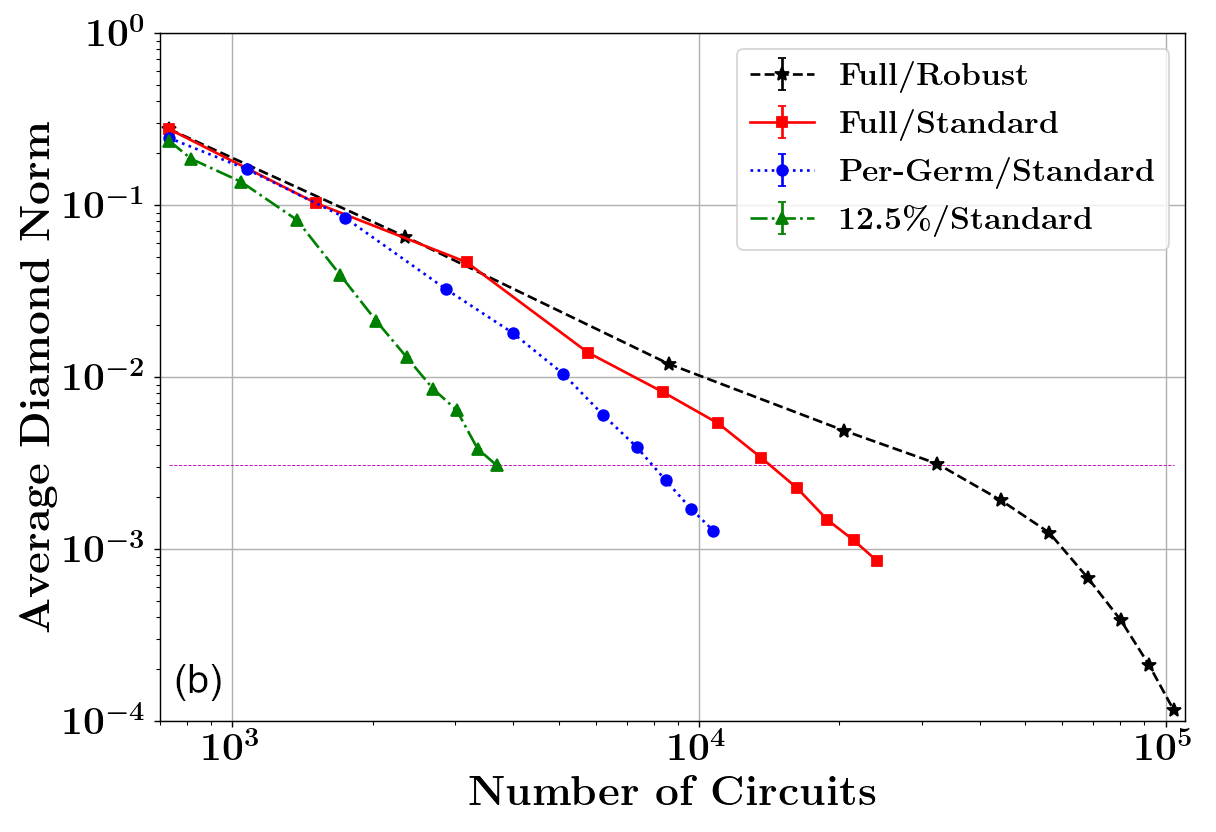}
    \end{minipage}
    \caption{Comparison of the average gate set diamond distance versus the total number of circuits used under a coherent-only error model for the (a) one-qubit XYI gate set and (b) two-qubit XYCPHASE gate set. Each point corresponds to a particular maximum circuit depth increasing from left-to-right as labeled.  We see that for both one-qubit and two-qubit GST, smaller experiments can achieve a particular level of accuracy with far fewer total circuits than the unreduced experiments.}
    \label{fig:one_and_two_qubit_ddists_vs_num_ckts}
\end{figure}

While understanding the sensitivity of our GST experiment designs as a function of the circuit depth is important, from a resource requirement standpoint it is also important to understand the cost in terms of the total number of circuits required overall for an experiment. In Figures \ref{fig:one_and_two_qubit_ddists_vs_num_ckts}(a) and (b) we compare the average gate set diamond distance as a function of total number of circuits for the one and two-qubit simulations respectively. These results are for the coherent-only error model and for a fixed number of shots per-circuit, independent of the experiment design. Figure \ref{fig:one_and_two_qubit_ddists_vs_num_ckts} reveals an as-yet unappreciated feature of using reduced experiment designs. While as a function of the maximum circuit depth reduced designs are less sensitive than their non-reduced counterparts, it is nonetheless possible to achieve comparable final precision using substantially fewer circuits overall.

For one-qubit GST, the \edesign{standard}{per-germ} design achieves a final mean average gate set diamond distance of $4.12\times 10^{-4}$ at $L=1024$. This is comparable to the mean average gate set diamond distance of $3.57 \times 10^{-4}$ the \edesign{robust}{full} design achieves at $L=64$. However, the \edesign{standard}{per-germ} design uses only $338$ circuits in total, while the \edesign{robust}{full} design up to $L=64$ uses $2122$ circuits, a factor of $\sim 6.3$ reduction, which increases all the way up to a factor of $\sim 7.7$ reduction if we instead compare to the \edesign{standard}{random} design with a keep fraction of $12.5\%$.

Similarly, for two-qubit GST the \edesign{standard}{per-germ} experiment design achieves a final average gate set diamond distance value of $1.28\times 10^{-3}$ at $L=1024$, which is comparable to the value of $1.12 \times 10^{-3}$ achieved by the \edesign{standard}{full} design at $L=512$ or the value of $1.24 \times 10^{-3}$ achieved by the \edesign{robust}{full} design at $L=64$. However, the \edesign{standard}{per-germ} utilizes only $10725$ circuits as compared to the $21433$ circuits used by the \edesign{standard}{full} design and $56254$ circuits used by the \edesign{robust}{full} design; a savings of $\sim 50$\% and $\sim 81$\%, respectively. For the two-qubit case we find even greater relative savings are possible when using the \edesign{standard}{random} design with a keep fraction of $12.5\%$, with that design achieving a final precision at $L=1024$ of $3.10 \times 10^{-3}$ in average gate set diamond distance, comparable to the values of $3.38 \times 10^{-3}$ at $L=64$ and $3.12 \times 10^{-3}$ at $L=16$ the \edesign{standard}{full} and \edesign{robust}{full} designs, respectively. The \edesign{standard}{random} design, however, uses only a total of $3683$ circuits to achieve this precision, while the \edesign{standard}{random} design uses $13606$ and the \edesign{robust}{full} design uses $32380$, giving a factor of $\sim 3.7$ and $\sim 8.8$ reduction, respectively.

While we've found significant experimental savings in the particular experiment design and noise parameter regimes investigated in this work there remain a number of rich and interesting questions regarding the generalizability of these results to new settings. For example, we can see from Figure \ref{fig:one_and_two_qubit_ddists_vs_num_ckts}(a) an inflection point in the scaling of the number of circuits required for a given precision for the experiment designs using the Standard germ set (consistent with the theory from Section \ref{sec:germ_reduction}) that does not appear for the Robust germ set, making the existence of a crossover point plausible at very large maximum circuit depths. Similarly, the impacts and limitations imposed by decoherence on practically achievable experimental savings for a given target precision remains to be investigated.

\section{Fisher Information}
\label{sec:fisher_information}

In the previous section we presented numerical evidence supporting our claims regarding the efficacy of FPR and relaxing the AC constraint, but could we have done so without having directly simulated GST, or potentially worse, running  experiments on quantum hardware and hoping for the best? In this section we will show how the Fisher information can be used to analyze the sensitivity of proposed experiment designs without ever running a single circuit, allowing us to evaluate \textit{a priori} the various germ and fiducial pair schemes considered in this paper.

The Fisher information \cite{ly2017tutorial} is a way of quantifying how sensitive a random variable is to a particular unknown parameter or parameters.  We may think of an experiment with a non-deterministic outcome (say, a quantum circuit) as a method for generating samples from a random variable; the random variable's distribution is governed by some underlying parameter or parameters (say, the model describing a gate set).  If we understand the relationship between the parameters and the experiments that generate samples from the random variables (e.g., Eq.~\eqref{eqn:prob}), then we can use the Fisher information to determine how sensitive our collection of experiments is to the underlying parameters we wish to estimate.  Thus the Fisher information is a good way of evaluating if a collection of experiments is sensitive to all estimable parameters.  Additionally, if an experiment design consists of nested collections of experiments (as is the case with GST \footnote{We may think of each increment in $L$ as ``growing'' the GST experiment design.  For a given germ and fiducial pair scheme, the experiment design for $L=2^{k+1}$ contains all the circuits in the $L=2^k$ experiment design.}), then we can examine the Fisher information of successive experiment designs to see how parameter sensitivity improves with each successive generation of experiments.  We therefore can use Fisher information to determine not just if a GST experiment design is sensitive to all the (non-gauge) parameters in a gate set, but also if that experiment design is maximally sensitive (i.e., achieving Heisenberg scaling in accuracy).

In Section \ref{sec:fisher_info_definition} we provide the definition of the Fisher information and describe its properties. In Section \ref{sec:fisher_information_gst} we explain exactly how the Fisher information is calculated with respect to GST experiment designs. Section \ref{sec:fisher_info_cramer_rao} introduces the Cram\'er-Rao bound, which underpins the subsequent Fisher information analysis of one- and two-qubit experiment designs in Section \ref{sec:one_and_two_q_fisher_info_spectra}.

\subsection{Definition and Properties}
\label{sec:fisher_info_definition}
We begin our discussion of the Fisher information with its formal definition.  Consider a random variable, $X$, whose probability density function is given by the conditional distribution $f(x|\theta)$ where $\theta \in \Theta$ is a parameter within the parameter space $\Theta$. Now, consider the task of estimating the true value of $\theta$, $\theta^*$, given observations $x$ from the probability distribution $f(x|\theta^*)$. At the core of maximum likelihood estimation (MLE) is the idea that if our estimate of $\theta^*$, $\hat{\theta}$, is equal to or close to $\theta^*$, then the value of the likelihood for the observation $x$, $\mathcal{L}(x|\hat{\theta})$, should be large. For the sake of convenience we typically choose to work with the log-likelihood $l(x|\theta)=\log(\mathcal{L}(x|\theta))$, and we likewise expect that estimates $\hat{\theta}$ close to $\theta^*$ have a large log-likelihood.

The quantity $\frac{\partial}{\partial \theta} l(x|\theta)$ is called the score, and it is a measure of the sensitivity of the log-likelihood to changes in the value of $\theta$. In one-dimension the Fisher information, $I(\theta)$, is defined as the variance of the score with respect to the possible observations $x$ evaluated at some value of $\theta$,

\begin{equation}
    I(\theta) = \mathbb{E}_x \left[ \left( \frac{\partial}{\partial \theta} l(x|\theta) \right)^2 \right].
\end{equation}

\noindent In the multi-dimensional case where $\vec{\theta}$ is instead a vector valued quantity, the score is given by the gradient of the log-likelihood, $\nabla l(x|\vec{\theta})$, and the Fisher information is now the covariance matrix of the score. Under suitable regularity conditions the elements of this covariance matrix can be written in terms of the Hessian of the log-likelihood evaluated at some point in parameter space $\vec{\theta}$

\begin{equation}
     I_{i,j}(\vec{\theta}) = -\mathbb{E}_x\left[\frac{\partial^2}{\partial \theta_i \partial \theta_j}  l(x|\vec{\theta}) \right].
     \label{eqn:fisher_info_def_hessian}
\end{equation}

\noindent The Fisher information tells us how sharply the log-likelihood varies as we perturb the values of the parameters. Put another way, the Fisher information measures how much information an observation of the random variable $X \sim f(x|\theta)$ tells us on average about the unknown value of $\theta$.

\subsection{Calculating the Fisher Information for GST}
\label{sec:fisher_information_gst}
 
At the highest level, a GST experiment consists simply of a set of circuits $\mathcal{C}$ where for each circuit $c \in \mathcal{C}$ we collect $N_c$ clicks. These $N_c$ clicks are arranged to form a vector $\vec{n}_c$, aggregated by the $N_o$ different possible circuit outcomes for $c$. Given a model for a gate set with some associated parameter vector $\vec{\theta}$, we can calculate the predicted outcome probability distribution for a circuit $p_c(\vec{\theta})$.  In this work we use a very simple model parameterization, in which each element of $\vec{\theta}$ corresponds to a single element of a gate's Pauli transfer matrix \cite{Greenbaum15}. However, we emphasize that other model parameterizations could be used; details of the model aren't all that important for this purpose except insofar as we're able to use that model to predict the outcome probabilities for every circuit in $\mathcal{C}$. The likelihood that the model with parameter vector $\vec{\theta}$ generated some observed data is the probability of sampling that set of counts from a multinomial distribution with outcome probabilities $p_c(\vec{\theta})$,

\begin{equation}
    \mathcal{L}_c (\vec{n}|\vec{\theta})= \frac{N_c!}{\vec{n}_c!} \prod_{i=0}^{N_o-1} p_{c,i}^{n_{c,i}}(\vec{\theta}),
    \label{eqn:ckt_likelihood}
\end{equation}

\noindent where $\vec{n}_c!= n_{c,0}! n_{c,1}! \cdots n_{c,N_o-1}!$, and where $p_{c,i}$ and $n_{c,i}$ are the predicted probability and empirical count for the $i$\textsuperscript{th} circuit outcome respectively. The corresponding log-likelihood for this circuit is given by

\begin{equation}
l_c(\vec{n}|\vec{\theta}) = \log(N_c!) - \sum_{i=0}^{N_o-1} \log(n_{c,i}!) +  \sum_{i=0}^{N_o-1} n_{c,i} \log(p_{c,i}).
\label{eqn:ckt_log_likelihood}
\end{equation}

\noindent Recall from Equation \ref{eqn:fisher_info_def_hessian} that the elements of the Fisher information matrix are given by the Hessian 
$I_{i,j}(\vec{\theta}) = -\mathbb{E}_{\vec{n}}\left[\frac{\partial^2}{\partial \theta_i \partial \theta_j}  l(\vec{n}|\vec{\theta}) \right]$. Note that the components of the Hessian of the log-likelihood related to the predicted outcome probabilities of the model are data-independent. Plugging in our expression for the log-likelihood in Equation \ref{eqn:ckt_log_likelihood} into the Hessian and taking the expectation value with respect to the observed data gives us the elements of the Fisher information matrix for circuit $c$,

\begin{equation}
    [I_c]_{i,j}= N_c \sum_{i=0}^{N_o-1} \left( \frac{1}{p_{c,i}} \frac{\partial p_{c,i}}{\partial \theta_m} \frac{\partial p_{c,i}}{\partial \theta_n} -  \frac{\partial^2 p_{c,i}}{\partial \theta_m \partial \theta_n} \right),
\end{equation}

\noindent where we have used $\mathbb{E}_{\vec{n}} [n_{c,i}]= N_c p_{c,i}$. The full Fisher information matrix can be written

\begin{equation}
    I_c= N_c \sum_{i=0}^{N_o-1} \left( \frac{1}{p_{c,i}} (\vec{\nabla} p_{c,i}) (\vec{\nabla}p_{c,i})^\top - H_i \right),
\end{equation}

\noindent where $H_i$ is the Hessian of $p_{c,i}$ with respect to the vector of parameters $\vec{\theta}$. As the Hessian is a linear operation, and the log-likelihood is additive with respect to multiple circuits, so too is the Fisher information matrix. That is, the Fisher information matrix for an entire experiment design is simply the sum of the Fisher information matrices for each constituent circuit,

\begin{equation}
    I_{\mathcal{C}}=\sum_{c \in \mathcal{C}} I_c.
\end{equation}

\subsection{The Cram\'er-Rao Bound}
\label{sec:fisher_info_cramer_rao}

What do we gain by evaluating our candidate experiment designs via the Fisher information? Among the most powerful properties of Fisher information based analysis is that it places strict lower-bounds on the minimum achievable variance on our estimates of model parameters for any given experiment design. Critically, these lower bounds inform us of our ability to learn the parameters of our model with a given experiment design \textit{before} running a single circuit in the lab. This all follows as a direct consequence of the Cram\'er-Rao bound which states that, if $\hat{\theta}$ is an unbiased estimator of the true value of the model parameters $\vec{\theta^*}$, then

\begin{equation}
    \Sigma(\hat{\theta}) \geq I(\vec{\theta}^*)^{-1},
    \label{eqn:cramer-rao}
\end{equation}

\noindent where $\Sigma(\hat{\theta})$ is the covariance matrix of our estimates and $I(\vec{\theta^*})^{-1}$ is the inverse of the Fisher information matrix. There is a minor catch, in that the Fisher information in the Cram\'er-Rao bound is evaluated at the true value of the model parameters $\vec{\theta}^*$, which is of course generally unknown. However, if $\vec{\theta^*}$ is sufficiently close to our model's target values $\vec{\theta}_{\textrm{target}}$, i.e. the noise in our gate set is small, then it can be shown that any corrections to Equation \ref{eqn:cramer-rao} are likewise small and can be safely ignored. 

The covariance matrix $\Sigma(\hat{\theta})$ quantifies our uncertainty in the estimate of $\vec{\theta^*}$, and in parameter space defines an ellipse with major and minor axis orientations given by the eigenvectors and lengths by the eigenvalues of the covariance matrix. This motivates the use of the spectra of the Fisher information to evaluate the sensitivity of experiment designs, as a large value for an eigenvalue of the Fisher information corresponds to a direction of parameter space about which we have low-uncertainty and vice versa. In Section \ref{sec:one_and_two_q_fisher_info_spectra} we will use the spectra of the Fisher information to evaluate the efficacy of a wide array of different combinations of FPR and GR for one- and two-qubit GST experiments.  

\subsection{One and Two-Qubit Fisher Information Spectra}
\label{sec:one_and_two_q_fisher_info_spectra}

The Cram\'er-Rao bound motivates the use of the spectra of the Fisher information matrix as a tool to analyze and compare the sensitivity of reduced experiment designs. In Section \ref{sec:circuit_reduction_techniques} we introduced the use of FPR and relaxation of the AC constraint as methods for significantly decreasing the size of experiment designs without substantial loss of sensitivity. In Section \ref{sec:1Q_2Q_ddist_results} we presented numerical evidence supporting these claims, but we can actually use the Fisher information to analyze the efficacy of these techniques directly.

\begin{figure}[tbp]
	\centering
	\begin{minipage}{.35\textwidth}
		\includegraphics[height=1.95in]{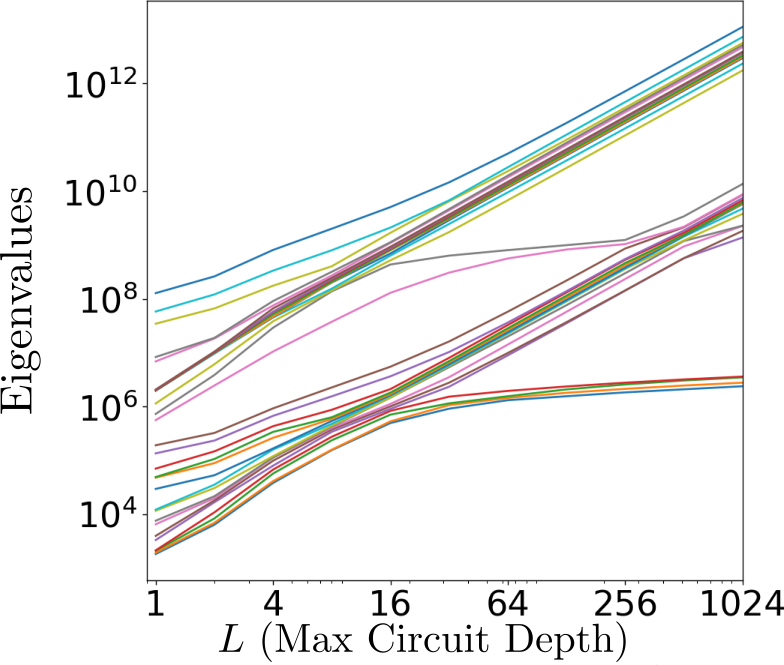}
	\end{minipage}
	\begin{minipage}{.31\textwidth}
		\includegraphics[height=1.95in]{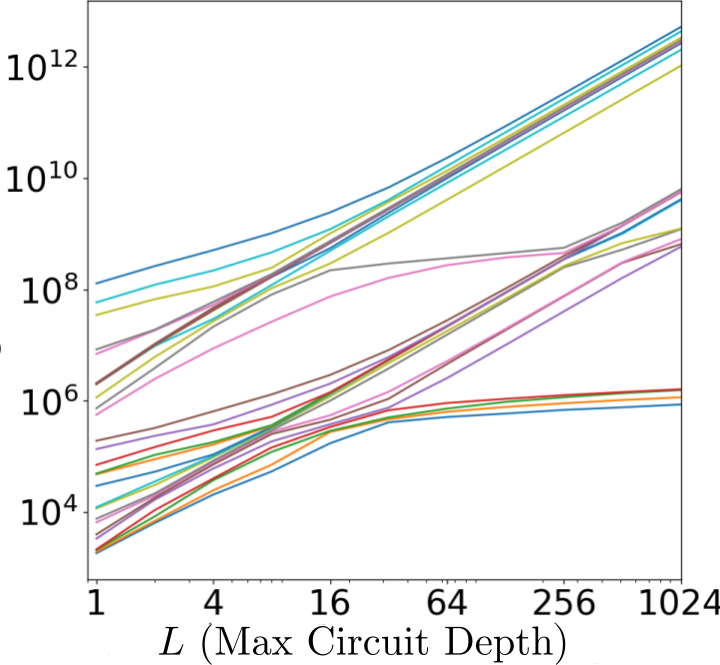}
	\end{minipage}
	\hspace{.25em}
	\begin{minipage}{.31\textwidth}
		\includegraphics[height=1.95in]{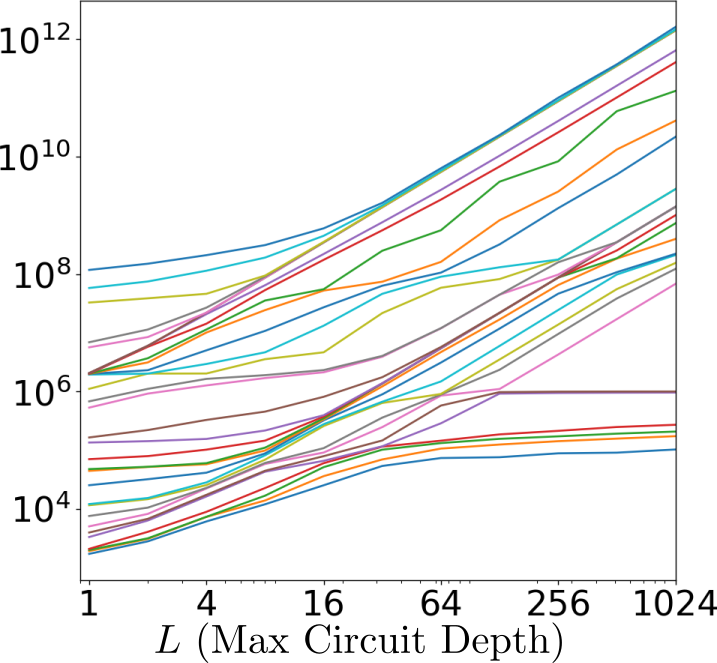}
	\end{minipage}
	\caption{Spectra of the Fisher information matrix as a function of course-grained circuit depth for three different experiment designs. The (a) full fiducials/robust germs, (b) full fiducials/standard germs and (c) per-germ fiducials/standard germs experiment designs respectively.  Eigenvalues increasing with maximum circuit depth correspond to parameters in the model that the experiment continues to probe more accurately with higher depth.  Plateauing eigenvalues correspond to parameters we do not learn more about with increased circuit depth.}
	\label{fig:fisher_info_vs_l_spectra_3_main}
\end{figure}

In Figure \ref{fig:fisher_info_vs_l_spectra_3_main} the spectrum of the Fisher information matrices for the \edesign{robust}{full}, \edesign{standard}{full}, and \edesign{standard}{per-germ} designs are plotted versus $L$, the maximum circuit depth of the experiment design---note that the Fisher information matrices are cumulative, and so they incorporate all circuits with a depth up to and including $L$. As per the Cram\'er-Rao bound, an increase in the magnitude of the eigenvalues of the Fisher information matrix corresponds to a decreases in the variance of our estimates in the direction of the corresponding eigenvector. Recall that the one-qubit $XYI$ gate set has $31$ non-gauge parameters. As such, in order for an experiment design to be well-constructed the Fisher information matrix must have $31$ non-zero eigenvalues, with the remaining eigenvalues associated with gauge directions. In Figure \ref{fig:fisher_info_vs_l_spectra_3_main} we can see that for all three experiment designs, all but $6$ of the Fisher information eigenvalues increase linearly with respect to maximum circuit depth. This corresponds to a reduction in the variance of our estimates in the corresponding directions in parameter space. The $6$ plateauing eigenvalues correspond to SPAM-dominated directions in parameter space which, as SPAM is only ever performed once in any given circuit, cannot be amplified. This demonstrates explicitly that even after relaxing the AC constraint and applying per-germ FPR, the reduced one-qubit experiment designs maintain high sensitivity to all of the parameters of our model we expect to be able to amplify.

\begin{figure}[htbp]
	\centering
	\begin{minipage}{.45\textwidth}
		\includegraphics[height=2.5in]{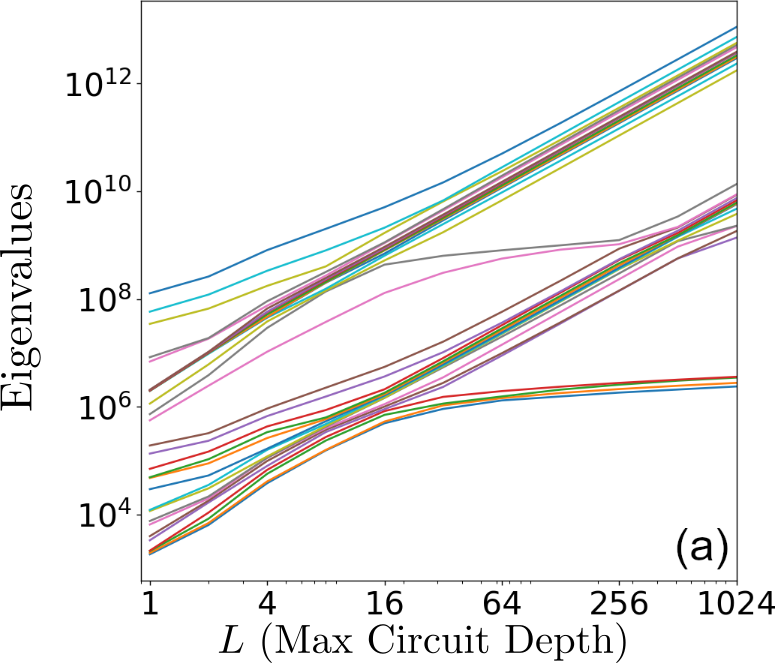}
	\end{minipage}
	\begin{minipage}{.45\textwidth}
		\includegraphics[height=2.5in]{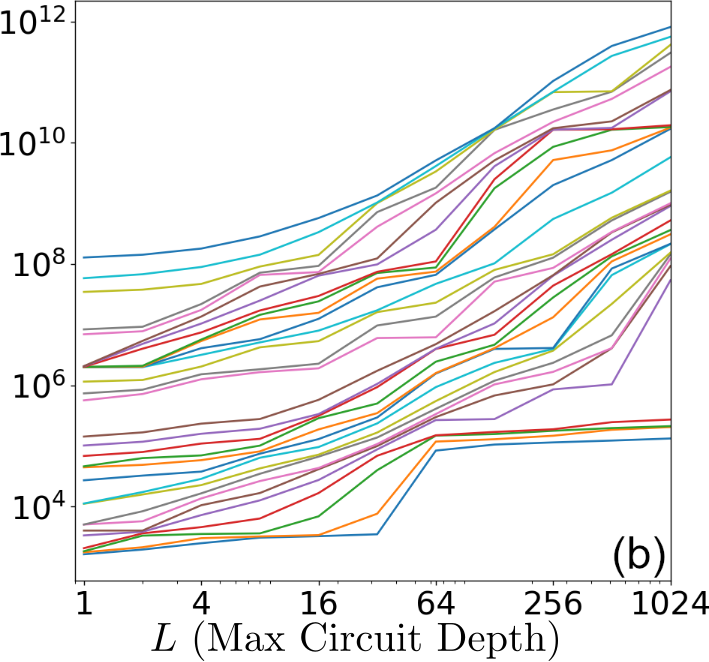}
	\end{minipage}\\[1em]
	\begin{minipage}{.45\textwidth}
		\includegraphics[height=2.5in]{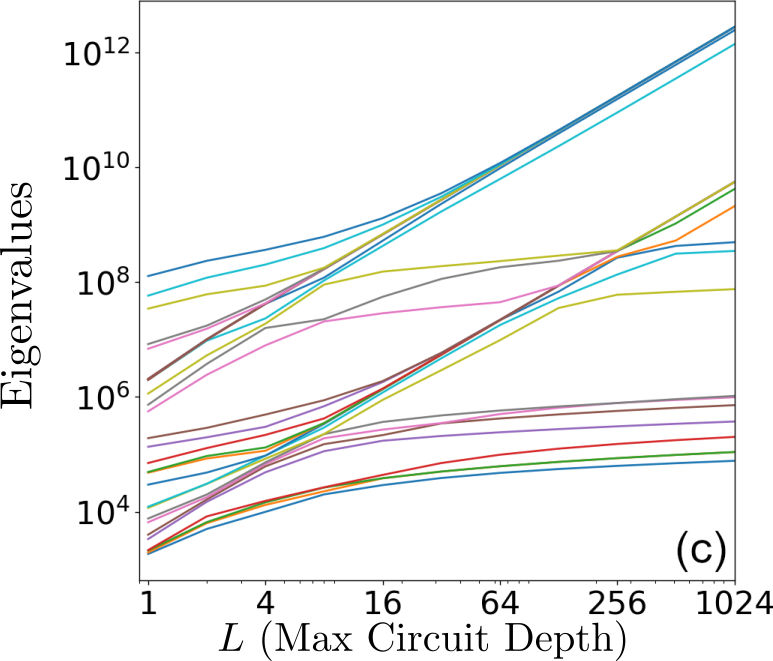}
	\end{minipage}
	\begin{minipage}{.45\textwidth}
		\includegraphics[height=2.5in]{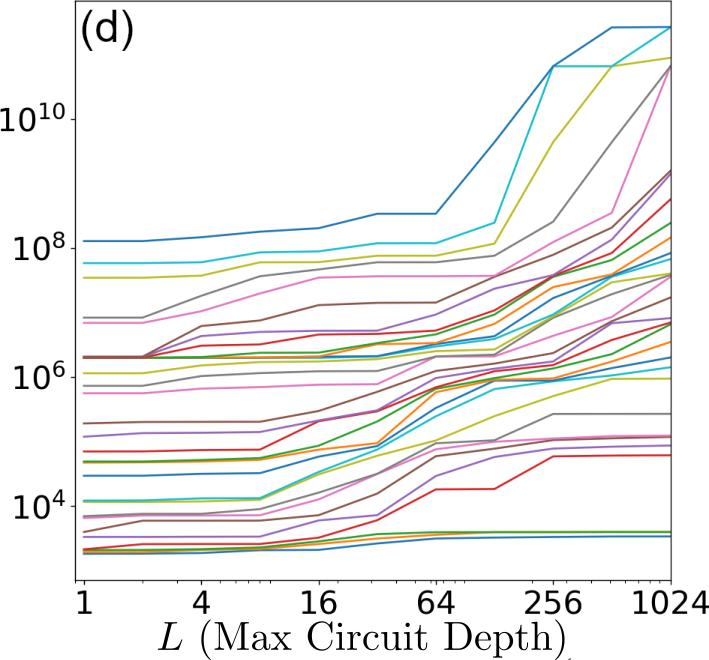}
	\end{minipage}
	\caption{Fisher information spectra vs L demonstrating failure modes for overly aggressive circuit reduction. (a) The baseline \edesign{robust}{full} design. (b) A \edesign{standard}{$\alpha=0.125$ random} design. (c) The \edesign{bare}{full} design.  Note the loss of sensitivity to additional parameters compared to the robust germ set. (d) A \edesign{standard}{$\alpha=.03$ random} design. As we remove more and more fiducial pairs we begin to gain only intermittent information about various parameters at each $L$ until, for sufficiently aggressive FPR, there are some parameters we learn almost nothing about.
	}
	\label{fig:fisher_info_failure_modes}
\end{figure}

However, the relaxation of the AC constraint and FPR can both fail when applied too aggressively. The failures modes for the relaxation of the AC constraint and FPR each have unique signatures in the spectra of the Fisher information matrix. In Figures \ref{fig:fisher_info_failure_modes}(a) and (c) we contrast behavior of the Fisher information matrices for an experiment design using the robust germ set with an experiment design using the bare germ set---the germ set consisting of only the base elements of the gate set---with a full set of fiducials for each. For the experiment design using the robust germ set, all of the expected directions in parameter space are amplified and we have a well-constructed experiment design. When using the bare germs, however, we can see in Figure \ref{fig:fisher_info_failure_modes}(c) that there are now at least $9$ eigenvalues which plateau at large $L$, more than the $6$ we expect coming from SPAM. These additional parameters are an indication of a lack of amplificational completeness for the bare-germ-only experiment design.

The signature of a failure due to overly aggressive FPR depends heavily on the details of the FPR algorithm used to generate the experiment design, so we'll focus here on failures resulting from random FPR. In Figures \ref{fig:fisher_info_failure_modes}(b) and (d) we plot the spectra of the FIMs for two \edesign{standard}{random} designs, using retention fractions of $12.5\%$ and $3\%$ respectively. For the one-qubit $XYI$ gate set there are $36$ fiducial pairs in the full set, so these retention fractions correspond to keeping $4$ and $1$ fiducial pairs at random at each germ-power, respectively. When applying random FPR with very low retention fractions, we generally no longer have an informationally complete set of fiducial pairs at at every germ-power. As such, we only gain sensitivity to most directions in parameter space intermittently and at random. This loss in sensitivity is to the point where, for a retention fraction of $3\%$, there are non-gauge directions in parameter space we are almost entirely insensitive to.

\begin{figure}[htbp]
    \centering
    \includegraphics[width=\textwidth]{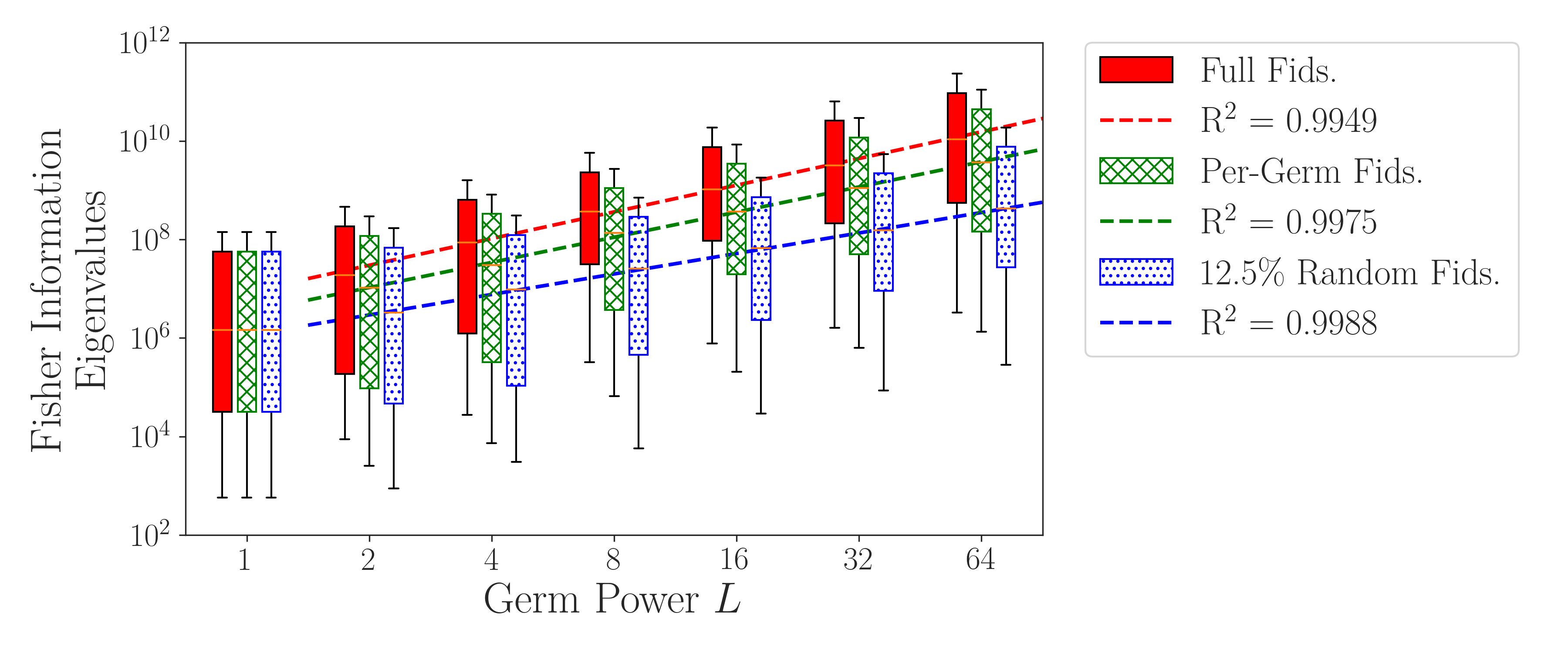}
    \caption{Distribution of Fisher information matrix eigenvalues as a function of maximum circuit depth $L$ for two-qubit GST.  Each color corresponds to a different choice of fiducial pair reduction for experiment design construction, in conjunction with the standard germ choice.  Trendlines are shown as a linear fits to distribution medians from $L=2$ onwards since all designs share the same $L=1$ circuits.}
    \label{fig:2Q_full_and_pergerm_lite_fisher_info}
\end{figure}

We also perform a comparable Fisher information analysis of reduced experiment designs for the two-qubit XYCPHASE gate set, though only for experiment designs up to a maximum depth of $L=64$ unlike in the simulations from Section \ref{sec:1Q_2Q_ddist_results}, as seen in Figure \ref{fig:2Q_full_and_pergerm_lite_fisher_info}. Here we plot the distributions of the Fisher information matrix eigenvalues as a function of $L$, as with over $1000$ non-gauge parameter looking at the traces directly is not tenable. 

Figure \ref{fig:2Q_full_and_pergerm_lite_fisher_info} compares the Fisher information spectra for the \edesign{standard}{full}, \edesign{standard}{per-germ} and \edesign{standard}{random} designs. All of these experiment designs can be categorized as well-constructed and achieve a linear increase in the magnitude of the distribution of the Fisher information eigenvalues overall, indicating Heisenberg-like scaling in accuracy for all of these experiment designs. 

\section{Methods}
\label{sec:methods}
 In this section we will give a more detailed description of the theory behind the construction of GST experiment designs and the techniques for reducing their size which were summarized in Section \ref{sec:standard_gst_edesign}. In Section \ref{sec:circuit_selection} we will discuss how informationally complete sets of fiducials and amplificationally complete sets of germs are constructed. In Section \ref{sec:circuit_reduction_techniques} we will then describe in detail the theory behind the two forms of fiducial pair reduction explored in this work, per-germ and per-germ power random. In Section \ref{sec:germ_reduction} we will discuss the relaxation of the AC constraint as a way to reduce the number of circuits in an experiment design, as well as the consequences of doing so when it comes to our precision scaling (which were previewed in Figure \ref{fig:per_gate_avg_ddist_full_lite}).

\subsection{Circuit selection}
\label{sec:circuit_selection}
In Section \ref{sec:standard_gst_edesign} we discussed the need for IC sets of fiducials and AC sets of germs for the performance of GST, how do we construct sets of circuits with these properties? An overview of the germ and fiducial selection methods from \cite{Nielsen2020_GST} is provided here, with more specific algorithms implemented in the \texttt{pyGSTi} \cite{nielsen2019python} software package provided in Appendix \ref{app:algorithms}.

\subsubsection{Fiducial selection}
\label{sec:fiducial_selection}

The purpose of fiducial selection is to find a good IC set of state prep and measurement fiducials, allowing us to probe all observable gate set parameters. A set of effective state preps is IC if and only if they span Hilbert-Schmidt space. Likewise, a set of effective measurements is IC if and only if they span the corresponding dual space. We can quantify this requirement by constructing gram matrices for the sets of effective state preparations and measurements,
\begin{equation}
    \begin{aligned}
    \tilde{\mathbbm{1}}^{\rho}_{ij} &= \bbraket{\rho_i^\prime}{\rho_j^\prime},\\
    \tilde{\mathbbm{1}}^{E}_{ij} &= \bbraket{E_i^\prime}{E_j^\prime}.\\
    \end{aligned} 
    \label{eqn:fiducial_gram_matrix}
\end{equation}

\noindent If either $\{ \kket{\rho_j'}\}$ or $\{\bbra{E_i'} \}$, as defined in Equations \ref{eqn:rho_j} and \ref{eqn:E_i}, fail to be informationally complete then the corresponding Gram matrix will fail to have sufficient rank. For $\{ \kket{\rho_j'}\}$ this requires that all $d^2$ eigenvalues, where $d$ is the dimension of the system's Hilbert space, are non-zero. For $\{\bbra{E_i'} \}$ we only require $d^2-1$ non-zero eigenvalues due to conservation of probability. Moreover, as the smallest non-trivial eigenvalue approaches $0$ this indicates that either $\{ \kket{\rho_j'}\}$ or $\{\bbra{E_i'} \}$ are becoming linearly dependent.
Fiducial selection can now be formulated as an optimization problem:
given a master list of prep and measure fiducials to choose from, select $\pfids$ and $\mfids$ such that the smallest non-trivial eigenvalue is maximized. Other objective functions can be, and are, used in practice, but the details of these are beyond the scope of this work.

\subsubsection{Germ selection}
\label{sec:germ_selection}
In germ selection we aim to find a set of germs such that together they amplify all possible observable parameters of the gate set. Given a set of germs, $\germs$, what directions in parameter space are amplified via repetition of those germs?

Let $\tau(g_k)^p$ denote the Pauli transfer matrix representation of the $k$\textsuperscript{th} germ for some germ-power $p$. Consider the Jacobian of $\tau(g_k)^p$ with respect to the model parameters,

\begin{equation}
    \vec{\nabla}_k^{(p)} \equiv \frac{1}{p} \frac{\partial (\tau(g_k)^p)}{\partial \vec{\theta}},
    \label{eqn:germ_jacobian}
\end{equation}

\noindent where $\vec{\theta}$ is a vector of gate set parameters of length $N_p$, and the derivatives are evaluated at the target value of $\vec{\theta}_0$ (as are all derivatives in this section). Note the inclusion of the normalizing term $\frac{1}{p}$, which is non-standard in the definition of the Jacobian, but the purpose of which will soon be clear. The Jacobian $\vec{\nabla}_k^{(p)}$ is a tensor-valued quantity with elements

\begin{equation}
    \left[\vec{\nabla}_{g_k}^{(p)}\right]_{lmn} = \frac{1}{p}\frac{\partial (\tau(g_k)^p)_{lm}}{\partial \theta_n}.
\end{equation}

\noindent Consider the slice through the tensor for a particular parameter $\theta_n$. Applying the product rule to $\frac{\partial (\tau(g_k)^p)}{\partial \theta_n}$ we can rewrite this slice as

\begin{equation}
    \frac{1}{p}\frac{\partial (\tau(g_k)^p)}{\partial \theta_n} = \frac{1}{p} \sum_{i=0}^{p-1} \left( (\tau(g_k))^i \frac{\partial \tau(g_k)}{\partial \theta_n} (\tau(g_k))^{-i} \right) \tau(g_k)^{p-1}.
    \label{eqn:tensor_slice_group_avg}
\end{equation}

\noindent In the limit as $p\rightarrow \infty$ the set of all powers of $\tau(g_k)$, $\set{\tau(g_k)^p}_p$, form an approximate representation of the cyclic group, and so written in this form we can recognize that Equation \ref{eqn:tensor_slice_group_avg} corresponds to a group average, or twirl, of $\frac{\partial \tau(g_k)}{\partial \theta_n}$. From Schur's lemma, this twirl can be shown to project $\frac{\partial \tau(g_k)}{\partial \theta_n}$ onto the commutant of $\tau(g_k)$ \cite{gambetta2012simultaneousRB}, the matrix subalgebra of all of the matrices which commute with $\tau(g_k)$. Consequently, for large $p$ the subspace of parameter space a germ is sensitive to, and can amplify, corresponds to the space spanned by the commutant of $g_k$. 

Let $\vec{\nabla}_k^{\infty}$ be the result of projecting each of the slices of $\vec{\nabla}_k^{(p)}$ onto the commutant of $g_k$. $\vec{\nabla}_k^{\infty}$ can be matricized, converting it from a $d^2\times d^2 \times N_p$ tensor to a $d^4\times  N_p$ matrix (which, for brevity, we'll likewise denote $\vec{\nabla}_k^{\infty}$). The right singular vectors of $\vec{\nabla}_k^{\infty}$ are directions in parameter space amplified by the germ $g_k$ in the large $p$ limit, and the corresponding singular values indicate how much amplification is done.

To consider the amplificational properties for an entire set of $N_g$ germs $\germs$, we concatenate the Jacobians for each of the individual germs,

\begin{equation}
    J_{\germs}^{\infty} = \begin{pmatrix}
    \vec{\nabla}_0^{\infty}\\
    \vec{\nabla}_1^{\infty}\\
    \vdots\\
    \vec{\nabla}_{N_g-1}^{\infty}
    \end{pmatrix}.
    \label{eqn:infl_germset_jacobian}
\end{equation}

The number of non-zero singular values, i.e. the rank of $J_{\germs}^{\infty}$, is the number of directions in parameter space amplified by the entire germ set in the large $p$ limit. When the rank $J_{\germs}^{\infty}$ equals the expected number of non-gauge model parameters then the germ set is AC.

Germ selection can now, like fiducial pair selection, be framed as an optimization problem: given a large set of candidate germs, select a subset of germs such that $J_{\germs}^{\infty}$ has a right singular rank equal to the number of amplifiable model parameters and is optimal with respect to an appropriate objective function. See Appendix \ref{app:algorithms} and \cite{Nielsen2020_GST} for more details on the germ selection algorithm.

\subsection{Circuit Reduction Techniques}
\label{sec:circuit_reduction_techniques}

In our description of the fiducial and germ selection procedures we required IC sets of prep and measure fiducials and AC sets of germs, respectively. It is, however, possible to weaken these requirements in order to reduce the overall cost of an experiment design. Intuitively, it makes sense that we should be able to do such a thing---a GST experiment design is massively overcomplete, i.e., the number of experimental degrees of freedom (independent circuit outcomes times number of circuits) is much larger than the number of parameters in a gate set. In this section, we will explain the theory behind fiducial pair reduction and the relaxation of the AC constraint, two methods that can (independently or jointly) shrink the GST experiment design without introducing large inaccuracies in the resulting gate set estimate.

In Section \ref{sec:FPR} we will discuss fiducial pair reduction (FPR), and in Section \ref{sec:germ_reduction} we will discuss the relaxation of the AC constraint.

\subsubsection{Fiducial Pair Reduction}
\label{sec:FPR}

When selecting fiducial pairs for GST we aim to identify IC sets of state preparation and measurement fiducials such that together they span all of Hilbert-Schmidt space. However, for any given germ we also know that only a subset of directions in parameter space are amplified. As such, for any given germ it is typically the case that there are some fiducial pairs which simply do not provide much, if any, practically useful information. This intuition forms the basis of a family of techniques called fiducial pair reduction (FPR), which aim to construct reduced sets of fiducial pairs which nonetheless allow for successfully running GST.

There are two ways to approach the process of applying FPR to an experiment design. In the first, called per-germ FPR, we utilize a structured approach based on an analysis of the commutant of each germ in our germ set. This analysis identifies the directions in parameter space amplified by a given germ and uses that information to search for reduced sets of fiducial pairs that provide good overlap with those directions. In the second, much simpler, approach we ignore this structure and simply discard, at random, a fraction of the fiducial pairs at each germ-power pair. We call this approach per-germ-power random FPR, or simply random FPR.

\begin{figure}[htbp]
    \centering
    \includegraphics[width=.8\textwidth]{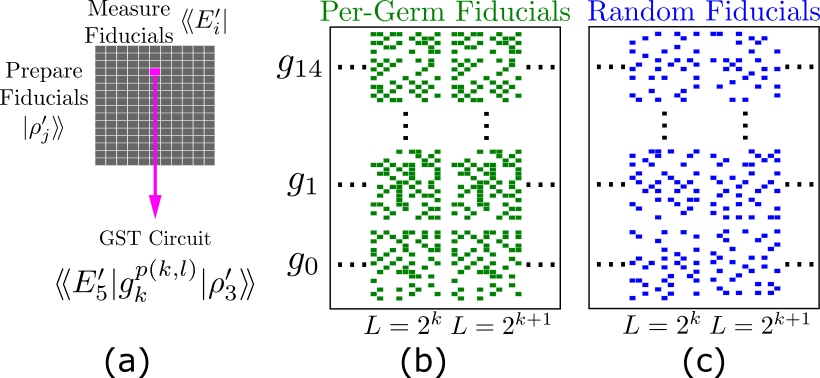}
    \caption{(a) Fiducial pair plaquette structure for a GST experiment.  Each row corresponds to a different preparation fiducial and each column corresponds to a different measurement fiducial.  Each box corresponds to a particular circuit, beginning and ending with the indicated preparation and measurement fiducials, respectively; sandwiched between the fiducials is a particular germ repeated some number of times.  A comparison of fiducial pair plaquette structures for GST experiment designs using per-germ FPR and random FPR is given in panels (b) and (c).  Reproduced (with minor modification) from Figure \ref{fig:GR-FPR-overview}.}
    \label{fig:per_germ_and_random_plaquettes}
\end{figure}

To showcase the differences, we turn our attention to Figure \ref{fig:per_germ_and_random_plaquettes}, which visualizes the circuits in an GST experiment design.
The experiment is organized into fiducial pair ``plaquettes'' for each germ (rows) and maximum length (columns). Within each plaquette a row corresponds to a single prep fiducial, while a column corresponds to a single measurement fiducial. Initially, fiducial selection determines a full set of fiducials for an experiment design, and so the number of rows and columns in each plaquette. Per-germ FPR uses the same sparsity pattern across each row corresponding to a particular germ, while in random FPR every plaquette has an independently random sparsity pattern.

\subsubsection*{Per-Germ FPR}

As discussed in Section \ref{sec:germ_selection}, at large germ powers the directions in parameter space a germ amplifies is determined by that germ's commutant. As such, it suffices in practice to, rather than utilize an informationally complete set of fiducials for each germ, use a set of fiducials with good overlap with the directions in parameter space amplified by that germ. Let $g\in \germs$ be an element of a GST experiment's germ set. The commutant of $g$ defines a matrix subalgebra with a particular structure that we refer to as a germ's ``kite structure.'' The parameters of a germ's kite structure correspond directly to the parameters amplified by a germ.

\begin{figure}
    \centering
    \begin{minipage}{.3\textwidth}
        \includegraphics[width=\textwidth]{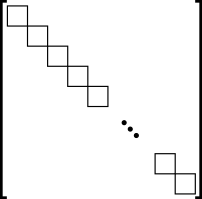}
    \end{minipage}
    \hspace{2em}
    \begin{minipage}{.3\textwidth}
        \includegraphics[width=\textwidth]{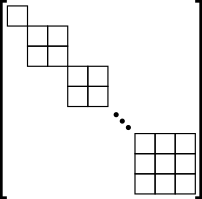}
    \end{minipage}
    \caption{(a) Kite structure for a germ with a non-degenerate eigenvalue spectrum corresponds simply to a matrix with $1\times 1$ blocks of parameters along the diagonal. (b) Kite structure example for a more general case in which a germ has a degnerate spectrum. The kite structure in this case is block diagonal with one block for each unique eigenvalue. The dimensions of a block is equal to the dimension of the degenerate eigenspace for the corresponding eigenvalue.}
    \label{fig:kite_structure}
\end{figure}

The kite structure of a germ can be identified by performing a generalized eigenvector decomposition on $\tau(g)$. When $\tau(g)$ has a non-degenerate spectrum the corresponding kite structure is simply a diagonal matrix of $1\times 1$ blocks. More generally, if $\tau(g)$ has a degenerate spectrum with unique eigenvalues $\{ \lambda_i \}$, then the kite structure of $\tau(g)$ will instead be block diagonal. The dimension of the $i$\textsuperscript{th} block is given by the dimension of the generalized eigenspace associated with $\lambda_i$. Any matrix which, when written in the generalized eigenbasis of $\tau(g)$, has non-zero entries within $\tau(g)$'s kite is an element of the commutant.  The number of parameters in a germ's kite structure is equal to the sum of the number of elements within each block\footnote{A "parameter" of a kite structure in this context is equivalent to a coordinate within one of its blocks.}.

The kite structure of $\tau(g)$ can be used to identify reduced sets of fiducial pairs with sensitivity to all of the directions amplified by $g$. To do so we search for sets of fiducial pairs that produce probabilities which are sensitive to changes to the values of parameters within $\tau(g)$'s kite. Let $\{(F_j,H_i)\}$ denote a candidate set of fiducial pairs where $F_j\in \pfids$ and $H_i \in \mfids$ are elements of the sets of available state preparation and measurement fiducials respectively, and let $\vec{p}_{\{(F_j,H_i)\}}$ be a vectorized list of probabilities for each of the circuits corresponding to a candidate set of fiducials. For each candidate fiducial set we construct a Jacobian of the form

\begin{equation}
    J_{\{(F_j,H_i) \} }= \frac{\partial \vec{p}_{\{(F_j,H_i)\}}}{\partial \vec{\theta}_K},
    \label{eqn:per_germ_fpr_jacobian}
\end{equation}

\noindent where $\vec{\theta}_K$ is the vector of parameters within the kite structure for a particular germ. In addition to the Jacobian in Equation \ref{eqn:per_germ_fpr_jacobian}, we also calculate an analogous Jacobian for the complete set of fiducial pairs, $J_{\{ \pfids, \mfids \}}$. This second Jacobian establishes a baseline for evaluating candidate sets. All that remains is to identify a reduced set of fiducial pairs $\{(f_i,h_i)\}$ such that $ J_{\{(F_j,H_i) \}}$ both has the same rank as  $J_{\{ \pfids, \mfids \}}$ and doesn't suffer too large a reduction in overall sensitivity compared to the full set of fiducial pairs. For additional details on the algorithm used to find reduced sets of fiducial pairs and the specific objective function used to evaluate them see Appendix \ref{app:algorithms}.

\subsubsection*{Random FPR}
\label{sec:per_germ_power_random_FPR}

Per-germ FPR, as described in the preceding section, provides a structured approach to generating reduced experiment designs which guarantee a certain level of sensitivity to all of the amplified parameters for each germ. At the other end of the spectrum is random FPR, where we ignore the structure in the problem and instead, for each germ-power pair (i.e. each of the plaquettes in Figure \ref{fig:GR-FPR-overview}), simply throw out all but some user specified fraction of the fiducial pairs. The general structure (or lack thereof) of a reduced experiment design produced this way can be seen in the last panel of Figure \ref{fig:GR-FPR-overview}. While random FPR doesn't guarantee a minimum sensitivity like per-germ FPR does, we have found empirically that with a sufficiently large fraction of fiducials this scheme performs well in practice. Moreover, these designs are incredibly inexpensive to generate when compared to the computational cost of performing per-germ FPR.

\subsubsection{Relaxing the AC Constraint}
\label{sec:germ_reduction}

The second key tool we have for generating reduced experiment designs is a relaxation of the amplificational completeness constraint, which decreases the size of a GST germ set. When generating what we call the ``robust germ'' set we augment the procedure described in Section \ref{sec:germ_selection} by adding the requirement that a set of germs not only amplifies all of the parameters of the target gate set, but also those of some number of gate sets perturbed about the target. The reason for such a choice is to account for perturbations to the gate set that modify the kite structure of a germ. The most stark example of perturbations changing the kite structure of a germ is the idle gate. The PTM for the ideal idle gate is simply the identity matrix, which commutes with every operator. The kite structure for the idle gate is consequently a single $d^2 \times d^2$ dimensional block, indicating that the commutant includes all of Hilbert-Schmidt space. A naive implementation of the procedure in Section \ref{sec:germ_selection} would thus conclude that perturbations to the idle gate in every direction in parameter space are amplified and therefore the bare idle gate suffices to amplify all of its own errors. The issue with this, however, is that this only works when the analysis is performed exactly at the target gate set, \textit{any} non-trivial perturbation to the idle gate will modify its kite structure and change the directions in parameter space amplified by this gate. While the idle gate is the most extreme example of this effect, it can also occur with other gates. When a perturbation to a gate changes its kite structure we say this perturbation ``breaks degeneracies'' in that gate.

Adding the requirement that a set of germs not only amplify perturbations to all of the parameters of the target gate set, but also those of some number of nearby gate sets, increases the overall cost of a GST experiment design by substantially increasing the number of germs required. This motivates the defintition and use of the ``standard'' germ set, which is simply the result of performing germ selection with respect only to the target gate set. Using the standard germ set results in a significant cost reduction in the size of an experiment design, but at the cost of relaxing the amplificational completeness requirement, the consequences of which were observed in Section \ref{sec:1Q_2Q_ddist_results} and will be revisited in Section \ref{sec:connection_to_fisher_info}.

\section{Discussion}

\subsection{Connection to the Fisher Information}
\label{sec:connection_to_fisher_info}

In Section \ref{sec:1Q_2Q_ddist_results} we discussed the observation that for the one-qubit XYI gate set the precision as measured in terms of the average gate set diamond distance with respect to the truth began to plateau at longer $L$ values for experiment designs using the standard germ set. This is a well-understood feature of the standard germ set, as discussed in Section \ref{sec:germ_reduction}. Among all gates the idle gate is the most susceptible to this degeneracy breaking effect, and as such it was found in Figure \ref{fig:per_gate_avg_ddist_full_lite} that the plateauing sensitivity was mostly attributable to the idle gate. Why were we unable to see this effect manifest itself in the Fisher information spectra for the one-qubit experiment designs using the standard germ set in Figures \ref{fig:fisher_info_vs_l_spectra_3_main}(b) and (c)? In short, because we've chosen to evaluate the Fisher information matrix with respect to the target gate set. This results in the Fisher information matrix being susceptible to the same issues with degeneracy that impact the analysis underpinning the standard germ set. It is nonetheless possible to, by slightly modifying our analysis, predict the plateauing sensitivity of the standard germ set experiment designs solely using the Fisher information. 

Similarly to how the germ selection routine is modified to account for degeneracies, we begin by calculating the per-circuit contributions to the Fisher information matrix with respect to one or more gate sets unitarily perturbed about the target gate set. Let $\mathcal{C}_{L} \subseteq \mathcal{C}$ denote the subset of circuits for a particular value of $L$. We sum the per-circuit Fisher information contributions for each subset $\mathcal{C}_{L}$ to form an incremental Fisher information matrix $I_{\mathcal{C}_{L}}$,

\begin{equation}
    I_{\mathcal{C}_{L}} = \sum_{c\in \mathcal{C}_L}I_c.
    \label{eqn:incremental_fisher_info}
\end{equation}

While it is possible to construct an analysis based on the cumulative Fisher information matrices, as was done in Section \ref{sec:one_and_two_q_fisher_info_spectra}, the incremental Fisher information matrices defined in Equation \ref{eqn:incremental_fisher_info} have a nice interpretation when working with coherent-only noise models (such as our unitarily perturbed gate sets). Specifically, when a gate set is subject to coherent-only errors it is the case that, as the coherent errors can be amplified indefinitely, longer circuits provide a greater amount of information about the coherent errors than short circuits. Consequently, for such a gate set the magnitude of the eigenvalues of the incremental Fisher information matrices with corresponding eigenvectors in amplifiable directions of parameter space should strictly increase. If a parameter is either non-amplifiable (e.g. SPAM), or otherwise fails to be amplified by a particular experiment design, we should find that the magnitude of the spectra of the incremental Fisher information matrices is either entirely flat over all $L$ values or plateaus at large $L$.

\begin{figure}[htbp]
    \centering
    \begin{minipage}{.45\textwidth}
        \includegraphics[height=2.5in]{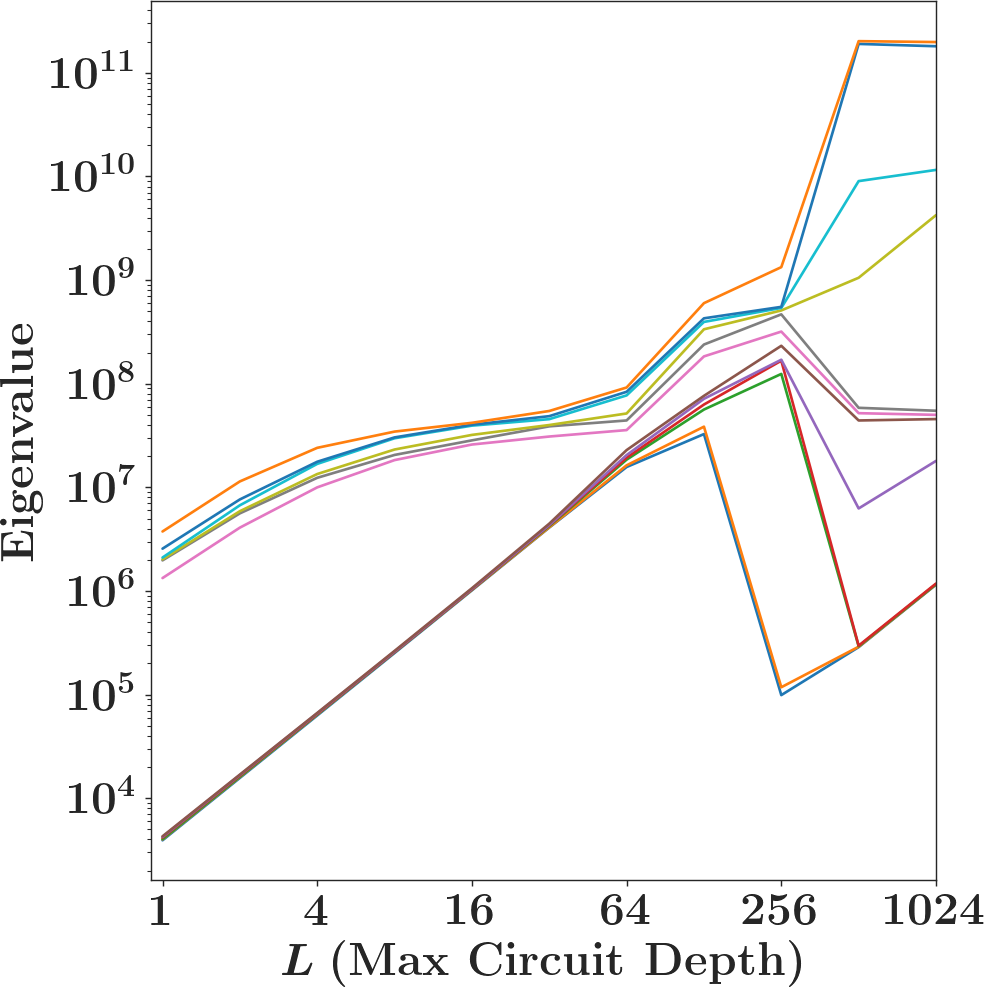}
    \end{minipage}
    \begin{minipage}{.45\textwidth}
        \includegraphics[height=2.5in]{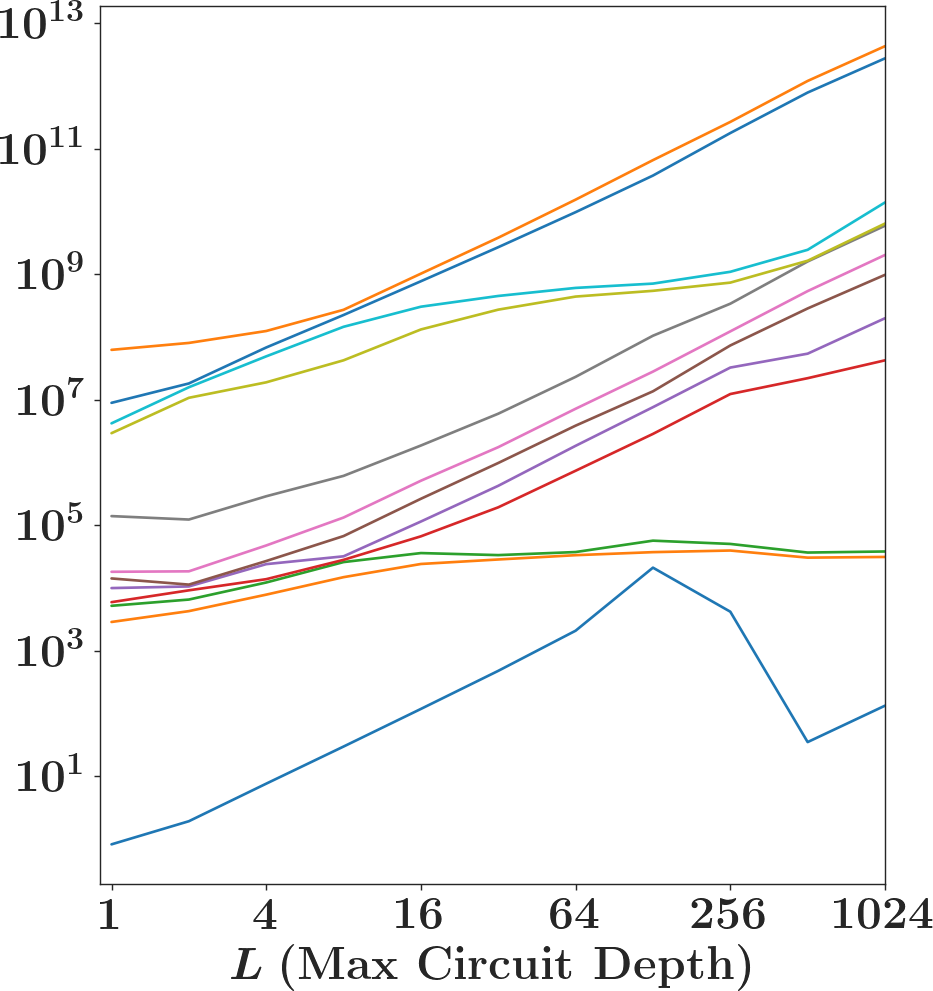}
    \end{minipage}\\[.25em]
    \begin{minipage}{.45\textwidth}
        \includegraphics[height=2.5in]{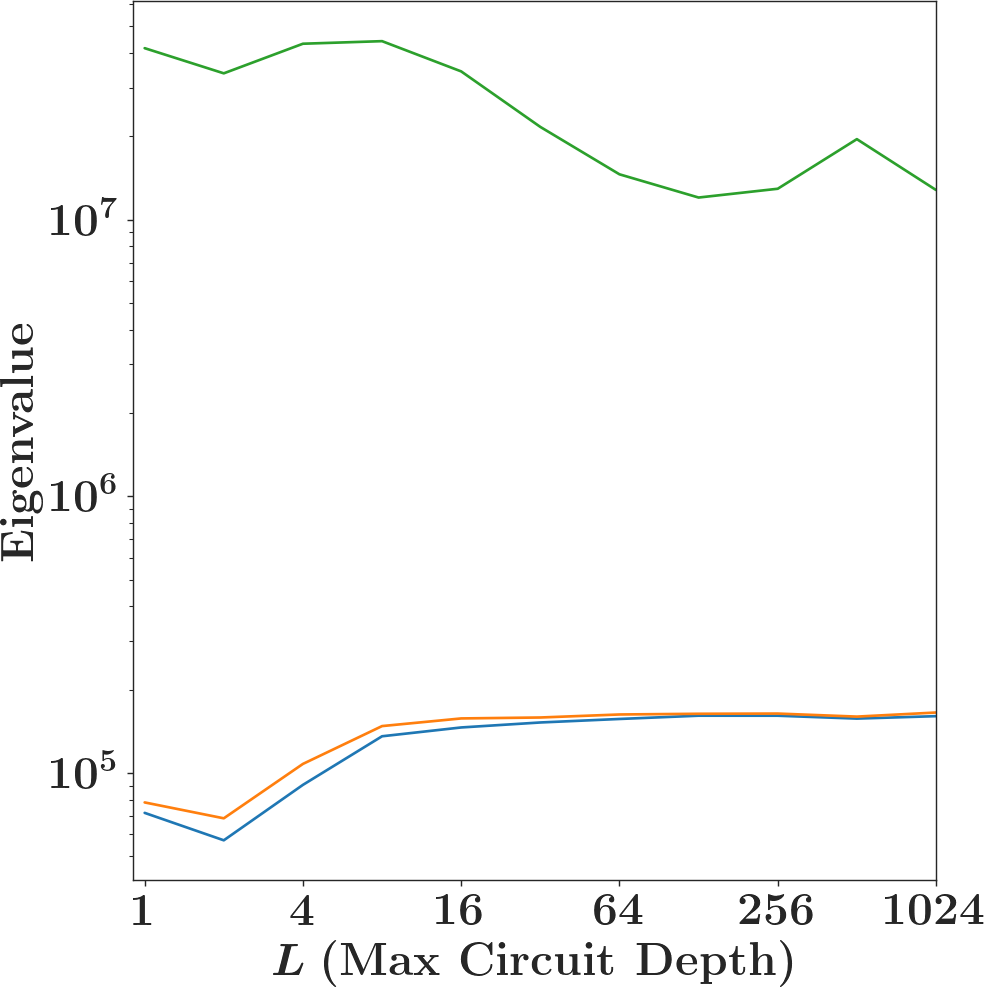}
    \end{minipage}
        \begin{minipage}{.45\textwidth}
        \includegraphics[height=2.5in]{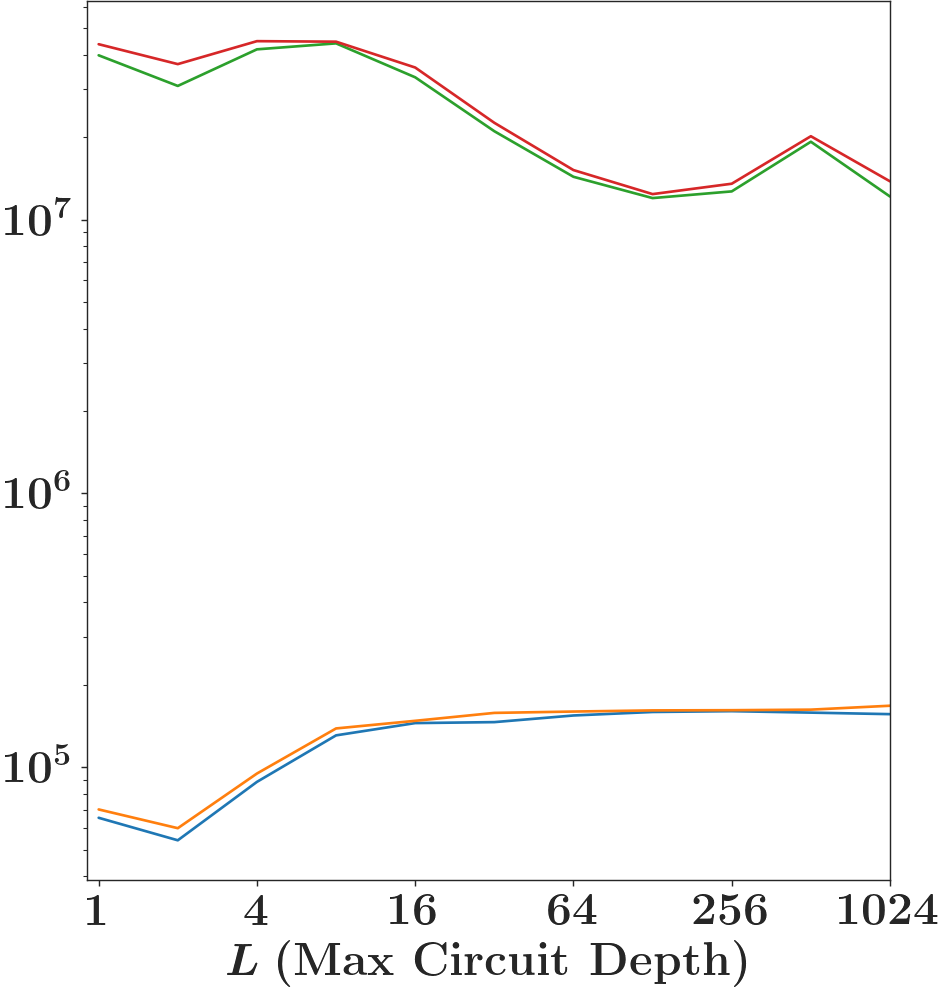}
    \end{minipage}
    \caption{Spectra of the projected incremental fisher information for the full-standard design as a function of circuit depth evaluated at a point unitarily perturbed from the target model. We've projected separately onto the subspaces of gate set parameter space corresponding to (a) $G_i$, (b) $G_{x_{\pi/2}}$, (c) state preparation and (d) measurement.}
    \label{fig:projected_incremental_fisher_information_spectra_full_lite}
\end{figure}

A small problem with this argument is that it is generally not the case that the eigenvectors of the Fisher information matrix for a gate set partition cleanly into directions that can be associated directly with any particular gate or SPAM operation. This is a problem insofar as we'd like to directly reproduce not only the plateau in the gate set precision overall seen in Figures \ref{fig:ddist_vs_L_full_lite_pergerm}(b) and (c), but also the ability to attribute this specifically to the idle gate as in Figure \ref{fig:per_gate_avg_ddist_full_lite}. This motivates the definition of the projected Fisher information matrices, which are constructed by explicitly projecting the incremental Fisher information matrices onto a subspace of the full parameter space associated with a particular gate or SPAM operation, 

\begin{equation}
I^{\alpha}_{\mathcal{C}_L} = \Pi_\alpha I_{\mathcal{C}_L} \Pi_\alpha,
\end{equation}

\noindent where $\Pi_\alpha$ is the projector onto the subspace for $\alpha \in \mathcal{G}$. In Figure \ref{fig:projected_incremental_fisher_information_spectra_full_lite} we have plotted spectra of the projected Fisher information matrices for the one-qubit full/standard experiment design, broken out into the projections onto $G_i$, $G_{x_{\pi/2}}$ (the results for $G_{y_{\pi/2}}$ are virtually identical to $G_{x_{\pi/2}}$ and so are omitted), as well as the two SPAM operations. The plotted spectra for the projection onto the idle gate in Figure \ref{fig:projected_incremental_fisher_information_spectra_full_lite}(a) shows that, while the eigenvalues for some directions in parameter space continue to increase monotonically, a large fraction of the eigenvalues plateau at large $L$ indicating that these directions in parameter space are no longer being amplified beyond this point. Moreover, the point at which we observe this transition, at $L\approx 128$, corresponds precisely to the inflection point in Figure \ref{fig:per_gate_avg_ddist_full_lite} whereafter the diamond distance to truth of the idle gate plateaus. Similar behavior of the projected Fisher information spectra is also found to a lesser extent for $G_{x_{\pi/2}}$ in \ref{fig:projected_incremental_fisher_information_spectra_full_lite}(b), again at the point in Figure \ref{fig:per_gate_avg_ddist_full_lite} where the diamond distance to truth is observed to undergo an inflection. Finally, the spectra of the projected Fisher information matrices for the state preparation and measurement operations plotted in Figure \ref{fig:projected_incremental_fisher_information_spectra_full_lite}(c) and (d) are essentially flat, as would be expected for the non-amplifiable SPAM parameters.

\begin{figure}[htbp]
    \centering
    \begin{minipage}{.45\textwidth}
        \includegraphics[height=2.5in]{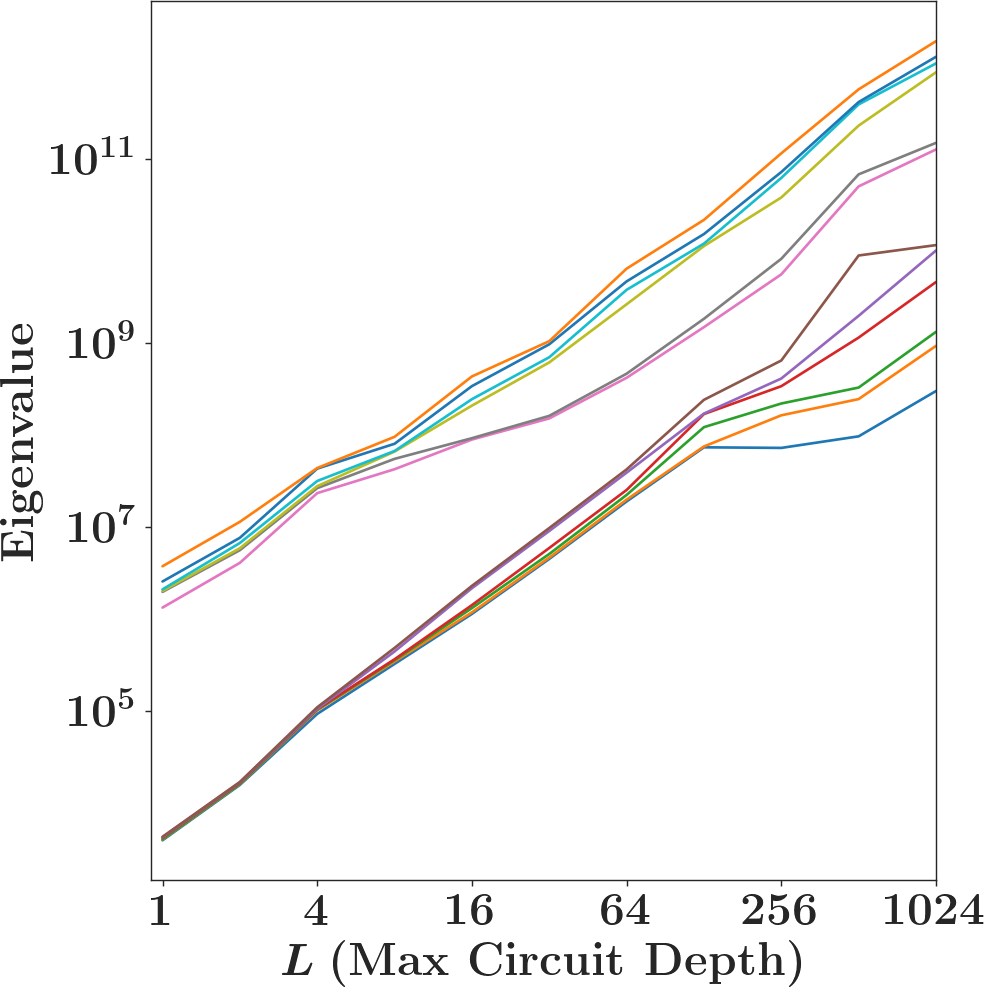}
    \end{minipage}
    \begin{minipage}{.45\textwidth}
        \includegraphics[height=2.5in]{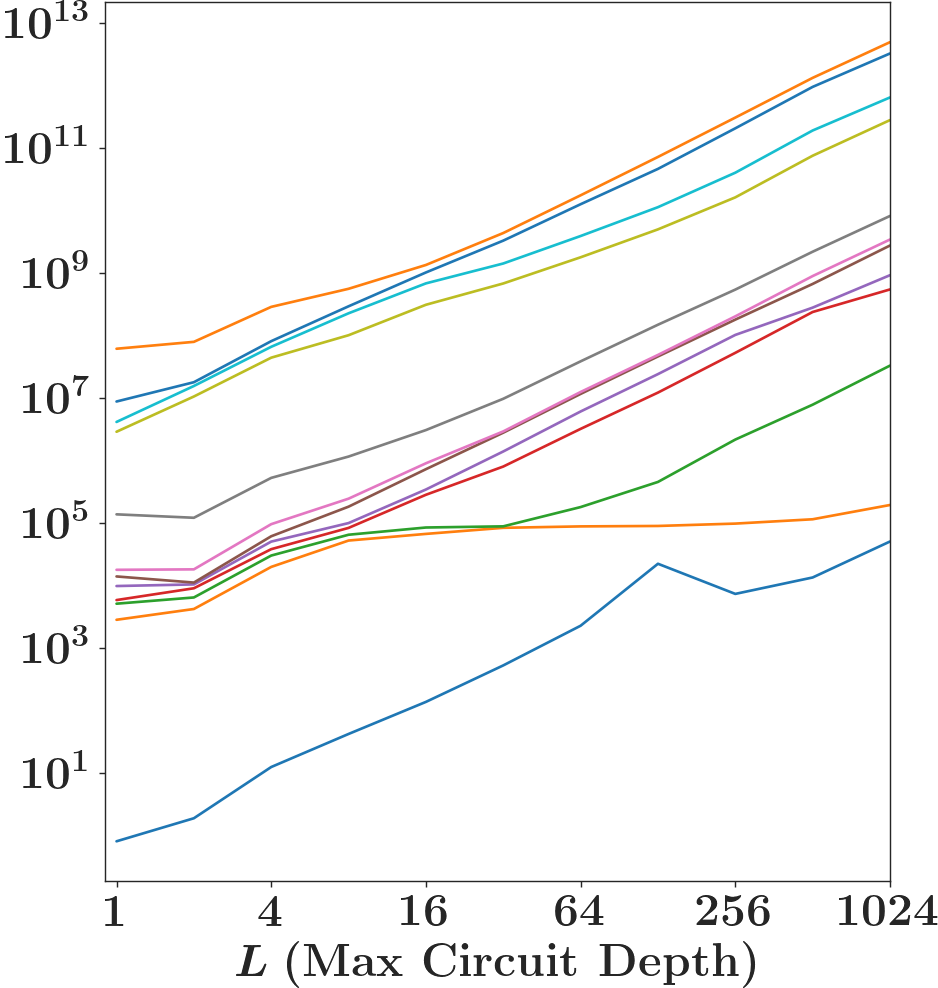}
    \end{minipage}
    \caption{Spectra of the projected incremental fisher information analogous to Figure \ref{fig:projected_incremental_fisher_information_spectra_full_lite}, but for the \edesign{robust}{full} experiment design. (a) $G_i$, (b) $G_{x_{\pi/2}}$.}
    \label{fig:projected_incremental_fisher_information_spectra_full_full}
\end{figure}

While rigorously justifying the use of the projected Fisher information is beyond the scope of this present work, we can confirm that the story it tells is genuinely meaningful by comparing the results for the projected Fisher information matrices for the \edesign{standard}{full} design to those for the \edesign{robust}{full} design. We expect that for the \edesign{robust}{standard} design with a coherent-only error model the spectra of the incremental Fisher information should continue to increase for arbitrarily long circuit depths, even for the idle gate. In Figure \ref{fig:projected_incremental_fisher_information_spectra_full_full}(a) and (b) we plot the spectra of the projected Fisher information matrices as a function of $L$ for the $G_i$ and $G_{x_{\pi/2}}$ gates, respectively. As expected, it is found that for both gates the the magnitudes of all the eigenvalues continues to grow as a function of $L$, consistent with the empirically observed behavior of the average gate set diamond diamond distance.

\subsection{Scaling Up GST Experiments}
\label{sec:scaling_up_GST}

In this work we have introduced techniques for reducing the size of GST experiment designs without significantly decreasing their precision. Historically, the size of the experiment designs were a key factor limiting the use of GST for multi-qubit systems. In this section we will discuss the significance of these results for the prospects of scaling up GST to larger systems.

In Section \ref{sec:densely_parameterized_GST} we will discuss the prospects for scaling up the traditional GST protocol to systems of three or more qubits. In Section \ref{sec:reduced_model_GST} we will talk about a proposal for scaling precision characterization up to multi-qubit systems, and the role that traditional GST could play in this setting. Finally, in Section \ref{sec:future_FPR} we will discuss a possible future direction for improving the per-germ fiducial pair reduction technique we have presented to generated even more heavily streamlined experiment designs.

\subsubsection{Densely-Parameterized GST}
\label{sec:densely_parameterized_GST}

The techniques for reducing the experimental cost of performing GST presented in this work have implicitly targeted making the traditional variant of GST in which the set of operations we are fitting experimental data to span the entire space of CPTP maps. We call any such parameterization enough sufficient expressivity to describe the entire space of CPTP maps \textit{densely parameterized}. The exponential scaling on the number of parameters required for a dense parameterization of CPTP maps fundamentally limits the size of the systems we can tomograph. 

None of the techniques introduced for reducing experimental costs, fiducial pair reduction, and relaxation of the AC constraint fundamentally change the exponential scaling for the cost of densely-parameterized GST. They do, however, push the frontier for feasibility slightly further out. Historically, two-qubit GST was considered to be the largest system size for which densely-parameterized GST was feasible. With the use of FPR and relaxed AC constraints, however, initial resource estimations suggest that dense characterizations of three-qubit and two-qutrit systems are now squarely within the range of what is feasible experimentally. For example, we've estimated that for a three-qubit system with a gate set consisting of single-qubit $G_{x_{\pi/2}}$ and $G_{y_{\pi/2}}$ gates along with a three-qubit controlled-SWAP gate, performing GST would require $\sim 30000$ circuits using a Per-Germ/Standard experiment design, well within the range of feasibility on some modern quantum computers.

\subsubsection{Reduced model GST}
\label{sec:reduced_model_GST}

Though the circuit reduction techniques in this paper are evidently not sufficient by themselves to make GST tractable for four or more qubits, these techniques do indicate a path forward for making such experiments feasible. In \cite{rbk2022taxonomy} the error generator formalism was introduced, which provides an alternative parameterization for the space of CPTP maps in terms of physically recognizable error processes. A key feature of this formalism is that it lends itself to the construction of so-called ``reduced models'' wherein we restrict our descriptions to lower-dimensional subspaces of the full space of CPTP maps by ``turning off'' parameters corresponding to error mechanisms deemed to be physically implausible. As reduced models have fewer parameters they accordingly require fewer circuits to learn them. For example, by restricting the allowed error processes to act on one or two qubits only, the total number of parameters in the model is reduced from exponential in qubit number to merely quadratic.

While such reduced models have the potential to tackle the exponential scaling requirements of GST head-on, constructing reduced models apriori from the bottom-up requires deeps knowledge and trust in the underlying device physics for a particular platform. In \cite{mkadzik2022precision} an alternative approach to the construction of reduced models was presented which utilized a top-down approach leveraging a dense characterization of a two-qubit gate set together with statistical model selection tools to identify the subset of error mechanisms which were physically relevant. The resulting reduced model successfully fit the data nearly as well as the dense model from which it was constructed using an order of magnitude fewer parameters overall. This success points to a path toward scalable characterization of many-qubit systems where we use densely-parameterized GST on smaller subsystems as a starting point from which to build lower-dimensional reduced models which can then be efficiently brought up to the scale of the entire system.

\subsubsection{Future Directions for Fiducial Pair Reduction}
\label{sec:future_FPR}

In fiducial pair reduction, specifically per-germ FPR, we leverage the fact that any individual germ gives us high-precision information about only a relatively small fraction of the total number of parameters. This allows us to identify and remove circuits for each germ which provide comparably little information about our gate set's parameters. Can we do any better than this? That is, is there any additional redundancy in the resulting GST experiment designs that still remains?

One possible candidate for where such redundancy could exist is in the germ set. In germ selection we generate germ sets which are amplificationally complete, but do not explicitly prioritize the selection germ sets which amplify mostly disjoint sets of parameters. As such, there can be non-trivial overlap between the parameters various germs amplify. In per-germ FPR each germ is treated independently of all of the others, and we aim for each germ to select a set of fiducial pairs that is sensitive to \textit{all} of the parameters a germ is sensitive too. As such, a path forward toward even more heavily streamlined experiment designs, and the subject of future work, would be to modify our FPR algorithms to take into account both the local properties of each germ and the collective properties of the entire germ set to account for these redundancies.

\section{Conclusion}
\label{sec:conclusion}

In order to increase the experimental feasibility of gate set tomography, we have developed and demonstrated techniques that drastically reduce the number of circuits required while maintaining Heisenberg-like scaling of accuracy. The two circuit reduction techniques are fiducial pair reduction and the relaxation of the amplificational completeness constraint. To relax the amplificational completeness constraint we use only the ideal gate set rather than the traditional choice of unitarily perturbed gate sets for germ selection. This choice fails to amplify "accidental" degeneracies in the gates, which are particularly relevant for idle operations. However, we show that these effects do not impact overall GST accuracy until large circuit depths, where experiments are more likely to be decoherence-limited anyway. Fiducial pair reduction removes pairs of state preparation and measurement fiducials that may not amplify any parameters for a given germ. In addition to a full (unreduced) set of fiducial pairs, we consider a per-germ scheme based on the commutant of each germ and a random scheme that performs well empirically for even relatively small retention fractions.

Numerical simulations show that experiment designs using germ and fiducial pair reduction can achieve similar accuracy while drastically reducing the number of circuits required. Using FPR together with a reduced germ set resulted in reductions in the number of circuits required to achieve an accuracy comparable to the full design by factors of $\sim 7.7$ and $\sim 3.7$ for one- and two-qubit GST, respectively. Our numerical simulations were backed up theoretical results based on an analysis of the various experiment designs' Fisher information matrices. Here, by looking at how the magnitude of the eigenvalues of the FIM scaled with circuit depth we were able to confirm the existence of Heisenberg-like precision scaling.

Gate set tomography is among the most powerful tools currently available to experimentalists for completely and precisely characterizing the dynamics of their systems, with the main impediment to its adoption being the heavy experimental cost associated with it. The techniques introduced in this work will help open up the door to using GST on many systems where its use had previously been out-of-reach.

\section{Acknowledgements}
This material was funded in part by the U.S. Department of Energy, Office of Science, Office of Advanced Scientific Computing Research Quantum Testbed Pathfinder Program. This material was also funded in part by the Office of the Director of National Intelligence (ODNI), Intelligence Advanced Research Projects Activity (IARPA).  Sandia National Laboratories is a multimission laboratory managed and operated by National Technology and Engineering Solutions of Sandia, LLC., a wholly owned subsidiary of Honeywell International, Inc., for the U.S. Department of Energy’s National Nuclear Security Administration under Contract No. DENA-0003525.  All statements of fact, opinion, or conclusions contained herein are those of the authors and should not be construed as representing the official views or policies of IARPA, the ODNI, the U.S. Department of Energy, or the U.S. Government.  We also thank Ashlyn Burch, Susan Clark, Akel Hashim, Dan Lobser, Holly Stemp, and Chris Yale for helpful conversations.  
\bibliography{references,referencesGSTQuantum}

\begin{thebibliography}{123}%
\makeatletter
\providecommand \@ifxundefined [1]{%
 \@ifx{#1\undefined}
}%
\providecommand \@ifnum [1]{%
 \ifnum #1\expandafter \@firstoftwo
 \else \expandafter \@secondoftwo
 \fi
}%
\providecommand \@ifx [1]{%
 \ifx #1\expandafter \@firstoftwo
 \else \expandafter \@secondoftwo
 \fi
}%
\providecommand \natexlab [1]{#1}%
\providecommand \enquote  [1]{``#1''}%
\providecommand \bibnamefont  [1]{#1}%
\providecommand \bibfnamefont [1]{#1}%
\providecommand \citenamefont [1]{#1}%
\providecommand \href@noop [0]{\@secondoftwo}%
\providecommand \href [0]{\begingroup \@sanitize@url \@href}%
\providecommand \@href[1]{\@@startlink{#1}\@@href}%
\providecommand \@@href[1]{\endgroup#1\@@endlink}%
\providecommand \@sanitize@url [0]{\catcode `\\12\catcode `\$12\catcode
  `\&12\catcode `\#12\catcode `\^12\catcode `\_12\catcode `\%12\relax}%
\providecommand \@@startlink[1]{}%
\providecommand \@@endlink[0]{}%
\providecommand \url  [0]{\begingroup\@sanitize@url \@url }%
\providecommand \@url [1]{\endgroup\@href {#1}{\urlprefix }}%
\providecommand \urlprefix  [0]{URL }%
\providecommand \Eprint [0]{\href }%
\providecommand \doibase [0]{https://doi.org/}%
\providecommand \selectlanguage [0]{\@gobble}%
\providecommand \bibinfo  [0]{\@secondoftwo}%
\providecommand \bibfield  [0]{\@secondoftwo}%
\providecommand \translation [1]{[#1]}%
\providecommand \BibitemOpen [0]{}%
\providecommand \bibitemStop [0]{}%
\providecommand \bibitemNoStop [0]{.\EOS\space}%
\providecommand \EOS [0]{\spacefactor3000\relax}%
\providecommand \BibitemShut  [1]{\csname bibitem#1\endcsname}%
\let\auto@bib@innerbib\@empty
\bibitem [{\citenamefont {Artiles}\ \emph {et~al.}(2005)\citenamefont
  {Artiles}, \citenamefont {Gill},\ and\ \citenamefont
  {Guta}}]{Artiles2005-ts}%
  \BibitemOpen
  \bibfield  {author} {\bibinfo {author} {\bibfnamefont {L.~M.}\ \bibnamefont
  {Artiles}}, \bibinfo {author} {\bibfnamefont {R.~D.}\ \bibnamefont {Gill}},\
  and\ \bibinfo {author} {\bibfnamefont {M.~I.}\ \bibnamefont {Guta}},\
  }\bibfield  {title} {\bibinfo {title} {An invitation to quantum tomography},\
  }\href {https://doi.org/10.1111/j.1467-9868.2005.00491.x} {\bibfield
  {journal} {\bibinfo  {journal} {J. R. Stat. Soc. Series B Stat. Methodol.}\
  }\textbf {\bibinfo {volume} {67}},\ \bibinfo {pages} {109} (\bibinfo {year}
  {2005})}\BibitemShut {NoStop}%
\bibitem [{\citenamefont {Altepeter}\ \emph {et~al.}(2003)\citenamefont
  {Altepeter}, \citenamefont {Branning}, \citenamefont {Jeffrey}, \citenamefont
  {Wei}, \citenamefont {Kwiat}, \citenamefont {Thew}, \citenamefont {O'Brien},
  \citenamefont {Nielsen},\ and\ \citenamefont {White}}]{Altepeter2003-fb}%
  \BibitemOpen
  \bibfield  {author} {\bibinfo {author} {\bibfnamefont {J.~B.}\ \bibnamefont
  {Altepeter}}, \bibinfo {author} {\bibfnamefont {D.}~\bibnamefont {Branning}},
  \bibinfo {author} {\bibfnamefont {E.}~\bibnamefont {Jeffrey}}, \bibinfo
  {author} {\bibfnamefont {T.~C.}\ \bibnamefont {Wei}}, \bibinfo {author}
  {\bibfnamefont {P.~G.}\ \bibnamefont {Kwiat}}, \bibinfo {author}
  {\bibfnamefont {R.~T.}\ \bibnamefont {Thew}}, \bibinfo {author}
  {\bibfnamefont {J.~L.}\ \bibnamefont {O'Brien}}, \bibinfo {author}
  {\bibfnamefont {M.~A.}\ \bibnamefont {Nielsen}},\ and\ \bibinfo {author}
  {\bibfnamefont {A.~G.}\ \bibnamefont {White}},\ }\bibfield  {title} {\bibinfo
  {title} {Ancilla-assisted quantum process tomography},\ }\href
  {https://doi.org/10.1103/PhysRevLett.90.193601} {\bibfield  {journal}
  {\bibinfo  {journal} {Phys. Rev. Lett.}\ }\textbf {\bibinfo {volume} {90}},\
  \bibinfo {pages} {193601} (\bibinfo {year} {2003})}\BibitemShut {NoStop}%
\bibitem [{\citenamefont {Banaszek}\ \emph {et~al.}(2013)\citenamefont
  {Banaszek}, \citenamefont {Cramer},\ and\ \citenamefont
  {Gross}}]{Banaszek2013-zx}%
  \BibitemOpen
  \bibfield  {author} {\bibinfo {author} {\bibfnamefont {K.}~\bibnamefont
  {Banaszek}}, \bibinfo {author} {\bibfnamefont {M.}~\bibnamefont {Cramer}},\
  and\ \bibinfo {author} {\bibfnamefont {D.}~\bibnamefont {Gross}},\ }\bibfield
   {title} {\bibinfo {title} {Focus on quantum tomography},\ }\href
  {https://doi.org/10.1088/1367-2630/15/12/125020} {\bibfield  {journal}
  {\bibinfo  {journal} {New J. Phys.}\ }\textbf {\bibinfo {volume} {15}},\
  \bibinfo {pages} {125020} (\bibinfo {year} {2013})}\BibitemShut {NoStop}%
\bibitem [{\citenamefont {Bendersky}\ \emph {et~al.}(2008)\citenamefont
  {Bendersky}, \citenamefont {Pastawski},\ and\ \citenamefont
  {Paz}}]{Bendersky2008-xa}%
  \BibitemOpen
  \bibfield  {author} {\bibinfo {author} {\bibfnamefont {A.}~\bibnamefont
  {Bendersky}}, \bibinfo {author} {\bibfnamefont {F.}~\bibnamefont
  {Pastawski}},\ and\ \bibinfo {author} {\bibfnamefont {J.~P.}\ \bibnamefont
  {Paz}},\ }\bibfield  {title} {\bibinfo {title} {Selective and efficient
  estimation of parameters for quantum process tomography},\ }\href
  {https://doi.org/10.1103/PhysRevLett.100.190403} {\bibfield  {journal}
  {\bibinfo  {journal} {Phys. Rev. Lett.}\ }\textbf {\bibinfo {volume} {100}},\
  \bibinfo {pages} {190403} (\bibinfo {year} {2008})}\BibitemShut {NoStop}%
\bibitem [{\citenamefont {Bialczak}\ \emph {et~al.}(2010)\citenamefont
  {Bialczak}, \citenamefont {Ansmann}, \citenamefont {Hofheinz}, \citenamefont
  {Lucero}, \citenamefont {Neeley}, \citenamefont {O'Connell}, \citenamefont
  {Sank}, \citenamefont {Wang}, \citenamefont {Wenner}, \citenamefont
  {Steffen}, \citenamefont {Cleland},\ and\ \citenamefont
  {Martinis}}]{Bialczak2010-et}%
  \BibitemOpen
  \bibfield  {author} {\bibinfo {author} {\bibfnamefont {R.~C.}\ \bibnamefont
  {Bialczak}}, \bibinfo {author} {\bibfnamefont {M.}~\bibnamefont {Ansmann}},
  \bibinfo {author} {\bibfnamefont {M.}~\bibnamefont {Hofheinz}}, \bibinfo
  {author} {\bibfnamefont {E.}~\bibnamefont {Lucero}}, \bibinfo {author}
  {\bibfnamefont {M.}~\bibnamefont {Neeley}}, \bibinfo {author} {\bibfnamefont
  {A.~D.}\ \bibnamefont {O'Connell}}, \bibinfo {author} {\bibfnamefont
  {D.}~\bibnamefont {Sank}}, \bibinfo {author} {\bibfnamefont {H.}~\bibnamefont
  {Wang}}, \bibinfo {author} {\bibfnamefont {J.}~\bibnamefont {Wenner}},
  \bibinfo {author} {\bibfnamefont {M.}~\bibnamefont {Steffen}}, \bibinfo
  {author} {\bibfnamefont {A.~N.}\ \bibnamefont {Cleland}},\ and\ \bibinfo
  {author} {\bibfnamefont {J.~M.}\ \bibnamefont {Martinis}},\ }\bibfield
  {title} {\bibinfo {title} {Quantum process tomography of a universal
  entangling gate implemented with josephson phase qubits},\ }\href
  {https://doi.org/10.1038/nphys1639} {\bibfield  {journal} {\bibinfo
  {journal} {Nat. Phys.}\ }\textbf {\bibinfo {volume} {6}},\ \bibinfo {pages}
  {409} (\bibinfo {year} {2010})}\BibitemShut {NoStop}%
\bibitem [{\citenamefont
  {Blume-Kohout}(2010{\natexlab{a}})}]{Blume-Kohout2010-hb}%
  \BibitemOpen
  \bibfield  {author} {\bibinfo {author} {\bibfnamefont {R.}~\bibnamefont
  {Blume-Kohout}},\ }\bibfield  {title} {\bibinfo {title} {Hedged maximum
  likelihood quantum state estimation},\ }\href
  {https://doi.org/10.1103/PhysRevLett.105.200504} {\bibfield  {journal}
  {\bibinfo  {journal} {Phys. Rev. Lett.}\ }\textbf {\bibinfo {volume} {105}},\
  \bibinfo {pages} {200504} (\bibinfo {year} {2010}{\natexlab{a}})}\BibitemShut
  {NoStop}%
\bibitem [{\citenamefont
  {Blume-Kohout}(2010{\natexlab{b}})}]{Blume-Kohout2010-vv}%
  \BibitemOpen
  \bibfield  {author} {\bibinfo {author} {\bibfnamefont {R.}~\bibnamefont
  {Blume-Kohout}},\ }\bibfield  {title} {\bibinfo {title} {Optimal, reliable
  estimation of quantum states},\ }\href
  {https://doi.org/10.1088/1367-2630/12/4/043034} {\bibfield  {journal}
  {\bibinfo  {journal} {New J. Phys.}\ }\textbf {\bibinfo {volume} {12}},\
  \bibinfo {pages} {043034} (\bibinfo {year} {2010}{\natexlab{b}})}\BibitemShut
  {NoStop}%
\bibitem [{\citenamefont {Childs}\ \emph {et~al.}(2001)\citenamefont {Childs},
  \citenamefont {Chuang},\ and\ \citenamefont {Leung}}]{Childs2001-en}%
  \BibitemOpen
  \bibfield  {author} {\bibinfo {author} {\bibfnamefont {A.~M.}\ \bibnamefont
  {Childs}}, \bibinfo {author} {\bibfnamefont {I.~L.}\ \bibnamefont {Chuang}},\
  and\ \bibinfo {author} {\bibfnamefont {D.~W.}\ \bibnamefont {Leung}},\
  }\bibfield  {title} {\bibinfo {title} {Realization of quantum process
  tomography in {NMR}},\ }\href {https://doi.org/10.1103/PhysRevA.64.012314}
  {\bibfield  {journal} {\bibinfo  {journal} {Phys. Rev. A}\ }\textbf {\bibinfo
  {volume} {64}},\ \bibinfo {pages} {012314} (\bibinfo {year}
  {2001})}\BibitemShut {NoStop}%
\bibitem [{\citenamefont {Christandl}\ and\ \citenamefont
  {Renner}(2012)}]{Christandl2012-am}%
  \BibitemOpen
  \bibfield  {author} {\bibinfo {author} {\bibfnamefont {M.}~\bibnamefont
  {Christandl}}\ and\ \bibinfo {author} {\bibfnamefont {R.}~\bibnamefont
  {Renner}},\ }\bibfield  {title} {\bibinfo {title} {Reliable quantum state
  tomography},\ }\href {https://doi.org/10.1103/PhysRevLett.109.120403}
  {\bibfield  {journal} {\bibinfo  {journal} {Phys. Rev. Lett.}\ }\textbf
  {\bibinfo {volume} {109}},\ \bibinfo {pages} {120403} (\bibinfo {year}
  {2012})}\BibitemShut {NoStop}%
\bibitem [{\citenamefont {Chuang}\ and\ \citenamefont
  {Nielsen}(1997)}]{Chuang1997-vf}%
  \BibitemOpen
  \bibfield  {author} {\bibinfo {author} {\bibfnamefont {I.~L.}\ \bibnamefont
  {Chuang}}\ and\ \bibinfo {author} {\bibfnamefont {M.~A.}\ \bibnamefont
  {Nielsen}},\ }\bibfield  {title} {\bibinfo {title} {Prescription for
  experimental determination of the dynamics of a quantum black box},\ }\href
  {https://doi.org/10.1080/09500349708231894} {\bibfield  {journal} {\bibinfo
  {journal} {J. Mod. Opt.}\ }\textbf {\bibinfo {volume} {44}},\ \bibinfo
  {pages} {2455} (\bibinfo {year} {1997})}\BibitemShut {NoStop}%
\bibitem [{\citenamefont {D'Ariano}\ and\ \citenamefont
  {Lo~Presti}(2001)}]{DAriano2001-fx}%
  \BibitemOpen
  \bibfield  {author} {\bibinfo {author} {\bibfnamefont {G.~M.}\ \bibnamefont
  {D'Ariano}}\ and\ \bibinfo {author} {\bibfnamefont {P.}~\bibnamefont
  {Lo~Presti}},\ }\bibfield  {title} {\bibinfo {title} {Quantum tomography for
  measuring experimentally the matrix elements of an arbitrary quantum
  operation},\ }\href {https://doi.org/10.1103/PhysRevLett.86.4195} {\bibfield
  {journal} {\bibinfo  {journal} {Phys. Rev. Lett.}\ }\textbf {\bibinfo
  {volume} {86}},\ \bibinfo {pages} {4195} (\bibinfo {year}
  {2001})}\BibitemShut {NoStop}%
\bibitem [{\citenamefont {Fiur\'a\ifmmode~\check{s}\else \v{s}\fi{}ek}\ and\
  \citenamefont {Hradil}(2001)}]{Fiurasek2001-mp}%
  \BibitemOpen
  \bibfield  {author} {\bibinfo {author} {\bibfnamefont {J.}~\bibnamefont
  {Fiur\'a\ifmmode~\check{s}\else \v{s}\fi{}ek}}\ and\ \bibinfo {author}
  {\bibfnamefont {Z.~c.~v.}\ \bibnamefont {Hradil}},\ }\bibfield  {title}
  {\bibinfo {title} {Maximum-likelihood estimation of quantum processes},\
  }\href {https://doi.org/10.1103/PhysRevA.63.020101} {\bibfield  {journal}
  {\bibinfo  {journal} {Phys. Rev. A}\ }\textbf {\bibinfo {volume} {63}},\
  \bibinfo {pages} {020101} (\bibinfo {year} {2001})}\BibitemShut {NoStop}%
\bibitem [{\citenamefont {Gale}\ \emph {et~al.}(1968)\citenamefont {Gale},
  \citenamefont {Guth},\ and\ \citenamefont {Trammell}}]{Gale1968-mk}%
  \BibitemOpen
  \bibfield  {author} {\bibinfo {author} {\bibfnamefont {W.}~\bibnamefont
  {Gale}}, \bibinfo {author} {\bibfnamefont {E.}~\bibnamefont {Guth}},\ and\
  \bibinfo {author} {\bibfnamefont {G.~T.}\ \bibnamefont {Trammell}},\
  }\bibfield  {title} {\bibinfo {title} {Determination of the quantum state by
  measurements},\ }\href {https://doi.org/10.1103/PhysRev.165.1434} {\bibfield
  {journal} {\bibinfo  {journal} {Phys. Rev.}\ }\textbf {\bibinfo {volume}
  {165}},\ \bibinfo {pages} {1434} (\bibinfo {year} {1968})}\BibitemShut
  {NoStop}%
\bibitem [{\citenamefont {Granade}\ \emph {et~al.}(2016)\citenamefont
  {Granade}, \citenamefont {Combes},\ and\ \citenamefont
  {Cory}}]{Granade2016-qy}%
  \BibitemOpen
  \bibfield  {author} {\bibinfo {author} {\bibfnamefont {C.}~\bibnamefont
  {Granade}}, \bibinfo {author} {\bibfnamefont {J.}~\bibnamefont {Combes}},\
  and\ \bibinfo {author} {\bibfnamefont {D.~G.}\ \bibnamefont {Cory}},\
  }\bibfield  {title} {\bibinfo {title} {Practical bayesian tomography},\
  }\href {https://doi.org/10.1088/1367-2630/18/3/033024} {\bibfield  {journal}
  {\bibinfo  {journal} {New Journal of Physics}\ }\textbf {\bibinfo {volume}
  {18}},\ \bibinfo {pages} {033024} (\bibinfo {year} {2016})}\BibitemShut
  {NoStop}%
\bibitem [{\citenamefont {Haah}\ \emph {et~al.}(2017)\citenamefont {Haah},
  \citenamefont {Harrow}, \citenamefont {Ji}, \citenamefont {Wu},\ and\
  \citenamefont {Yu}}]{Haah2017-jw}%
  \BibitemOpen
  \bibfield  {author} {\bibinfo {author} {\bibfnamefont {J.}~\bibnamefont
  {Haah}}, \bibinfo {author} {\bibfnamefont {A.~W.}\ \bibnamefont {Harrow}},
  \bibinfo {author} {\bibfnamefont {Z.}~\bibnamefont {Ji}}, \bibinfo {author}
  {\bibfnamefont {X.}~\bibnamefont {Wu}},\ and\ \bibinfo {author}
  {\bibfnamefont {N.}~\bibnamefont {Yu}},\ }\bibfield  {title} {\bibinfo
  {title} {{Sample-Optimal} tomography of quantum states},\ }\href
  {https://doi.org/10.1109/TIT.2017.2719044} {\bibfield  {journal} {\bibinfo
  {journal} {IEEE Trans. Inf. Theory}\ }\textbf {\bibinfo {volume} {63}},\
  \bibinfo {pages} {5628} (\bibinfo {year} {2017})}\BibitemShut {NoStop}%
\bibitem [{\citenamefont {Hradil}(1997)}]{Hradil1997-fv}%
  \BibitemOpen
  \bibfield  {author} {\bibinfo {author} {\bibfnamefont {Z.}~\bibnamefont
  {Hradil}},\ }\bibfield  {title} {\bibinfo {title} {Quantum-state
  estimation},\ }\href {https://doi.org/10.1103/PhysRevA.55.R1561} {\bibfield
  {journal} {\bibinfo  {journal} {Phys. Rev. A}\ }\textbf {\bibinfo {volume}
  {55}},\ \bibinfo {pages} {R1561} (\bibinfo {year} {1997})}\BibitemShut
  {NoStop}%
\bibitem [{\citenamefont {Kim}\ \emph {et~al.}(2014)\citenamefont {Kim},
  \citenamefont {Shi}, \citenamefont {Simmons}, \citenamefont {Ward},
  \citenamefont {Prance}, \citenamefont {Koh}, \citenamefont {Gamble},
  \citenamefont {Savage}, \citenamefont {Lagally}, \citenamefont {Friesen},
  \citenamefont {Coppersmith},\ and\ \citenamefont {Eriksson}}]{Kim2014-ix}%
  \BibitemOpen
  \bibfield  {author} {\bibinfo {author} {\bibfnamefont {D.}~\bibnamefont
  {Kim}}, \bibinfo {author} {\bibfnamefont {Z.}~\bibnamefont {Shi}}, \bibinfo
  {author} {\bibfnamefont {C.~B.}\ \bibnamefont {Simmons}}, \bibinfo {author}
  {\bibfnamefont {D.~R.}\ \bibnamefont {Ward}}, \bibinfo {author}
  {\bibfnamefont {J.~R.}\ \bibnamefont {Prance}}, \bibinfo {author}
  {\bibfnamefont {T.~S.}\ \bibnamefont {Koh}}, \bibinfo {author} {\bibfnamefont
  {J.~K.}\ \bibnamefont {Gamble}}, \bibinfo {author} {\bibfnamefont {D.~E.}\
  \bibnamefont {Savage}}, \bibinfo {author} {\bibfnamefont {M.~G.}\
  \bibnamefont {Lagally}}, \bibinfo {author} {\bibfnamefont {M.}~\bibnamefont
  {Friesen}}, \bibinfo {author} {\bibfnamefont {S.~N.}\ \bibnamefont
  {Coppersmith}},\ and\ \bibinfo {author} {\bibfnamefont {M.~A.}\ \bibnamefont
  {Eriksson}},\ }\bibfield  {title} {\bibinfo {title} {Quantum control and
  process tomography of a semiconductor quantum dot hybrid qubit},\ }\href
  {https://doi.org/10.1038/nature13407} {\bibfield  {journal} {\bibinfo
  {journal} {Nature}\ }\textbf {\bibinfo {volume} {511}},\ \bibinfo {pages}
  {70} (\bibinfo {year} {2014})}\BibitemShut {NoStop}%
\bibitem [{\citenamefont {Kimmel}\ \emph {et~al.}(2014)\citenamefont {Kimmel},
  \citenamefont {da~Silva}, \citenamefont {Ryan}, \citenamefont {Johnson},\
  and\ \citenamefont {Ohki}}]{Kimmel2014-nx}%
  \BibitemOpen
  \bibfield  {author} {\bibinfo {author} {\bibfnamefont {S.}~\bibnamefont
  {Kimmel}}, \bibinfo {author} {\bibfnamefont {M.~P.}\ \bibnamefont
  {da~Silva}}, \bibinfo {author} {\bibfnamefont {C.~A.}\ \bibnamefont {Ryan}},
  \bibinfo {author} {\bibfnamefont {B.~R.}\ \bibnamefont {Johnson}},\ and\
  \bibinfo {author} {\bibfnamefont {T.}~\bibnamefont {Ohki}},\ }\bibfield
  {title} {\bibinfo {title} {Robust extraction of tomographic information via
  randomized benchmarking},\ }\href {https://doi.org/10.1103/PhysRevX.4.011050}
  {\bibfield  {journal} {\bibinfo  {journal} {Phys. Rev. X}\ }\textbf {\bibinfo
  {volume} {4}},\ \bibinfo {pages} {011050} (\bibinfo {year}
  {2014})}\BibitemShut {NoStop}%
\bibitem [{\citenamefont {Lobino}\ \emph {et~al.}(2008)\citenamefont {Lobino},
  \citenamefont {Korystov}, \citenamefont {Kupchak}, \citenamefont {Figueroa},
  \citenamefont {Sanders},\ and\ \citenamefont {Lvovsky}}]{Lobino2008-ud}%
  \BibitemOpen
  \bibfield  {author} {\bibinfo {author} {\bibfnamefont {M.}~\bibnamefont
  {Lobino}}, \bibinfo {author} {\bibfnamefont {D.}~\bibnamefont {Korystov}},
  \bibinfo {author} {\bibfnamefont {C.}~\bibnamefont {Kupchak}}, \bibinfo
  {author} {\bibfnamefont {E.}~\bibnamefont {Figueroa}}, \bibinfo {author}
  {\bibfnamefont {B.~C.}\ \bibnamefont {Sanders}},\ and\ \bibinfo {author}
  {\bibfnamefont {A.~I.}\ \bibnamefont {Lvovsky}},\ }\bibfield  {title}
  {\bibinfo {title} {Complete characterization of quantum-optical processes},\
  }\href {https://doi.org/10.1126/science.1162086} {\bibfield  {journal}
  {\bibinfo  {journal} {Science}\ }\textbf {\bibinfo {volume} {322}},\ \bibinfo
  {pages} {563} (\bibinfo {year} {2008})}\BibitemShut {NoStop}%
\bibitem [{\citenamefont {James}\ \emph {et~al.}(2001)\citenamefont {James},
  \citenamefont {Kwiat}, \citenamefont {Munro},\ and\ \citenamefont
  {White}}]{James2001-hz}%
  \BibitemOpen
  \bibfield  {author} {\bibinfo {author} {\bibfnamefont {D.~F.~V.}\
  \bibnamefont {James}}, \bibinfo {author} {\bibfnamefont {P.~G.}\ \bibnamefont
  {Kwiat}}, \bibinfo {author} {\bibfnamefont {W.~J.}\ \bibnamefont {Munro}},\
  and\ \bibinfo {author} {\bibfnamefont {A.~G.}\ \bibnamefont {White}},\
  }\bibfield  {title} {\bibinfo {title} {Measurement of qubits},\ }\href
  {https://doi.org/10.1103/PhysRevA.64.052312} {\bibfield  {journal} {\bibinfo
  {journal} {Phys. Rev. A}\ }\textbf {\bibinfo {volume} {64}},\ \bibinfo
  {pages} {052312} (\bibinfo {year} {2001})}\BibitemShut {NoStop}%
\bibitem [{\citenamefont {O'Brien}\ \emph {et~al.}(2004)\citenamefont
  {O'Brien}, \citenamefont {Pryde}, \citenamefont {Gilchrist}, \citenamefont
  {James}, \citenamefont {Langford}, \citenamefont {Ralph},\ and\ \citenamefont
  {White}}]{OBrien2004-tr}%
  \BibitemOpen
  \bibfield  {author} {\bibinfo {author} {\bibfnamefont {J.~L.}\ \bibnamefont
  {O'Brien}}, \bibinfo {author} {\bibfnamefont {G.~J.}\ \bibnamefont {Pryde}},
  \bibinfo {author} {\bibfnamefont {A.}~\bibnamefont {Gilchrist}}, \bibinfo
  {author} {\bibfnamefont {D.~F.~V.}\ \bibnamefont {James}}, \bibinfo {author}
  {\bibfnamefont {N.~K.}\ \bibnamefont {Langford}}, \bibinfo {author}
  {\bibfnamefont {T.~C.}\ \bibnamefont {Ralph}},\ and\ \bibinfo {author}
  {\bibfnamefont {A.~G.}\ \bibnamefont {White}},\ }\bibfield  {title} {\bibinfo
  {title} {Quantum process tomography of a controlled-not gate},\ }\href
  {https://doi.org/10.1103/PhysRevLett.93.080502} {\bibfield  {journal}
  {\bibinfo  {journal} {Phys. Rev. Lett.}\ }\textbf {\bibinfo {volume} {93}},\
  \bibinfo {pages} {080502} (\bibinfo {year} {2004})}\BibitemShut {NoStop}%
\bibitem [{\citenamefont {Poyatos}\ \emph {et~al.}(1997)\citenamefont
  {Poyatos}, \citenamefont {Cirac},\ and\ \citenamefont
  {Zoller}}]{Poyatos1997-mz}%
  \BibitemOpen
  \bibfield  {author} {\bibinfo {author} {\bibfnamefont {J.~F.}\ \bibnamefont
  {Poyatos}}, \bibinfo {author} {\bibfnamefont {J.~I.}\ \bibnamefont {Cirac}},\
  and\ \bibinfo {author} {\bibfnamefont {P.}~\bibnamefont {Zoller}},\
  }\bibfield  {title} {\bibinfo {title} {Complete characterization of a quantum
  process: The {Two-Bit} quantum gate},\ }\href
  {https://doi.org/10.1103/PhysRevLett.78.390} {\bibfield  {journal} {\bibinfo
  {journal} {Phys. Rev. Lett.}\ }\textbf {\bibinfo {volume} {78}},\ \bibinfo
  {pages} {390} (\bibinfo {year} {1997})}\BibitemShut {NoStop}%
\bibitem [{\citenamefont {Riebe}\ \emph {et~al.}(2006)\citenamefont {Riebe},
  \citenamefont {Kim}, \citenamefont {Schindler}, \citenamefont {Monz},
  \citenamefont {Schmidt}, \citenamefont {K{\"o}rber}, \citenamefont
  {H{\"a}nsel}, \citenamefont {H{\"a}ffner}, \citenamefont {Roos},\ and\
  \citenamefont {Blatt}}]{Riebe2006-tc}%
  \BibitemOpen
  \bibfield  {author} {\bibinfo {author} {\bibfnamefont {M.}~\bibnamefont
  {Riebe}}, \bibinfo {author} {\bibfnamefont {K.}~\bibnamefont {Kim}}, \bibinfo
  {author} {\bibfnamefont {P.}~\bibnamefont {Schindler}}, \bibinfo {author}
  {\bibfnamefont {T.}~\bibnamefont {Monz}}, \bibinfo {author} {\bibfnamefont
  {P.~O.}\ \bibnamefont {Schmidt}}, \bibinfo {author} {\bibfnamefont {T.~K.}\
  \bibnamefont {K{\"o}rber}}, \bibinfo {author} {\bibfnamefont
  {W.}~\bibnamefont {H{\"a}nsel}}, \bibinfo {author} {\bibfnamefont
  {H.}~\bibnamefont {H{\"a}ffner}}, \bibinfo {author} {\bibfnamefont {C.~F.}\
  \bibnamefont {Roos}},\ and\ \bibinfo {author} {\bibfnamefont
  {R.}~\bibnamefont {Blatt}},\ }\bibfield  {title} {\bibinfo {title} {Process
  tomography of ion trap quantum gates},\ }\href
  {https://doi.org/10.1103/PhysRevLett.97.220407} {\bibfield  {journal}
  {\bibinfo  {journal} {Phys. Rev. Lett.}\ }\textbf {\bibinfo {volume} {97}},\
  \bibinfo {pages} {220407} (\bibinfo {year} {2006})}\BibitemShut {NoStop}%
\bibitem [{\citenamefont {Smolin}\ \emph {et~al.}(2012)\citenamefont {Smolin},
  \citenamefont {Gambetta},\ and\ \citenamefont {Smith}}]{Smolin2012-yv}%
  \BibitemOpen
  \bibfield  {author} {\bibinfo {author} {\bibfnamefont {J.~A.}\ \bibnamefont
  {Smolin}}, \bibinfo {author} {\bibfnamefont {J.~M.}\ \bibnamefont
  {Gambetta}},\ and\ \bibinfo {author} {\bibfnamefont {G.}~\bibnamefont
  {Smith}},\ }\bibfield  {title} {\bibinfo {title} {Efficient method for
  computing the maximum-likelihood quantum state from measurements with
  additive gaussian noise},\ }\href
  {https://doi.org/10.1103/PhysRevLett.108.070502} {\bibfield  {journal}
  {\bibinfo  {journal} {Phys. Rev. Lett.}\ }\textbf {\bibinfo {volume} {108}},\
  \bibinfo {pages} {070502} (\bibinfo {year} {2012})}\BibitemShut {NoStop}%
\bibitem [{\citenamefont {Vogel}\ and\ \citenamefont
  {Risken}(1989)}]{Vogel1989-vt}%
  \BibitemOpen
  \bibfield  {author} {\bibinfo {author} {\bibfnamefont {K.}~\bibnamefont
  {Vogel}}\ and\ \bibinfo {author} {\bibfnamefont {H.}~\bibnamefont {Risken}},\
  }\bibfield  {title} {\bibinfo {title} {Determination of quasiprobability
  distributions in terms of probability distributions for the rotated
  quadrature phase},\ }\href {https://doi.org/10.1103/physreva.40.2847}
  {\bibfield  {journal} {\bibinfo  {journal} {Phys. Rev. A Gen. Phys.}\
  }\textbf {\bibinfo {volume} {40}},\ \bibinfo {pages} {2847} (\bibinfo {year}
  {1989})}\BibitemShut {NoStop}%
\bibitem [{\citenamefont {Weinstein}\ \emph {et~al.}(2004)\citenamefont
  {Weinstein}, \citenamefont {Havel}, \citenamefont {Emerson}, \citenamefont
  {Boulant}, \citenamefont {Saraceno}, \citenamefont {Lloyd},\ and\
  \citenamefont {Cory}}]{Weinstein2004-vn}%
  \BibitemOpen
  \bibfield  {author} {\bibinfo {author} {\bibfnamefont {Y.~S.}\ \bibnamefont
  {Weinstein}}, \bibinfo {author} {\bibfnamefont {T.~F.}\ \bibnamefont
  {Havel}}, \bibinfo {author} {\bibfnamefont {J.}~\bibnamefont {Emerson}},
  \bibinfo {author} {\bibfnamefont {N.}~\bibnamefont {Boulant}}, \bibinfo
  {author} {\bibfnamefont {M.}~\bibnamefont {Saraceno}}, \bibinfo {author}
  {\bibfnamefont {S.}~\bibnamefont {Lloyd}},\ and\ \bibinfo {author}
  {\bibfnamefont {D.~G.}\ \bibnamefont {Cory}},\ }\bibfield  {title} {\bibinfo
  {title} {Quantum process tomography of the quantum fourier transform},\
  }\href {https://doi.org/10.1063/1.1785151} {\bibfield  {journal} {\bibinfo
  {journal} {The Journal of Chemical Physics}\ }\textbf {\bibinfo {volume}
  {121}},\ \bibinfo {pages} {6117} (\bibinfo {year} {2004})}\BibitemShut
  {NoStop}%
\bibitem [{\citenamefont {Nielsen}\ \emph {et~al.}(2021)\citenamefont
  {Nielsen}, \citenamefont {Gamble}, \citenamefont {Rudinger}, \citenamefont
  {Scholten}, \citenamefont {Young},\ and\ \citenamefont
  {Blume-Kohout}}]{Nielsen2020_GST}%
  \BibitemOpen
  \bibfield  {author} {\bibinfo {author} {\bibfnamefont {E.}~\bibnamefont
  {Nielsen}}, \bibinfo {author} {\bibfnamefont {J.~K.}\ \bibnamefont {Gamble}},
  \bibinfo {author} {\bibfnamefont {K.}~\bibnamefont {Rudinger}}, \bibinfo
  {author} {\bibfnamefont {T.}~\bibnamefont {Scholten}}, \bibinfo {author}
  {\bibfnamefont {K.}~\bibnamefont {Young}},\ and\ \bibinfo {author}
  {\bibfnamefont {R.}~\bibnamefont {Blume-Kohout}},\ }\bibfield  {title}
  {\bibinfo {title} {Gate {S}et {T}omography},\ }\href
  {https://doi.org/10.22331/q-2021-10-05-557} {\bibfield  {journal} {\bibinfo
  {journal} {{Quantum}}\ }\textbf {\bibinfo {volume} {5}},\ \bibinfo {pages}
  {557} (\bibinfo {year} {2021})}\BibitemShut {NoStop}%
\bibitem [{\citenamefont {Merkel}\ \emph {et~al.}(2013)\citenamefont {Merkel},
  \citenamefont {Gambetta}, \citenamefont {Smolin}, \citenamefont {Poletto},
  \citenamefont {C\'orcoles}, \citenamefont {Johnson}, \citenamefont {Ryan},\
  and\ \citenamefont {Steffen}}]{MerkelPRA13}%
  \BibitemOpen
  \bibfield  {author} {\bibinfo {author} {\bibfnamefont {S.~T.}\ \bibnamefont
  {Merkel}}, \bibinfo {author} {\bibfnamefont {J.~M.}\ \bibnamefont
  {Gambetta}}, \bibinfo {author} {\bibfnamefont {J.~A.}\ \bibnamefont
  {Smolin}}, \bibinfo {author} {\bibfnamefont {S.}~\bibnamefont {Poletto}},
  \bibinfo {author} {\bibfnamefont {A.~D.}\ \bibnamefont {C\'orcoles}},
  \bibinfo {author} {\bibfnamefont {B.~R.}\ \bibnamefont {Johnson}}, \bibinfo
  {author} {\bibfnamefont {C.~A.}\ \bibnamefont {Ryan}},\ and\ \bibinfo
  {author} {\bibfnamefont {M.}~\bibnamefont {Steffen}},\ }\bibfield  {title}
  {\bibinfo {title} {Self-consistent quantum process tomography},\ }\href
  {https://doi.org/10.1103/PhysRevA.87.062119} {\bibfield  {journal} {\bibinfo
  {journal} {Phys. Rev. A}\ }\textbf {\bibinfo {volume} {87}},\ \bibinfo
  {pages} {062119} (\bibinfo {year} {2013})}\BibitemShut {NoStop}%
\bibitem [{\citenamefont {Blume-Kohout}\ \emph {et~al.}(2013)\citenamefont
  {Blume-Kohout}, \citenamefont {Gamble}, \citenamefont {Nielsen},
  \citenamefont {Mizrahi}, \citenamefont {Sterk},\ and\ \citenamefont
  {Maunz}}]{GST2013}%
  \BibitemOpen
  \bibfield  {author} {\bibinfo {author} {\bibfnamefont {R.}~\bibnamefont
  {Blume-Kohout}}, \bibinfo {author} {\bibfnamefont {J.}~\bibnamefont
  {Gamble}}, \bibinfo {author} {\bibfnamefont {E.}~\bibnamefont {Nielsen}},
  \bibinfo {author} {\bibfnamefont {J.}~\bibnamefont {Mizrahi}}, \bibinfo
  {author} {\bibfnamefont {J.}~\bibnamefont {Sterk}},\ and\ \bibinfo {author}
  {\bibfnamefont {P.}~\bibnamefont {Maunz}},\ }\bibfield  {title} {\bibinfo
  {title} {Robust, self-consistent, closed-form tomography of quantum logic
  gates on a trapped ion qubit},\ }\href@noop {} {\bibfield  {journal}
  {\bibinfo  {journal} {arXiv preprint arXiv:1310.4492}\ } (\bibinfo {year}
  {2013})}\BibitemShut {NoStop}%
\bibitem [{\citenamefont {Greenbaum}(2015)}]{Greenbaum15}%
  \BibitemOpen
  \bibfield  {author} {\bibinfo {author} {\bibfnamefont {D.}~\bibnamefont
  {Greenbaum}},\ }\bibfield  {title} {\bibinfo {title} {Introduction to quantum
  gate set tomography},\ }\href@noop {} {\bibfield  {journal} {\bibinfo
  {journal} {arXiv:1509.02921}\ } (\bibinfo {year} {2015})}\BibitemShut
  {NoStop}%
\bibitem [{\citenamefont {Blume-Kohout}\ \emph {et~al.}(2017)\citenamefont
  {Blume-Kohout}, \citenamefont {Gamble}, \citenamefont {Nielsen},
  \citenamefont {Rudinger}, \citenamefont {Mizrahi}, \citenamefont {Fortier},\
  and\ \citenamefont {Maunz}}]{Blume-Kohout2017-kn}%
  \BibitemOpen
  \bibfield  {author} {\bibinfo {author} {\bibfnamefont {R.}~\bibnamefont
  {Blume-Kohout}}, \bibinfo {author} {\bibfnamefont {J.~K.}\ \bibnamefont
  {Gamble}}, \bibinfo {author} {\bibfnamefont {E.}~\bibnamefont {Nielsen}},
  \bibinfo {author} {\bibfnamefont {K.}~\bibnamefont {Rudinger}}, \bibinfo
  {author} {\bibfnamefont {J.}~\bibnamefont {Mizrahi}}, \bibinfo {author}
  {\bibfnamefont {K.}~\bibnamefont {Fortier}},\ and\ \bibinfo {author}
  {\bibfnamefont {P.}~\bibnamefont {Maunz}},\ }\bibfield  {title} {\bibinfo
  {title} {Demonstration of qubit operations below a rigorous fault tolerance
  threshold with gate set tomography},\ }\bibfield  {journal} {\bibinfo
  {journal} {Nat. Commun.}\ }\textbf {\bibinfo {volume} {8}},\ \href
  {https://doi.org/10.1038/ncomms14485} {10.1038/ncomms14485} (\bibinfo {year}
  {2017})\BibitemShut {NoStop}%
\bibitem [{\citenamefont {Kim}\ \emph {et~al.}(2015)\citenamefont {Kim},
  \citenamefont {Ward}, \citenamefont {Simmons}, \citenamefont {Gamble},
  \citenamefont {Blume-Kohout}, \citenamefont {Nielsen}, \citenamefont
  {Savage}, \citenamefont {Lagally}, \citenamefont {Friesen}, \citenamefont
  {Coppersmith},\ and\ \citenamefont {Eriksson}}]{GST2015}%
  \BibitemOpen
  \bibfield  {author} {\bibinfo {author} {\bibfnamefont {D.}~\bibnamefont
  {Kim}}, \bibinfo {author} {\bibfnamefont {D.~R.}\ \bibnamefont {Ward}},
  \bibinfo {author} {\bibfnamefont {C.~B.}\ \bibnamefont {Simmons}}, \bibinfo
  {author} {\bibfnamefont {J.~K.}\ \bibnamefont {Gamble}}, \bibinfo {author}
  {\bibfnamefont {R.}~\bibnamefont {Blume-Kohout}}, \bibinfo {author}
  {\bibfnamefont {E.}~\bibnamefont {Nielsen}}, \bibinfo {author} {\bibfnamefont
  {D.~E.}\ \bibnamefont {Savage}}, \bibinfo {author} {\bibfnamefont {M.~G.}\
  \bibnamefont {Lagally}}, \bibinfo {author} {\bibfnamefont {M.}~\bibnamefont
  {Friesen}}, \bibinfo {author} {\bibfnamefont {S.~N.}\ \bibnamefont
  {Coppersmith}},\ and\ \bibinfo {author} {\bibfnamefont {M.~A.}\ \bibnamefont
  {Eriksson}},\ }\bibfield  {title} {\bibinfo {title} {Microwave-driven
  coherent operation of a semiconductor quantum dot charge qubit},\ }\href
  {https://doi.org/10.1038/nnano.2014.336} {\bibfield  {journal} {\bibinfo
  {journal} {Nat. Nanotechnol.}\ }\textbf {\bibinfo {volume} {10}},\ \bibinfo
  {pages} {243} (\bibinfo {year} {2015})}\BibitemShut {NoStop}%
\bibitem [{\citenamefont {Dehollain}\ \emph
  {et~al.}(2016{\natexlab{a}})\citenamefont {Dehollain}, \citenamefont
  {Muhonen}, \citenamefont {Blume-Kohout}, \citenamefont {Rudinger},
  \citenamefont {Gamble}, \citenamefont {Nielsen}, \citenamefont {Laucht},
  \citenamefont {Simmons}, \citenamefont {Kalra}, \citenamefont {Dzurak},\ and\
  \citenamefont {Morello}}]{Dehollain2016-zt}%
  \BibitemOpen
  \bibfield  {author} {\bibinfo {author} {\bibfnamefont {J.~P.}\ \bibnamefont
  {Dehollain}}, \bibinfo {author} {\bibfnamefont {J.~T.}\ \bibnamefont
  {Muhonen}}, \bibinfo {author} {\bibfnamefont {R.}~\bibnamefont
  {Blume-Kohout}}, \bibinfo {author} {\bibfnamefont {K.~M.}\ \bibnamefont
  {Rudinger}}, \bibinfo {author} {\bibfnamefont {J.~K.}\ \bibnamefont
  {Gamble}}, \bibinfo {author} {\bibfnamefont {E.}~\bibnamefont {Nielsen}},
  \bibinfo {author} {\bibfnamefont {A.}~\bibnamefont {Laucht}}, \bibinfo
  {author} {\bibfnamefont {S.}~\bibnamefont {Simmons}}, \bibinfo {author}
  {\bibfnamefont {R.}~\bibnamefont {Kalra}}, \bibinfo {author} {\bibfnamefont
  {A.~S.}\ \bibnamefont {Dzurak}},\ and\ \bibinfo {author} {\bibfnamefont
  {A.}~\bibnamefont {Morello}},\ }\bibfield  {title} {\bibinfo {title}
  {Optimization of a solid-state electron spin qubit using gate set
  tomography},\ }\href {https://doi.org/10.1088/1367-2630/18/10/103018}
  {\bibfield  {journal} {\bibinfo  {journal} {New J. Phys.}\ }\textbf {\bibinfo
  {volume} {18}},\ \bibinfo {pages} {103018} (\bibinfo {year}
  {2016}{\natexlab{a}})}\BibitemShut {NoStop}%
\bibitem [{\citenamefont {Di~Matteo}\ \emph {et~al.}(2020)\citenamefont
  {Di~Matteo}, \citenamefont {Gamble}, \citenamefont {Granade}, \citenamefont
  {Rudinger},\ and\ \citenamefont {Wiebe}}]{matteo2020operational}%
  \BibitemOpen
  \bibfield  {author} {\bibinfo {author} {\bibfnamefont {O.}~\bibnamefont
  {Di~Matteo}}, \bibinfo {author} {\bibfnamefont {J.}~\bibnamefont {Gamble}},
  \bibinfo {author} {\bibfnamefont {C.}~\bibnamefont {Granade}}, \bibinfo
  {author} {\bibfnamefont {K.}~\bibnamefont {Rudinger}},\ and\ \bibinfo
  {author} {\bibfnamefont {N.}~\bibnamefont {Wiebe}},\ }\href
  {https://doi.org/10.22331/q-2020-11-17-364} {\bibinfo {title} {Operational,
  gauge-free quantum tomography}} (\bibinfo {year} {2020})\BibitemShut
  {NoStop}%
\bibitem [{\citenamefont {Hong}\ \emph
  {et~al.}(2020{\natexlab{a}})\citenamefont {Hong}, \citenamefont {Papageorge},
  \citenamefont {Sivarajah}, \citenamefont {Crossman}, \citenamefont {Didier},
  \citenamefont {Polloreno}, \citenamefont {Sete}, \citenamefont {Turkowski},
  \citenamefont {da~Silva},\ and\ \citenamefont {Johnson}}]{Hong2020-vc}%
  \BibitemOpen
  \bibfield  {author} {\bibinfo {author} {\bibfnamefont {S.~S.}\ \bibnamefont
  {Hong}}, \bibinfo {author} {\bibfnamefont {A.~T.}\ \bibnamefont
  {Papageorge}}, \bibinfo {author} {\bibfnamefont {P.}~\bibnamefont
  {Sivarajah}}, \bibinfo {author} {\bibfnamefont {G.}~\bibnamefont {Crossman}},
  \bibinfo {author} {\bibfnamefont {N.}~\bibnamefont {Didier}}, \bibinfo
  {author} {\bibfnamefont {A.~M.}\ \bibnamefont {Polloreno}}, \bibinfo {author}
  {\bibfnamefont {E.~A.}\ \bibnamefont {Sete}}, \bibinfo {author}
  {\bibfnamefont {S.~W.}\ \bibnamefont {Turkowski}}, \bibinfo {author}
  {\bibfnamefont {M.~P.}\ \bibnamefont {da~Silva}},\ and\ \bibinfo {author}
  {\bibfnamefont {B.~R.}\ \bibnamefont {Johnson}},\ }\bibfield  {title}
  {\bibinfo {title} {Demonstration of a parametrically activated entangling
  gate protected from flux noise},\ }\href
  {https://doi.org/10.1103/PhysRevA.101.012302} {\bibfield  {journal} {\bibinfo
   {journal} {Phys. Rev. A}\ }\textbf {\bibinfo {volume} {101}},\ \bibinfo
  {pages} {012302} (\bibinfo {year} {2020}{\natexlab{a}})}\BibitemShut
  {NoStop}%
\bibitem [{\citenamefont {Joshi}\ \emph {et~al.}(2020)\citenamefont {Joshi},
  \citenamefont {Elben}, \citenamefont {Vermersch}, \citenamefont {Brydges},
  \citenamefont {Maier}, \citenamefont {Zoller}, \citenamefont {Blatt},\ and\
  \citenamefont {Roos}}]{Joshi2020-wo}%
  \BibitemOpen
  \bibfield  {author} {\bibinfo {author} {\bibfnamefont {M.~K.}\ \bibnamefont
  {Joshi}}, \bibinfo {author} {\bibfnamefont {A.}~\bibnamefont {Elben}},
  \bibinfo {author} {\bibfnamefont {B.}~\bibnamefont {Vermersch}}, \bibinfo
  {author} {\bibfnamefont {T.}~\bibnamefont {Brydges}}, \bibinfo {author}
  {\bibfnamefont {C.}~\bibnamefont {Maier}}, \bibinfo {author} {\bibfnamefont
  {P.}~\bibnamefont {Zoller}}, \bibinfo {author} {\bibfnamefont
  {R.}~\bibnamefont {Blatt}},\ and\ \bibinfo {author} {\bibfnamefont {C.~F.}\
  \bibnamefont {Roos}},\ }\bibfield  {title} {\bibinfo {title} {Quantum
  information scrambling in a trapped-ion quantum simulator with tunable range
  interactions},\ }\href {https://doi.org/10.1103/PhysRevLett.124.240505}
  {\bibfield  {journal} {\bibinfo  {journal} {Phys. Rev. Lett.}\ }\textbf
  {\bibinfo {volume} {124}},\ \bibinfo {pages} {240505} (\bibinfo {year}
  {2020})}\BibitemShut {NoStop}%
\bibitem [{\citenamefont {Proctor}\ \emph {et~al.}(2020)\citenamefont
  {Proctor}, \citenamefont {Revelle}, \citenamefont {Nielsen}, \citenamefont
  {Rudinger}, \citenamefont {Lobser}, \citenamefont {Maunz}, \citenamefont
  {Blume-Kohout},\ and\ \citenamefont {Young}}]{Proctor2019-oi}%
  \BibitemOpen
  \bibfield  {author} {\bibinfo {author} {\bibfnamefont {T.}~\bibnamefont
  {Proctor}}, \bibinfo {author} {\bibfnamefont {M.}~\bibnamefont {Revelle}},
  \bibinfo {author} {\bibfnamefont {E.}~\bibnamefont {Nielsen}}, \bibinfo
  {author} {\bibfnamefont {K.}~\bibnamefont {Rudinger}}, \bibinfo {author}
  {\bibfnamefont {D.}~\bibnamefont {Lobser}}, \bibinfo {author} {\bibfnamefont
  {P.}~\bibnamefont {Maunz}}, \bibinfo {author} {\bibfnamefont
  {R.}~\bibnamefont {Blume-Kohout}},\ and\ \bibinfo {author} {\bibfnamefont
  {K.}~\bibnamefont {Young}},\ }\bibfield  {title} {\bibinfo {title} {Detecting
  and tracking drift in quantum information processors},\ }\href
  {https://doi.org/10.1038/s41467-020-19074-4} {\bibfield  {journal} {\bibinfo
  {journal} {Nature Communications}\ }\textbf {\bibinfo {volume} {11}},\
  \bibinfo {pages} {5396} (\bibinfo {year} {2020})}\BibitemShut {NoStop}%
\bibitem [{\citenamefont {Song}\ \emph {et~al.}(2019)\citenamefont {Song},
  \citenamefont {Cui}, \citenamefont {Wang}, \citenamefont {Hao}, \citenamefont
  {Feng},\ and\ \citenamefont {Li}}]{Song2019-fg}%
  \BibitemOpen
  \bibfield  {author} {\bibinfo {author} {\bibfnamefont {C.}~\bibnamefont
  {Song}}, \bibinfo {author} {\bibfnamefont {J.}~\bibnamefont {Cui}}, \bibinfo
  {author} {\bibfnamefont {H.}~\bibnamefont {Wang}}, \bibinfo {author}
  {\bibfnamefont {J.}~\bibnamefont {Hao}}, \bibinfo {author} {\bibfnamefont
  {H.}~\bibnamefont {Feng}},\ and\ \bibinfo {author} {\bibfnamefont
  {Y.}~\bibnamefont {Li}},\ }\bibfield  {title} {\bibinfo {title} {Quantum
  computation with universal error mitigation on a superconducting quantum
  processor},\ }\href {https://doi.org/10.1126/sciadv.aaw5686} {\bibfield
  {journal} {\bibinfo  {journal} {Sci Adv}\ }\textbf {\bibinfo {volume} {5}},\
  \bibinfo {pages} {eaaw5686} (\bibinfo {year} {2019})}\BibitemShut {NoStop}%
\bibitem [{\citenamefont {Ware}\ \emph {et~al.}(2021)\citenamefont {Ware},
  \citenamefont {Ribeill}, \citenamefont {Rist\`e}, \citenamefont {Ryan},
  \citenamefont {Johnson},\ and\ \citenamefont {da~Silva}}]{Ware2018-cq}%
  \BibitemOpen
  \bibfield  {author} {\bibinfo {author} {\bibfnamefont {M.}~\bibnamefont
  {Ware}}, \bibinfo {author} {\bibfnamefont {G.}~\bibnamefont {Ribeill}},
  \bibinfo {author} {\bibfnamefont {D.}~\bibnamefont {Rist\`e}}, \bibinfo
  {author} {\bibfnamefont {C.~A.}\ \bibnamefont {Ryan}}, \bibinfo {author}
  {\bibfnamefont {B.}~\bibnamefont {Johnson}},\ and\ \bibinfo {author}
  {\bibfnamefont {M.~P.}\ \bibnamefont {da~Silva}},\ }\bibfield  {title}
  {\bibinfo {title} {Experimental pauli-frame randomization on a
  superconducting qubit},\ }\href {https://doi.org/10.1103/PhysRevA.103.042604}
  {\bibfield  {journal} {\bibinfo  {journal} {Phys. Rev. A}\ }\textbf {\bibinfo
  {volume} {103}},\ \bibinfo {pages} {042604} (\bibinfo {year}
  {2021})}\BibitemShut {NoStop}%
\bibitem [{\citenamefont {Zhang}\ \emph
  {et~al.}(2020{\natexlab{a}})\citenamefont {Zhang}, \citenamefont {Lu},
  \citenamefont {Zhang}, \citenamefont {Chen}, \citenamefont {Li},
  \citenamefont {Zhang},\ and\ \citenamefont {Kim}}]{Zhang2020-ux}%
  \BibitemOpen
  \bibfield  {author} {\bibinfo {author} {\bibfnamefont {S.}~\bibnamefont
  {Zhang}}, \bibinfo {author} {\bibfnamefont {Y.}~\bibnamefont {Lu}}, \bibinfo
  {author} {\bibfnamefont {K.}~\bibnamefont {Zhang}}, \bibinfo {author}
  {\bibfnamefont {W.}~\bibnamefont {Chen}}, \bibinfo {author} {\bibfnamefont
  {Y.}~\bibnamefont {Li}}, \bibinfo {author} {\bibfnamefont {J.-N.}\
  \bibnamefont {Zhang}},\ and\ \bibinfo {author} {\bibfnamefont
  {K.}~\bibnamefont {Kim}},\ }\bibfield  {title} {\bibinfo {title}
  {Error-mitigated quantum gates exceeding physical fidelities in a trapped-ion
  system},\ }\href {https://doi.org/10.1038/s41467-020-14376-z} {\bibfield
  {journal} {\bibinfo  {journal} {Nat. Commun.}\ }\textbf {\bibinfo {volume}
  {11}},\ \bibinfo {pages} {587} (\bibinfo {year}
  {2020}{\natexlab{a}})}\BibitemShut {NoStop}%
\bibitem [{\citenamefont {Mavadia}\ \emph {et~al.}(2018)\citenamefont
  {Mavadia}, \citenamefont {Edmunds}, \citenamefont {Hempel}, \citenamefont
  {Ball}, \citenamefont {Roy}, \citenamefont {Stace},\ and\ \citenamefont
  {Biercuk}}]{Mavadia2018-al}%
  \BibitemOpen
  \bibfield  {author} {\bibinfo {author} {\bibfnamefont {S.}~\bibnamefont
  {Mavadia}}, \bibinfo {author} {\bibfnamefont {C.~L.}\ \bibnamefont
  {Edmunds}}, \bibinfo {author} {\bibfnamefont {C.}~\bibnamefont {Hempel}},
  \bibinfo {author} {\bibfnamefont {H.}~\bibnamefont {Ball}}, \bibinfo {author}
  {\bibfnamefont {F.}~\bibnamefont {Roy}}, \bibinfo {author} {\bibfnamefont
  {T.~M.}\ \bibnamefont {Stace}},\ and\ \bibinfo {author} {\bibfnamefont
  {M.~J.}\ \bibnamefont {Biercuk}},\ }\bibfield  {title} {\bibinfo {title}
  {Experimental quantum verification in the presence of temporally correlated
  noise},\ }\href {https://doi.org/10.1038/s41534-017-0052-0} {\bibfield
  {journal} {\bibinfo  {journal} {npj Quantum Information}\ }\textbf {\bibinfo
  {volume} {4}},\ \bibinfo {pages} {7} (\bibinfo {year} {2018})}\BibitemShut
  {NoStop}%
\bibitem [{\citenamefont {Rol}\ \emph {et~al.}(2017)\citenamefont {Rol},
  \citenamefont {Bultink}, \citenamefont {O'Brien}, \citenamefont {de~Jong},
  \citenamefont {Theis}, \citenamefont {Fu}, \citenamefont {Luthi},
  \citenamefont {Vermeulen}, \citenamefont {de~Sterke}, \citenamefont {Bruno},
  \citenamefont {Deurloo}, \citenamefont {Schouten}, \citenamefont {Wilhelm},\
  and\ \citenamefont {DiCarlo}}]{Rol2017-wn}%
  \BibitemOpen
  \bibfield  {author} {\bibinfo {author} {\bibfnamefont {M.~A.}\ \bibnamefont
  {Rol}}, \bibinfo {author} {\bibfnamefont {C.~C.}\ \bibnamefont {Bultink}},
  \bibinfo {author} {\bibfnamefont {T.~E.}\ \bibnamefont {O'Brien}}, \bibinfo
  {author} {\bibfnamefont {S.~R.}\ \bibnamefont {de~Jong}}, \bibinfo {author}
  {\bibfnamefont {L.~S.}\ \bibnamefont {Theis}}, \bibinfo {author}
  {\bibfnamefont {X.}~\bibnamefont {Fu}}, \bibinfo {author} {\bibfnamefont
  {F.}~\bibnamefont {Luthi}}, \bibinfo {author} {\bibfnamefont {R.~F.~L.}\
  \bibnamefont {Vermeulen}}, \bibinfo {author} {\bibfnamefont {J.~C.}\
  \bibnamefont {de~Sterke}}, \bibinfo {author} {\bibfnamefont {A.}~\bibnamefont
  {Bruno}}, \bibinfo {author} {\bibfnamefont {D.}~\bibnamefont {Deurloo}},
  \bibinfo {author} {\bibfnamefont {R.~N.}\ \bibnamefont {Schouten}}, \bibinfo
  {author} {\bibfnamefont {F.~K.}\ \bibnamefont {Wilhelm}},\ and\ \bibinfo
  {author} {\bibfnamefont {L.}~\bibnamefont {DiCarlo}},\ }\bibfield  {title}
  {\bibinfo {title} {Restless tuneup of {High-Fidelity} qubit gates},\ }\href
  {https://doi.org/10.1103/PhysRevApplied.7.041001} {\bibfield  {journal}
  {\bibinfo  {journal} {Phys. Rev. Applied}\ }\textbf {\bibinfo {volume} {7}},\
  \bibinfo {pages} {041001} (\bibinfo {year} {2017})},\ \Eprint
  {https://arxiv.org/abs/1611.04815} {arXiv:1611.04815 [quant-ph]} \BibitemShut
  {NoStop}%
\bibitem [{\citenamefont {White}\ \emph
  {et~al.}(2021{\natexlab{a}})\citenamefont {White}, \citenamefont {Hill},\
  and\ \citenamefont {Hollenberg}}]{White2019-ls}%
  \BibitemOpen
  \bibfield  {author} {\bibinfo {author} {\bibfnamefont {G.}~\bibnamefont
  {White}}, \bibinfo {author} {\bibfnamefont {C.}~\bibnamefont {Hill}},\ and\
  \bibinfo {author} {\bibfnamefont {L.}~\bibnamefont {Hollenberg}},\ }\bibfield
   {title} {\bibinfo {title} {Performance optimization for drift-robust
  fidelity improvement of two-qubit gates},\ }\href
  {https://doi.org/10.1103/PhysRevApplied.15.014023} {\bibfield  {journal}
  {\bibinfo  {journal} {Phys. Rev. Applied}\ }\textbf {\bibinfo {volume}
  {15}},\ \bibinfo {pages} {014023} (\bibinfo {year}
  {2021}{\natexlab{a}})}\BibitemShut {NoStop}%
\bibitem [{\citenamefont {M{\k{a}}dzik}\ \emph {et~al.}(2022)\citenamefont
  {M{\k{a}}dzik}, \citenamefont {Asaad}, \citenamefont {Youssry}, \citenamefont
  {Joecker}, \citenamefont {Rudinger}, \citenamefont {Nielsen}, \citenamefont
  {Young}, \citenamefont {Proctor}, \citenamefont {Baczewski}, \citenamefont
  {Laucht} \emph {et~al.}}]{mkadzik2022precision}%
  \BibitemOpen
  \bibfield  {author} {\bibinfo {author} {\bibfnamefont {M.~T.}\ \bibnamefont
  {M{\k{a}}dzik}}, \bibinfo {author} {\bibfnamefont {S.}~\bibnamefont {Asaad}},
  \bibinfo {author} {\bibfnamefont {A.}~\bibnamefont {Youssry}}, \bibinfo
  {author} {\bibfnamefont {B.}~\bibnamefont {Joecker}}, \bibinfo {author}
  {\bibfnamefont {K.~M.}\ \bibnamefont {Rudinger}}, \bibinfo {author}
  {\bibfnamefont {E.}~\bibnamefont {Nielsen}}, \bibinfo {author} {\bibfnamefont
  {K.~C.}\ \bibnamefont {Young}}, \bibinfo {author} {\bibfnamefont {T.~J.}\
  \bibnamefont {Proctor}}, \bibinfo {author} {\bibfnamefont {A.~D.}\
  \bibnamefont {Baczewski}}, \bibinfo {author} {\bibfnamefont {A.}~\bibnamefont
  {Laucht}}, \emph {et~al.},\ }\bibfield  {title} {\bibinfo {title} {Precision
  tomography of a three-qubit donor quantum processor in silicon},\ }\href@noop
  {} {\bibfield  {journal} {\bibinfo  {journal} {Nature}\ }\textbf {\bibinfo
  {volume} {601}},\ \bibinfo {pages} {348} (\bibinfo {year}
  {2022})}\BibitemShut {NoStop}%
\bibitem [{\citenamefont {Hashim}\ \emph {et~al.}(2022)\citenamefont {Hashim},
  \citenamefont {Seritan}, \citenamefont {Proctor}, \citenamefont {Rudinger},
  \citenamefont {Goss}, \citenamefont {Naik}, \citenamefont {Kreikebaum},
  \citenamefont {Santiago},\ and\ \citenamefont
  {Siddiqi}}]{hashim2022benchmarking}%
  \BibitemOpen
  \bibfield  {author} {\bibinfo {author} {\bibfnamefont {A.}~\bibnamefont
  {Hashim}}, \bibinfo {author} {\bibfnamefont {S.}~\bibnamefont {Seritan}},
  \bibinfo {author} {\bibfnamefont {T.}~\bibnamefont {Proctor}}, \bibinfo
  {author} {\bibfnamefont {K.}~\bibnamefont {Rudinger}}, \bibinfo {author}
  {\bibfnamefont {N.}~\bibnamefont {Goss}}, \bibinfo {author} {\bibfnamefont
  {R.~K.}\ \bibnamefont {Naik}}, \bibinfo {author} {\bibfnamefont {J.~M.}\
  \bibnamefont {Kreikebaum}}, \bibinfo {author} {\bibfnamefont {D.~I.}\
  \bibnamefont {Santiago}},\ and\ \bibinfo {author} {\bibfnamefont
  {I.}~\bibnamefont {Siddiqi}},\ }\bibfield  {title} {\bibinfo {title}
  {Benchmarking verified logic operations for fault tolerance},\ }\href@noop {}
  {\bibfield  {journal} {\bibinfo  {journal} {arXiv preprint arXiv:2207.08786}\
  } (\bibinfo {year} {2022})}\BibitemShut {NoStop}%
\bibitem [{\citenamefont {Rudinger}\ \emph {et~al.}(2022)\citenamefont
  {Rudinger}, \citenamefont {Ribeill}, \citenamefont {Govia}, \citenamefont
  {Ware}, \citenamefont {Nielsen}, \citenamefont {Young}, \citenamefont {Ohki},
  \citenamefont {Blume-Kohout},\ and\ \citenamefont
  {Proctor}}]{rudinger2022characterizing}%
  \BibitemOpen
  \bibfield  {author} {\bibinfo {author} {\bibfnamefont {K.}~\bibnamefont
  {Rudinger}}, \bibinfo {author} {\bibfnamefont {G.~J.}\ \bibnamefont
  {Ribeill}}, \bibinfo {author} {\bibfnamefont {L.~C.}\ \bibnamefont {Govia}},
  \bibinfo {author} {\bibfnamefont {M.}~\bibnamefont {Ware}}, \bibinfo {author}
  {\bibfnamefont {E.}~\bibnamefont {Nielsen}}, \bibinfo {author} {\bibfnamefont
  {K.}~\bibnamefont {Young}}, \bibinfo {author} {\bibfnamefont {T.~A.}\
  \bibnamefont {Ohki}}, \bibinfo {author} {\bibfnamefont {R.}~\bibnamefont
  {Blume-Kohout}},\ and\ \bibinfo {author} {\bibfnamefont {T.}~\bibnamefont
  {Proctor}},\ }\bibfield  {title} {\bibinfo {title} {Characterizing midcircuit
  measurements on a superconducting qubit using gate set tomography},\
  }\href@noop {} {\bibfield  {journal} {\bibinfo  {journal} {Physical Review
  Applied}\ }\textbf {\bibinfo {volume} {17}},\ \bibinfo {pages} {014014}
  (\bibinfo {year} {2022})}\BibitemShut {NoStop}%
\bibitem [{\citenamefont {Rudinger}\ \emph {et~al.}(2021)\citenamefont
  {Rudinger}, \citenamefont {Hogle}, \citenamefont {Naik}, \citenamefont
  {Hashim}, \citenamefont {Lobser}, \citenamefont {Santiago}, \citenamefont
  {Grace}, \citenamefont {Nielsen}, \citenamefont {Proctor}, \citenamefont
  {Seritan} \emph {et~al.}}]{rudinger2021experimental}%
  \BibitemOpen
  \bibfield  {author} {\bibinfo {author} {\bibfnamefont {K.}~\bibnamefont
  {Rudinger}}, \bibinfo {author} {\bibfnamefont {C.~W.}\ \bibnamefont {Hogle}},
  \bibinfo {author} {\bibfnamefont {R.~K.}\ \bibnamefont {Naik}}, \bibinfo
  {author} {\bibfnamefont {A.}~\bibnamefont {Hashim}}, \bibinfo {author}
  {\bibfnamefont {D.}~\bibnamefont {Lobser}}, \bibinfo {author} {\bibfnamefont
  {D.~I.}\ \bibnamefont {Santiago}}, \bibinfo {author} {\bibfnamefont {M.~D.}\
  \bibnamefont {Grace}}, \bibinfo {author} {\bibfnamefont {E.}~\bibnamefont
  {Nielsen}}, \bibinfo {author} {\bibfnamefont {T.}~\bibnamefont {Proctor}},
  \bibinfo {author} {\bibfnamefont {S.}~\bibnamefont {Seritan}}, \emph
  {et~al.},\ }\bibfield  {title} {\bibinfo {title} {Experimental
  characterization of crosstalk errors with simultaneous gate set tomography},\
  }\href@noop {} {\bibfield  {journal} {\bibinfo  {journal} {PRX Quantum}\
  }\textbf {\bibinfo {volume} {2}},\ \bibinfo {pages} {040338} (\bibinfo {year}
  {2021})}\BibitemShut {NoStop}%
\bibitem [{\citenamefont {Emerson}\ \emph {et~al.}(2007)\citenamefont
  {Emerson}, \citenamefont {Silva}, \citenamefont {Moussa}, \citenamefont
  {Ryan}, \citenamefont {Laforest}, \citenamefont {Baugh}, \citenamefont
  {Cory},\ and\ \citenamefont {Laflamme}}]{EmersonScience2007}%
  \BibitemOpen
  \bibfield  {author} {\bibinfo {author} {\bibfnamefont {J.}~\bibnamefont
  {Emerson}}, \bibinfo {author} {\bibfnamefont {M.}~\bibnamefont {Silva}},
  \bibinfo {author} {\bibfnamefont {O.}~\bibnamefont {Moussa}}, \bibinfo
  {author} {\bibfnamefont {C.}~\bibnamefont {Ryan}}, \bibinfo {author}
  {\bibfnamefont {M.}~\bibnamefont {Laforest}}, \bibinfo {author}
  {\bibfnamefont {J.}~\bibnamefont {Baugh}}, \bibinfo {author} {\bibfnamefont
  {D.~G.}\ \bibnamefont {Cory}},\ and\ \bibinfo {author} {\bibfnamefont
  {R.}~\bibnamefont {Laflamme}},\ }\bibfield  {title} {\bibinfo {title}
  {Symmetrized characterization of noisy quantum processes},\ }\href
  {https://doi.org/10.1126/science.1145699} {\bibfield  {journal} {\bibinfo
  {journal} {Science}\ }\textbf {\bibinfo {volume} {317}},\ \bibinfo {pages}
  {1893} (\bibinfo {year} {2007})},\ \Eprint
  {https://arxiv.org/abs/https://science.sciencemag.org/content/317/5846/1893.full.pdf}
  {https://science.sciencemag.org/content/317/5846/1893.full.pdf} \BibitemShut
  {NoStop}%
\bibitem [{\citenamefont {Knill}\ \emph {et~al.}(2008)\citenamefont {Knill},
  \citenamefont {Leibfried}, \citenamefont {Reichle}, \citenamefont {Britton},
  \citenamefont {Blakestad}, \citenamefont {Jost}, \citenamefont {Langer},
  \citenamefont {Ozeri}, \citenamefont {Seidelin},\ and\ \citenamefont
  {Wineland}}]{Knill2008}%
  \BibitemOpen
  \bibfield  {author} {\bibinfo {author} {\bibfnamefont {E.}~\bibnamefont
  {Knill}}, \bibinfo {author} {\bibfnamefont {D.}~\bibnamefont {Leibfried}},
  \bibinfo {author} {\bibfnamefont {R.}~\bibnamefont {Reichle}}, \bibinfo
  {author} {\bibfnamefont {J.}~\bibnamefont {Britton}}, \bibinfo {author}
  {\bibfnamefont {R.~B.}\ \bibnamefont {Blakestad}}, \bibinfo {author}
  {\bibfnamefont {J.~D.}\ \bibnamefont {Jost}}, \bibinfo {author}
  {\bibfnamefont {C.}~\bibnamefont {Langer}}, \bibinfo {author} {\bibfnamefont
  {R.}~\bibnamefont {Ozeri}}, \bibinfo {author} {\bibfnamefont
  {S.}~\bibnamefont {Seidelin}},\ and\ \bibinfo {author} {\bibfnamefont
  {D.~J.}\ \bibnamefont {Wineland}},\ }\bibfield  {title} {\bibinfo {title}
  {Randomized benchmarking of quantum gates},\ }\href
  {https://doi.org/10.1103/PhysRevA.77.012307} {\bibfield  {journal} {\bibinfo
  {journal} {Phys. Rev. A}\ }\textbf {\bibinfo {volume} {77}},\ \bibinfo
  {pages} {012307} (\bibinfo {year} {2008})}\BibitemShut {NoStop}%
\bibitem [{\citenamefont {Wallman}\ and\ \citenamefont
  {Flammia}(2014)}]{wallman_randomized_2014}%
  \BibitemOpen
  \bibfield  {author} {\bibinfo {author} {\bibfnamefont {J.~J.}\ \bibnamefont
  {Wallman}}\ and\ \bibinfo {author} {\bibfnamefont {S.~T.}\ \bibnamefont
  {Flammia}},\ }\bibfield  {title} {\bibinfo {title} {Randomized benchmarking
  with confidence},\ }\href {https://doi.org/10.1088/1367-2630/16/10/103032}
  {\bibfield  {journal} {\bibinfo  {journal} {New J. Phys.}\ }\textbf {\bibinfo
  {volume} {16}},\ \bibinfo {pages} {103032} (\bibinfo {year}
  {2014})}\BibitemShut {NoStop}%
\bibitem [{\citenamefont {Magesan}\ \emph {et~al.}(2011)\citenamefont
  {Magesan}, \citenamefont {Gambetta},\ and\ \citenamefont
  {Emerson}}]{Magesan2011-ra}%
  \BibitemOpen
  \bibfield  {author} {\bibinfo {author} {\bibfnamefont {E.}~\bibnamefont
  {Magesan}}, \bibinfo {author} {\bibfnamefont {J.~M.}\ \bibnamefont
  {Gambetta}},\ and\ \bibinfo {author} {\bibfnamefont {J.}~\bibnamefont
  {Emerson}},\ }\bibfield  {title} {\bibinfo {title} {Scalable and robust
  randomized benchmarking of quantum processes},\ }\href
  {https://doi.org/10.1103/PhysRevLett.106.180504} {\bibfield  {journal}
  {\bibinfo  {journal} {Phys. Rev. Lett.}\ }\textbf {\bibinfo {volume} {106}},\
  \bibinfo {pages} {180504} (\bibinfo {year} {2011})}\BibitemShut {NoStop}%
\bibitem [{\citenamefont {Magesan}\ \emph
  {et~al.}(2012{\natexlab{a}})\citenamefont {Magesan}, \citenamefont
  {Gambetta},\ and\ \citenamefont {Emerson}}]{Magesan2012-bo}%
  \BibitemOpen
  \bibfield  {author} {\bibinfo {author} {\bibfnamefont {E.}~\bibnamefont
  {Magesan}}, \bibinfo {author} {\bibfnamefont {J.~M.}\ \bibnamefont
  {Gambetta}},\ and\ \bibinfo {author} {\bibfnamefont {J.}~\bibnamefont
  {Emerson}},\ }\bibfield  {title} {\bibinfo {title} {Characterizing quantum
  gates via randomized benchmarking},\ }\href
  {https://doi.org/10.1103/PhysRevA.85.042311} {\bibfield  {journal} {\bibinfo
  {journal} {Phys. Rev. A}\ }\textbf {\bibinfo {volume} {85}},\ \bibinfo
  {pages} {042311} (\bibinfo {year} {2012}{\natexlab{a}})},\ \Eprint
  {https://arxiv.org/abs/1109.6887} {arXiv:1109.6887 [quant-ph]} \BibitemShut
  {NoStop}%
\bibitem [{\citenamefont {Gaebler}\ \emph {et~al.}(2012)\citenamefont
  {Gaebler}, \citenamefont {Meier}, \citenamefont {Tan}, \citenamefont
  {Bowler}, \citenamefont {Lin}, \citenamefont {Hanneke}, \citenamefont {Jost},
  \citenamefont {Home}, \citenamefont {Knill}, \citenamefont {Leibfried},\ and\
  \citenamefont {Wineland}}]{Gaebler2012-vq}%
  \BibitemOpen
  \bibfield  {author} {\bibinfo {author} {\bibfnamefont {J.~P.}\ \bibnamefont
  {Gaebler}}, \bibinfo {author} {\bibfnamefont {A.~M.}\ \bibnamefont {Meier}},
  \bibinfo {author} {\bibfnamefont {T.~R.}\ \bibnamefont {Tan}}, \bibinfo
  {author} {\bibfnamefont {R.}~\bibnamefont {Bowler}}, \bibinfo {author}
  {\bibfnamefont {Y.}~\bibnamefont {Lin}}, \bibinfo {author} {\bibfnamefont
  {D.}~\bibnamefont {Hanneke}}, \bibinfo {author} {\bibfnamefont {J.~D.}\
  \bibnamefont {Jost}}, \bibinfo {author} {\bibfnamefont {J.~P.}\ \bibnamefont
  {Home}}, \bibinfo {author} {\bibfnamefont {E.}~\bibnamefont {Knill}},
  \bibinfo {author} {\bibfnamefont {D.}~\bibnamefont {Leibfried}},\ and\
  \bibinfo {author} {\bibfnamefont {D.~J.}\ \bibnamefont {Wineland}},\
  }\bibfield  {title} {\bibinfo {title} {Randomized benchmarking of multiqubit
  gates},\ }\href {https://doi.org/10.1103/PhysRevLett.108.260503} {\bibfield
  {journal} {\bibinfo  {journal} {Phys. Rev. Lett.}\ }\textbf {\bibinfo
  {volume} {108}},\ \bibinfo {pages} {260503} (\bibinfo {year}
  {2012})}\BibitemShut {NoStop}%
\bibitem [{\citenamefont {Magesan}\ \emph
  {et~al.}(2012{\natexlab{b}})\citenamefont {Magesan}, \citenamefont
  {Gambetta}, \citenamefont {Johnson}, \citenamefont {Ryan}, \citenamefont
  {Chow}, \citenamefont {Merkel}, \citenamefont {da~Silva}, \citenamefont
  {Keefe}, \citenamefont {Rothwell}, \citenamefont {Ohki}, \citenamefont
  {Ketchen},\ and\ \citenamefont {Steffen}}]{Magesan2012-sg}%
  \BibitemOpen
  \bibfield  {author} {\bibinfo {author} {\bibfnamefont {E.}~\bibnamefont
  {Magesan}}, \bibinfo {author} {\bibfnamefont {J.~M.}\ \bibnamefont
  {Gambetta}}, \bibinfo {author} {\bibfnamefont {B.~R.}\ \bibnamefont
  {Johnson}}, \bibinfo {author} {\bibfnamefont {C.~A.}\ \bibnamefont {Ryan}},
  \bibinfo {author} {\bibfnamefont {J.~M.}\ \bibnamefont {Chow}}, \bibinfo
  {author} {\bibfnamefont {S.~T.}\ \bibnamefont {Merkel}}, \bibinfo {author}
  {\bibfnamefont {M.~P.}\ \bibnamefont {da~Silva}}, \bibinfo {author}
  {\bibfnamefont {G.~A.}\ \bibnamefont {Keefe}}, \bibinfo {author}
  {\bibfnamefont {M.~B.}\ \bibnamefont {Rothwell}}, \bibinfo {author}
  {\bibfnamefont {T.~A.}\ \bibnamefont {Ohki}}, \bibinfo {author}
  {\bibfnamefont {M.~B.}\ \bibnamefont {Ketchen}},\ and\ \bibinfo {author}
  {\bibfnamefont {M.}~\bibnamefont {Steffen}},\ }\bibfield  {title} {\bibinfo
  {title} {Efficient measurement of quantum gate error by interleaved
  randomized benchmarking},\ }\href
  {https://doi.org/10.1103/PhysRevLett.109.080505} {\bibfield  {journal}
  {\bibinfo  {journal} {Phys. Rev. Lett.}\ }\textbf {\bibinfo {volume} {109}},\
  \bibinfo {pages} {080505} (\bibinfo {year} {2012}{\natexlab{b}})}\BibitemShut
  {NoStop}%
\bibitem [{\citenamefont {Gambetta}\ \emph
  {et~al.}(2012{\natexlab{a}})\citenamefont {Gambetta}, \citenamefont
  {C{\'o}rcoles}, \citenamefont {Merkel}, \citenamefont {Johnson},
  \citenamefont {Smolin}, \citenamefont {Chow}, \citenamefont {Ryan},
  \citenamefont {Rigetti}, \citenamefont {Poletto}, \citenamefont {Ohki},
  \citenamefont {Ketchen},\ and\ \citenamefont {Steffen}}]{Gambetta2012-yu}%
  \BibitemOpen
  \bibfield  {author} {\bibinfo {author} {\bibfnamefont {J.~M.}\ \bibnamefont
  {Gambetta}}, \bibinfo {author} {\bibfnamefont {A.~D.}\ \bibnamefont
  {C{\'o}rcoles}}, \bibinfo {author} {\bibfnamefont {S.~T.}\ \bibnamefont
  {Merkel}}, \bibinfo {author} {\bibfnamefont {B.~R.}\ \bibnamefont {Johnson}},
  \bibinfo {author} {\bibfnamefont {J.~A.}\ \bibnamefont {Smolin}}, \bibinfo
  {author} {\bibfnamefont {J.~M.}\ \bibnamefont {Chow}}, \bibinfo {author}
  {\bibfnamefont {C.~A.}\ \bibnamefont {Ryan}}, \bibinfo {author}
  {\bibfnamefont {C.}~\bibnamefont {Rigetti}}, \bibinfo {author} {\bibfnamefont
  {S.}~\bibnamefont {Poletto}}, \bibinfo {author} {\bibfnamefont {T.~A.}\
  \bibnamefont {Ohki}}, \bibinfo {author} {\bibfnamefont {M.~B.}\ \bibnamefont
  {Ketchen}},\ and\ \bibinfo {author} {\bibfnamefont {M.}~\bibnamefont
  {Steffen}},\ }\bibfield  {title} {\bibinfo {title} {Characterization of
  addressability by simultaneous randomized benchmarking},\ }\href
  {https://doi.org/10.1103/PhysRevLett.109.240504} {\bibfield  {journal}
  {\bibinfo  {journal} {Phys. Rev. Lett.}\ }\textbf {\bibinfo {volume} {109}},\
  \bibinfo {pages} {240504} (\bibinfo {year} {2012}{\natexlab{a}})}\BibitemShut
  {NoStop}%
\bibitem [{\citenamefont {C{\'o}rcoles}\ \emph {et~al.}(2013)\citenamefont
  {C{\'o}rcoles}, \citenamefont {Gambetta}, \citenamefont {Chow}, \citenamefont
  {Smolin}, \citenamefont {Ware}, \citenamefont {Strand}, \citenamefont
  {Plourde},\ and\ \citenamefont {Steffen}}]{Corcoles2013-zs}%
  \BibitemOpen
  \bibfield  {author} {\bibinfo {author} {\bibfnamefont {A.~D.}\ \bibnamefont
  {C{\'o}rcoles}}, \bibinfo {author} {\bibfnamefont {J.~M.}\ \bibnamefont
  {Gambetta}}, \bibinfo {author} {\bibfnamefont {J.~M.}\ \bibnamefont {Chow}},
  \bibinfo {author} {\bibfnamefont {J.~A.}\ \bibnamefont {Smolin}}, \bibinfo
  {author} {\bibfnamefont {M.}~\bibnamefont {Ware}}, \bibinfo {author}
  {\bibfnamefont {J.}~\bibnamefont {Strand}}, \bibinfo {author} {\bibfnamefont
  {B.~L.~T.}\ \bibnamefont {Plourde}},\ and\ \bibinfo {author} {\bibfnamefont
  {M.}~\bibnamefont {Steffen}},\ }\bibfield  {title} {\bibinfo {title} {Process
  verification of two-qubit quantum gates by randomized benchmarking},\ }\href
  {https://doi.org/10.1103/PhysRevA.87.030301} {\bibfield  {journal} {\bibinfo
  {journal} {Phys. Rev. A}\ }\textbf {\bibinfo {volume} {87}},\ \bibinfo
  {pages} {030301} (\bibinfo {year} {2013})},\ \Eprint
  {https://arxiv.org/abs/1210.7011} {arXiv:1210.7011 [quant-ph]} \BibitemShut
  {NoStop}%
\bibitem [{\citenamefont {Barends}\ \emph {et~al.}(2014)\citenamefont
  {Barends}, \citenamefont {Kelly}, \citenamefont {Megrant}, \citenamefont
  {Veitia}, \citenamefont {Sank}, \citenamefont {Jeffrey}, \citenamefont
  {White}, \citenamefont {Mutus}, \citenamefont {Fowler}, \citenamefont
  {Campbell}, \citenamefont {Chen}, \citenamefont {Chen}, \citenamefont
  {Chiaro}, \citenamefont {Dunsworth}, \citenamefont {Neill}, \citenamefont
  {O'Malley}, \citenamefont {Roushan}, \citenamefont {Vainsencher},
  \citenamefont {Wenner}, \citenamefont {Korotkov}, \citenamefont {Cleland},\
  and\ \citenamefont {Martinis}}]{Barends2014-ap}%
  \BibitemOpen
  \bibfield  {author} {\bibinfo {author} {\bibfnamefont {R.}~\bibnamefont
  {Barends}}, \bibinfo {author} {\bibfnamefont {J.}~\bibnamefont {Kelly}},
  \bibinfo {author} {\bibfnamefont {A.}~\bibnamefont {Megrant}}, \bibinfo
  {author} {\bibfnamefont {A.}~\bibnamefont {Veitia}}, \bibinfo {author}
  {\bibfnamefont {D.}~\bibnamefont {Sank}}, \bibinfo {author} {\bibfnamefont
  {E.}~\bibnamefont {Jeffrey}}, \bibinfo {author} {\bibfnamefont {T.~C.}\
  \bibnamefont {White}}, \bibinfo {author} {\bibfnamefont {J.}~\bibnamefont
  {Mutus}}, \bibinfo {author} {\bibfnamefont {A.~G.}\ \bibnamefont {Fowler}},
  \bibinfo {author} {\bibfnamefont {B.}~\bibnamefont {Campbell}}, \bibinfo
  {author} {\bibfnamefont {Y.}~\bibnamefont {Chen}}, \bibinfo {author}
  {\bibfnamefont {Z.}~\bibnamefont {Chen}}, \bibinfo {author} {\bibfnamefont
  {B.}~\bibnamefont {Chiaro}}, \bibinfo {author} {\bibfnamefont
  {A.}~\bibnamefont {Dunsworth}}, \bibinfo {author} {\bibfnamefont
  {C.}~\bibnamefont {Neill}}, \bibinfo {author} {\bibfnamefont
  {P.}~\bibnamefont {O'Malley}}, \bibinfo {author} {\bibfnamefont
  {P.}~\bibnamefont {Roushan}}, \bibinfo {author} {\bibfnamefont
  {A.}~\bibnamefont {Vainsencher}}, \bibinfo {author} {\bibfnamefont
  {J.}~\bibnamefont {Wenner}}, \bibinfo {author} {\bibfnamefont {A.~N.}\
  \bibnamefont {Korotkov}}, \bibinfo {author} {\bibfnamefont {A.~N.}\
  \bibnamefont {Cleland}},\ and\ \bibinfo {author} {\bibfnamefont {J.~M.}\
  \bibnamefont {Martinis}},\ }\bibfield  {title} {\bibinfo {title}
  {Superconducting quantum circuits at the surface code threshold for fault
  tolerance},\ }\href {https://doi.org/10.1038/nature13171} {\bibfield
  {journal} {\bibinfo  {journal} {Nature}\ }\textbf {\bibinfo {volume} {508}},\
  \bibinfo {pages} {500} (\bibinfo {year} {2014})}\BibitemShut {NoStop}%
\bibitem [{\citenamefont {Wallman}\ \emph
  {et~al.}(2015{\natexlab{a}})\citenamefont {Wallman}, \citenamefont {Granade},
  \citenamefont {Harper},\ and\ \citenamefont {Flammia}}]{Wallman2015-pa}%
  \BibitemOpen
  \bibfield  {author} {\bibinfo {author} {\bibfnamefont {J.}~\bibnamefont
  {Wallman}}, \bibinfo {author} {\bibfnamefont {C.}~\bibnamefont {Granade}},
  \bibinfo {author} {\bibfnamefont {R.}~\bibnamefont {Harper}},\ and\ \bibinfo
  {author} {\bibfnamefont {S.~T.}\ \bibnamefont {Flammia}},\ }\bibfield
  {title} {\bibinfo {title} {Estimating the coherence of noise},\ }\href
  {https://doi.org/10.1088/1367-2630/17/11/113020} {\bibfield  {journal}
  {\bibinfo  {journal} {New Journal of Physics}\ }\textbf {\bibinfo {volume}
  {17}},\ \bibinfo {pages} {113020} (\bibinfo {year}
  {2015}{\natexlab{a}})}\BibitemShut {NoStop}%
\bibitem [{\citenamefont {Carignan-Dugas}\ \emph {et~al.}(2015)\citenamefont
  {Carignan-Dugas}, \citenamefont {Wallman},\ and\ \citenamefont
  {Emerson}}]{Carignan-Dugas2015-pz}%
  \BibitemOpen
  \bibfield  {author} {\bibinfo {author} {\bibfnamefont {A.}~\bibnamefont
  {Carignan-Dugas}}, \bibinfo {author} {\bibfnamefont {J.~J.}\ \bibnamefont
  {Wallman}},\ and\ \bibinfo {author} {\bibfnamefont {J.}~\bibnamefont
  {Emerson}},\ }\bibfield  {title} {\bibinfo {title} {Characterizing universal
  gate sets via dihedral benchmarking},\ }\bibfield  {journal} {\bibinfo
  {journal} {Phys. Rev. A}\ }\href {https://doi.org/10.1103/PhysRevA.92.060302}
  {10.1103/PhysRevA.92.060302} (\bibinfo {year} {2015})\BibitemShut {NoStop}%
\bibitem [{\citenamefont {Wallman}\ \emph
  {et~al.}(2015{\natexlab{b}})\citenamefont {Wallman}, \citenamefont
  {Barnhill},\ and\ \citenamefont {Emerson}}]{Wallman2015-uq}%
  \BibitemOpen
  \bibfield  {author} {\bibinfo {author} {\bibfnamefont {J.~J.}\ \bibnamefont
  {Wallman}}, \bibinfo {author} {\bibfnamefont {M.}~\bibnamefont {Barnhill}},\
  and\ \bibinfo {author} {\bibfnamefont {J.}~\bibnamefont {Emerson}},\
  }\bibfield  {title} {\bibinfo {title} {Robust characterization of loss
  rates},\ }\href {https://doi.org/10.1103/PhysRevLett.115.060501} {\bibfield
  {journal} {\bibinfo  {journal} {Phys. Rev. Lett.}\ }\textbf {\bibinfo
  {volume} {115}},\ \bibinfo {pages} {060501} (\bibinfo {year}
  {2015}{\natexlab{b}})}\BibitemShut {NoStop}%
\bibitem [{\citenamefont {Chasseur}\ and\ \citenamefont
  {Wilhelm}(2015)}]{Chasseur2015-zz}%
  \BibitemOpen
  \bibfield  {author} {\bibinfo {author} {\bibfnamefont {T.}~\bibnamefont
  {Chasseur}}\ and\ \bibinfo {author} {\bibfnamefont {F.~K.}\ \bibnamefont
  {Wilhelm}},\ }\bibfield  {title} {\bibinfo {title} {Complete randomized
  benchmarking protocol accounting for leakage errors},\ }\href
  {https://doi.org/10.1103/PhysRevA.92.042333} {\bibfield  {journal} {\bibinfo
  {journal} {Phys. Rev. A}\ }\textbf {\bibinfo {volume} {92}},\ \bibinfo
  {pages} {042333} (\bibinfo {year} {2015})}\BibitemShut {NoStop}%
\bibitem [{\citenamefont {Sheldon}\ \emph {et~al.}(2016)\citenamefont
  {Sheldon}, \citenamefont {Bishop}, \citenamefont {Magesan}, \citenamefont
  {Filipp}, \citenamefont {Chow},\ and\ \citenamefont
  {Gambetta}}]{Sheldon2016-sj}%
  \BibitemOpen
  \bibfield  {author} {\bibinfo {author} {\bibfnamefont {S.}~\bibnamefont
  {Sheldon}}, \bibinfo {author} {\bibfnamefont {L.~S.}\ \bibnamefont {Bishop}},
  \bibinfo {author} {\bibfnamefont {E.}~\bibnamefont {Magesan}}, \bibinfo
  {author} {\bibfnamefont {S.}~\bibnamefont {Filipp}}, \bibinfo {author}
  {\bibfnamefont {J.~M.}\ \bibnamefont {Chow}},\ and\ \bibinfo {author}
  {\bibfnamefont {J.~M.}\ \bibnamefont {Gambetta}},\ }\bibfield  {title}
  {\bibinfo {title} {Characterizing errors on qubit operations via iterative
  randomized benchmarking},\ }\href
  {https://doi.org/10.1103/PhysRevA.93.012301} {\bibfield  {journal} {\bibinfo
  {journal} {Phys. Rev. A}\ }\textbf {\bibinfo {volume} {93}},\ \bibinfo
  {pages} {012301} (\bibinfo {year} {2016})},\ \Eprint
  {https://arxiv.org/abs/1504.06597} {arXiv:1504.06597 [quant-ph]} \BibitemShut
  {NoStop}%
\bibitem [{\citenamefont {Wallman}\ \emph {et~al.}(2016)\citenamefont
  {Wallman}, \citenamefont {Barnhill},\ and\ \citenamefont
  {Emerson}}]{Wallman2016-kx}%
  \BibitemOpen
  \bibfield  {author} {\bibinfo {author} {\bibfnamefont {J.~J.}\ \bibnamefont
  {Wallman}}, \bibinfo {author} {\bibfnamefont {M.}~\bibnamefont {Barnhill}},\
  and\ \bibinfo {author} {\bibfnamefont {J.}~\bibnamefont {Emerson}},\
  }\bibfield  {title} {\bibinfo {title} {Robust characterization of leakage
  errors},\ }\href {https://doi.org/10.1088/1367-2630/18/4/043021} {\bibfield
  {journal} {\bibinfo  {journal} {New J. Phys.}\ }\textbf {\bibinfo {volume}
  {18}},\ \bibinfo {pages} {043021} (\bibinfo {year} {2016})}\BibitemShut
  {NoStop}%
\bibitem [{\citenamefont {Proctor}\ \emph {et~al.}(2017)\citenamefont
  {Proctor}, \citenamefont {Rudinger}, \citenamefont {Young}, \citenamefont
  {Sarovar},\ and\ \citenamefont {Blume-Kohout}}]{Proctor2017-ru}%
  \BibitemOpen
  \bibfield  {author} {\bibinfo {author} {\bibfnamefont {T.}~\bibnamefont
  {Proctor}}, \bibinfo {author} {\bibfnamefont {K.}~\bibnamefont {Rudinger}},
  \bibinfo {author} {\bibfnamefont {K.}~\bibnamefont {Young}}, \bibinfo
  {author} {\bibfnamefont {M.}~\bibnamefont {Sarovar}},\ and\ \bibinfo {author}
  {\bibfnamefont {R.}~\bibnamefont {Blume-Kohout}},\ }\bibfield  {title}
  {\bibinfo {title} {What randomized benchmarking actually measures},\ }\href
  {https://doi.org/10.1103/PhysRevLett.119.130502} {\bibfield  {journal}
  {\bibinfo  {journal} {Phys. Rev. Lett.}\ }\textbf {\bibinfo {volume} {119}},\
  \bibinfo {pages} {130502} (\bibinfo {year} {2017})}\BibitemShut {NoStop}%
\bibitem [{\citenamefont {Harper}\ and\ \citenamefont
  {Flammia}(2017)}]{Harper2017-oa}%
  \BibitemOpen
  \bibfield  {author} {\bibinfo {author} {\bibfnamefont {R.}~\bibnamefont
  {Harper}}\ and\ \bibinfo {author} {\bibfnamefont {S.~T.}\ \bibnamefont
  {Flammia}},\ }\bibfield  {title} {\bibinfo {title} {Estimating the fidelity
  of {T} gates using standard interleaved randomized benchmarking},\ }\href
  {https://doi.org/10.1088/2058-9565/aa5f8d} {\bibfield  {journal} {\bibinfo
  {journal} {Quantum Sci. Technol.}\ }\textbf {\bibinfo {volume} {2}},\
  \bibinfo {pages} {015008} (\bibinfo {year} {2017})}\BibitemShut {NoStop}%
\bibitem [{\citenamefont {Wallman}(2018)}]{Wallman2018-wy}%
  \BibitemOpen
  \bibfield  {author} {\bibinfo {author} {\bibfnamefont {J.~J.}\ \bibnamefont
  {Wallman}},\ }\bibfield  {title} {\bibinfo {title} {Randomized benchmarking
  with gate-dependent noise},\ }\href
  {https://doi.org/10.22331/q-2018-01-29-47} {\bibfield  {journal} {\bibinfo
  {journal} {{Quantum}}\ }\textbf {\bibinfo {volume} {2}},\ \bibinfo {pages}
  {47} (\bibinfo {year} {2018})}\BibitemShut {NoStop}%
\bibitem [{\citenamefont {Huang}\ \emph {et~al.}(2019)\citenamefont {Huang},
  \citenamefont {Yang}, \citenamefont {Chan}, \citenamefont {Tanttu},
  \citenamefont {Hensen}, \citenamefont {Leon}, \citenamefont {Fogarty},
  \citenamefont {Hwang}, \citenamefont {Hudson}, \citenamefont {Itoh},
  \citenamefont {Morello}, \citenamefont {Laucht},\ and\ \citenamefont
  {Dzurak}}]{Huang2019-zj}%
  \BibitemOpen
  \bibfield  {author} {\bibinfo {author} {\bibfnamefont {W.}~\bibnamefont
  {Huang}}, \bibinfo {author} {\bibfnamefont {C.~H.}\ \bibnamefont {Yang}},
  \bibinfo {author} {\bibfnamefont {K.~W.}\ \bibnamefont {Chan}}, \bibinfo
  {author} {\bibfnamefont {T.}~\bibnamefont {Tanttu}}, \bibinfo {author}
  {\bibfnamefont {B.}~\bibnamefont {Hensen}}, \bibinfo {author} {\bibfnamefont
  {R.~C.~C.}\ \bibnamefont {Leon}}, \bibinfo {author} {\bibfnamefont {M.~A.}\
  \bibnamefont {Fogarty}}, \bibinfo {author} {\bibfnamefont {J.~C.~C.}\
  \bibnamefont {Hwang}}, \bibinfo {author} {\bibfnamefont {F.~E.}\ \bibnamefont
  {Hudson}}, \bibinfo {author} {\bibfnamefont {K.~M.}\ \bibnamefont {Itoh}},
  \bibinfo {author} {\bibfnamefont {A.}~\bibnamefont {Morello}}, \bibinfo
  {author} {\bibfnamefont {A.}~\bibnamefont {Laucht}},\ and\ \bibinfo {author}
  {\bibfnamefont {A.~S.}\ \bibnamefont {Dzurak}},\ }\bibfield  {title}
  {\bibinfo {title} {Fidelity benchmarks for two-qubit gates in silicon},\
  }\href {https://doi.org/10.1038/s41586-019-1197-0} {\bibfield  {journal}
  {\bibinfo  {journal} {Nature}\ }\textbf {\bibinfo {volume} {569}},\ \bibinfo
  {pages} {532} (\bibinfo {year} {2019})}\BibitemShut {NoStop}%
\bibitem [{\citenamefont {McKay}\ \emph {et~al.}(2019)\citenamefont {McKay},
  \citenamefont {Sheldon}, \citenamefont {Smolin}, \citenamefont {Chow},\ and\
  \citenamefont {Gambetta}}]{McKay2019-kf}%
  \BibitemOpen
  \bibfield  {author} {\bibinfo {author} {\bibfnamefont {D.~C.}\ \bibnamefont
  {McKay}}, \bibinfo {author} {\bibfnamefont {S.}~\bibnamefont {Sheldon}},
  \bibinfo {author} {\bibfnamefont {J.~A.}\ \bibnamefont {Smolin}}, \bibinfo
  {author} {\bibfnamefont {J.~M.}\ \bibnamefont {Chow}},\ and\ \bibinfo
  {author} {\bibfnamefont {J.~M.}\ \bibnamefont {Gambetta}},\ }\bibfield
  {title} {\bibinfo {title} {{Three-Qubit} randomized benchmarking},\ }\href
  {https://doi.org/10.1103/PhysRevLett.122.200502} {\bibfield  {journal}
  {\bibinfo  {journal} {Phys. Rev. Lett.}\ }\textbf {\bibinfo {volume} {122}},\
  \bibinfo {pages} {200502} (\bibinfo {year} {2019})}\BibitemShut {NoStop}%
\bibitem [{\citenamefont {Proctor}\ \emph {et~al.}(2019)\citenamefont
  {Proctor}, \citenamefont {Carignan-Dugas}, \citenamefont {Rudinger},
  \citenamefont {Nielsen}, \citenamefont {Blume-Kohout},\ and\ \citenamefont
  {Young}}]{Proctor2019-ma}%
  \BibitemOpen
  \bibfield  {author} {\bibinfo {author} {\bibfnamefont {T.~J.}\ \bibnamefont
  {Proctor}}, \bibinfo {author} {\bibfnamefont {A.}~\bibnamefont
  {Carignan-Dugas}}, \bibinfo {author} {\bibfnamefont {K.}~\bibnamefont
  {Rudinger}}, \bibinfo {author} {\bibfnamefont {E.}~\bibnamefont {Nielsen}},
  \bibinfo {author} {\bibfnamefont {R.}~\bibnamefont {Blume-Kohout}},\ and\
  \bibinfo {author} {\bibfnamefont {K.}~\bibnamefont {Young}},\ }\bibfield
  {title} {\bibinfo {title} {Direct randomized benchmarking for multiqubit
  devices},\ }\href {https://doi.org/10.1103/PhysRevLett.123.030503} {\bibfield
   {journal} {\bibinfo  {journal} {Phys. Rev. Lett.}\ }\textbf {\bibinfo
  {volume} {123}},\ \bibinfo {pages} {030503} (\bibinfo {year}
  {2019})}\BibitemShut {NoStop}%
\bibitem [{\citenamefont {Erhard}\ \emph {et~al.}(2019)\citenamefont {Erhard},
  \citenamefont {Wallman}, \citenamefont {Postler}, \citenamefont {Meth},
  \citenamefont {Stricker}, \citenamefont {Martinez}, \citenamefont
  {Schindler}, \citenamefont {Monz}, \citenamefont {Emerson},\ and\
  \citenamefont {Blatt}}]{Erhard2019-ig}%
  \BibitemOpen
  \bibfield  {author} {\bibinfo {author} {\bibfnamefont {A.}~\bibnamefont
  {Erhard}}, \bibinfo {author} {\bibfnamefont {J.~J.}\ \bibnamefont {Wallman}},
  \bibinfo {author} {\bibfnamefont {L.}~\bibnamefont {Postler}}, \bibinfo
  {author} {\bibfnamefont {M.}~\bibnamefont {Meth}}, \bibinfo {author}
  {\bibfnamefont {R.}~\bibnamefont {Stricker}}, \bibinfo {author}
  {\bibfnamefont {E.~A.}\ \bibnamefont {Martinez}}, \bibinfo {author}
  {\bibfnamefont {P.}~\bibnamefont {Schindler}}, \bibinfo {author}
  {\bibfnamefont {T.}~\bibnamefont {Monz}}, \bibinfo {author} {\bibfnamefont
  {J.}~\bibnamefont {Emerson}},\ and\ \bibinfo {author} {\bibfnamefont
  {R.}~\bibnamefont {Blatt}},\ }\bibfield  {title} {\bibinfo {title}
  {Characterizing large-scale quantum computers via cycle benchmarking},\
  }\href {https://doi.org/10.1038/s41467-019-13068-7} {\bibfield  {journal}
  {\bibinfo  {journal} {Nat. Commun.}\ }\textbf {\bibinfo {volume} {10}},\
  \bibinfo {pages} {5347} (\bibinfo {year} {2019})}\BibitemShut {NoStop}%
\bibitem [{\citenamefont {Ekert}\ \emph {et~al.}(2002)\citenamefont {Ekert},
  \citenamefont {Alves}, \citenamefont {Oi}, \citenamefont {Horodecki},
  \citenamefont {Horodecki},\ and\ \citenamefont {Kwek}}]{Ekert2002-ma}%
  \BibitemOpen
  \bibfield  {author} {\bibinfo {author} {\bibfnamefont {A.~K.}\ \bibnamefont
  {Ekert}}, \bibinfo {author} {\bibfnamefont {C.~M.}\ \bibnamefont {Alves}},
  \bibinfo {author} {\bibfnamefont {D.~K.~L.}\ \bibnamefont {Oi}}, \bibinfo
  {author} {\bibfnamefont {M.}~\bibnamefont {Horodecki}}, \bibinfo {author}
  {\bibfnamefont {P.}~\bibnamefont {Horodecki}},\ and\ \bibinfo {author}
  {\bibfnamefont {L.~C.}\ \bibnamefont {Kwek}},\ }\bibfield  {title} {\bibinfo
  {title} {Direct estimations of linear and nonlinear functionals of a quantum
  state},\ }\href {https://doi.org/10.1103/PhysRevLett.88.217901} {\bibfield
  {journal} {\bibinfo  {journal} {Phys. Rev. Lett.}\ }\textbf {\bibinfo
  {volume} {88}},\ \bibinfo {pages} {217901} (\bibinfo {year}
  {2002})}\BibitemShut {NoStop}%
\bibitem [{\citenamefont {L{\'e}vi}\ \emph {et~al.}(2007)\citenamefont
  {L{\'e}vi}, \citenamefont {L{\'o}pez}, \citenamefont {Emerson},\ and\
  \citenamefont {Cory}}]{Levi2007-rb}%
  \BibitemOpen
  \bibfield  {author} {\bibinfo {author} {\bibfnamefont {B.}~\bibnamefont
  {L{\'e}vi}}, \bibinfo {author} {\bibfnamefont {C.~C.}\ \bibnamefont
  {L{\'o}pez}}, \bibinfo {author} {\bibfnamefont {J.}~\bibnamefont {Emerson}},\
  and\ \bibinfo {author} {\bibfnamefont {D.~G.}\ \bibnamefont {Cory}},\
  }\bibfield  {title} {\bibinfo {title} {Efficient error characterization in
  quantum information processing},\ }\href
  {https://doi.org/10.1103/PhysRevA.75.022314} {\bibfield  {journal} {\bibinfo
  {journal} {Phys. Rev. A}\ }\textbf {\bibinfo {volume} {75}},\ \bibinfo
  {pages} {022314} (\bibinfo {year} {2007})}\BibitemShut {NoStop}%
\bibitem [{\citenamefont {T{\'o}th}\ \emph {et~al.}(2010)\citenamefont
  {T{\'o}th}, \citenamefont {Wieczorek}, \citenamefont {Gross}, \citenamefont
  {Krischek}, \citenamefont {Schwemmer},\ and\ \citenamefont
  {Weinfurter}}]{Toth2010-xi}%
  \BibitemOpen
  \bibfield  {author} {\bibinfo {author} {\bibfnamefont {G.}~\bibnamefont
  {T{\'o}th}}, \bibinfo {author} {\bibfnamefont {W.}~\bibnamefont {Wieczorek}},
  \bibinfo {author} {\bibfnamefont {D.}~\bibnamefont {Gross}}, \bibinfo
  {author} {\bibfnamefont {R.}~\bibnamefont {Krischek}}, \bibinfo {author}
  {\bibfnamefont {C.}~\bibnamefont {Schwemmer}},\ and\ \bibinfo {author}
  {\bibfnamefont {H.}~\bibnamefont {Weinfurter}},\ }\bibfield  {title}
  {\bibinfo {title} {Permutationally invariant quantum tomography},\ }\href
  {https://doi.org/10.1103/PhysRevLett.105.250403} {\bibfield  {journal}
  {\bibinfo  {journal} {Phys. Rev. Lett.}\ }\textbf {\bibinfo {volume} {105}},\
  \bibinfo {pages} {250403} (\bibinfo {year} {2010})}\BibitemShut {NoStop}%
\bibitem [{\citenamefont {Flammia}\ and\ \citenamefont
  {Liu}(2011)}]{Flammia2011-rw}%
  \BibitemOpen
  \bibfield  {author} {\bibinfo {author} {\bibfnamefont {S.~T.}\ \bibnamefont
  {Flammia}}\ and\ \bibinfo {author} {\bibfnamefont {Y.-K.}\ \bibnamefont
  {Liu}},\ }\bibfield  {title} {\bibinfo {title} {Direct fidelity estimation
  from few pauli measurements},\ }\href
  {https://doi.org/10.1103/PhysRevLett.106.230501} {\bibfield  {journal}
  {\bibinfo  {journal} {Phys. Rev. Lett.}\ }\textbf {\bibinfo {volume} {106}},\
  \bibinfo {pages} {230501} (\bibinfo {year} {2011})}\BibitemShut {NoStop}%
\bibitem [{\citenamefont {da~Silva}\ \emph {et~al.}(2011)\citenamefont
  {da~Silva}, \citenamefont {Landon-Cardinal},\ and\ \citenamefont
  {Poulin}}]{Da_Silva2011-jv}%
  \BibitemOpen
  \bibfield  {author} {\bibinfo {author} {\bibfnamefont {M.~P.}\ \bibnamefont
  {da~Silva}}, \bibinfo {author} {\bibfnamefont {O.}~\bibnamefont
  {Landon-Cardinal}},\ and\ \bibinfo {author} {\bibfnamefont {D.}~\bibnamefont
  {Poulin}},\ }\bibfield  {title} {\bibinfo {title} {Practical characterization
  of quantum devices without tomography},\ }\href
  {https://doi.org/10.1103/PhysRevLett.107.210404} {\bibfield  {journal}
  {\bibinfo  {journal} {Phys. Rev. Lett.}\ }\textbf {\bibinfo {volume} {107}},\
  \bibinfo {pages} {210404} (\bibinfo {year} {2011})}\BibitemShut {NoStop}%
\bibitem [{\citenamefont {Moussa}\ \emph {et~al.}(2012)\citenamefont {Moussa},
  \citenamefont {da~Silva}, \citenamefont {Ryan},\ and\ \citenamefont
  {Laflamme}}]{Moussa2012-kd}%
  \BibitemOpen
  \bibfield  {author} {\bibinfo {author} {\bibfnamefont {O.}~\bibnamefont
  {Moussa}}, \bibinfo {author} {\bibfnamefont {M.~P.}\ \bibnamefont
  {da~Silva}}, \bibinfo {author} {\bibfnamefont {C.~A.}\ \bibnamefont {Ryan}},\
  and\ \bibinfo {author} {\bibfnamefont {R.}~\bibnamefont {Laflamme}},\
  }\bibfield  {title} {\bibinfo {title} {Practical experimental certification
  of computational quantum gates using a twirling procedure},\ }\href
  {https://doi.org/10.1103/PhysRevLett.109.070504} {\bibfield  {journal}
  {\bibinfo  {journal} {Phys. Rev. Lett.}\ }\textbf {\bibinfo {volume} {109}},\
  \bibinfo {pages} {070504} (\bibinfo {year} {2012})}\BibitemShut {NoStop}%
\bibitem [{\citenamefont {Reich}\ \emph {et~al.}(2013)\citenamefont {Reich},
  \citenamefont {Gualdi},\ and\ \citenamefont {Koch}}]{Reich2013-oj}%
  \BibitemOpen
  \bibfield  {author} {\bibinfo {author} {\bibfnamefont {D.~M.}\ \bibnamefont
  {Reich}}, \bibinfo {author} {\bibfnamefont {G.}~\bibnamefont {Gualdi}},\ and\
  \bibinfo {author} {\bibfnamefont {C.~P.}\ \bibnamefont {Koch}},\ }\bibfield
  {title} {\bibinfo {title} {Optimal strategies for estimating the average
  fidelity of quantum gates},\ }\href
  {https://doi.org/10.1103/PhysRevLett.111.200401} {\bibfield  {journal}
  {\bibinfo  {journal} {Phys. Rev. Lett.}\ }\textbf {\bibinfo {volume} {111}},\
  \bibinfo {pages} {200401} (\bibinfo {year} {2013})}\BibitemShut {NoStop}%
\bibitem [{\citenamefont {Kimmel}\ \emph {et~al.}(2015)\citenamefont {Kimmel},
  \citenamefont {Low},\ and\ \citenamefont {Yoder}}]{Kimmel2015-tj}%
  \BibitemOpen
  \bibfield  {author} {\bibinfo {author} {\bibfnamefont {S.}~\bibnamefont
  {Kimmel}}, \bibinfo {author} {\bibfnamefont {G.~H.}\ \bibnamefont {Low}},\
  and\ \bibinfo {author} {\bibfnamefont {T.~J.}\ \bibnamefont {Yoder}},\
  }\bibfield  {title} {\bibinfo {title} {Robust calibration of a universal
  single-qubit gate set via robust phase estimation},\ }\href
  {https://doi.org/10.1103/PhysRevA.92.062315} {\bibfield  {journal} {\bibinfo
  {journal} {Phys. Rev. A}\ }\textbf {\bibinfo {volume} {92}},\ \bibinfo
  {pages} {062315} (\bibinfo {year} {2015})},\ \Eprint
  {https://arxiv.org/abs/1502.02677} {arXiv:1502.02677 [quant-ph]} \BibitemShut
  {NoStop}%
\bibitem [{\citenamefont {Rudinger}\ \emph
  {et~al.}(2017{\natexlab{a}})\citenamefont {Rudinger}, \citenamefont {Kimmel},
  \citenamefont {Lobser},\ and\ \citenamefont {Maunz}}]{Rudinger2017-vy}%
  \BibitemOpen
  \bibfield  {author} {\bibinfo {author} {\bibfnamefont {K.}~\bibnamefont
  {Rudinger}}, \bibinfo {author} {\bibfnamefont {S.}~\bibnamefont {Kimmel}},
  \bibinfo {author} {\bibfnamefont {D.}~\bibnamefont {Lobser}},\ and\ \bibinfo
  {author} {\bibfnamefont {P.}~\bibnamefont {Maunz}},\ }\bibfield  {title}
  {\bibinfo {title} {Experimental demonstration of a cheap and accurate phase
  estimation},\ }\href {https://doi.org/10.1103/PhysRevLett.118.190502}
  {\bibfield  {journal} {\bibinfo  {journal} {Phys. Rev. Lett.}\ }\textbf
  {\bibinfo {volume} {118}},\ \bibinfo {pages} {190502} (\bibinfo {year}
  {2017}{\natexlab{a}})}\BibitemShut {NoStop}%
\bibitem [{\citenamefont {Aaronson}(2018)}]{Aaronson2018-tj}%
  \BibitemOpen
  \bibfield  {author} {\bibinfo {author} {\bibfnamefont {S.}~\bibnamefont
  {Aaronson}},\ }\bibfield  {title} {\bibinfo {title} {Shadow tomography of
  quantum states},\ }in\ \href@noop {} {\emph {\bibinfo {booktitle}
  {Proceedings of the 50th Annual {ACM} {SIGACT} Symposium on Theory of
  Computing}}}\ (\bibinfo  {publisher} {dl.acm.org},\ \bibinfo {year} {2018})\
  pp.\ \bibinfo {pages} {325--338}\BibitemShut {NoStop}%
\bibitem [{\citenamefont {Mayer}\ and\ \citenamefont
  {Knill}(2018)}]{Mayer2018-zt}%
  \BibitemOpen
  \bibfield  {author} {\bibinfo {author} {\bibfnamefont {K.}~\bibnamefont
  {Mayer}}\ and\ \bibinfo {author} {\bibfnamefont {E.}~\bibnamefont {Knill}},\
  }\bibfield  {title} {\bibinfo {title} {Quantum process fidelity bounds from
  sets of input states},\ }\href {https://doi.org/10.1103/PhysRevA.98.052326}
  {\bibfield  {journal} {\bibinfo  {journal} {Phys. Rev. A}\ }\textbf {\bibinfo
  {volume} {98}},\ \bibinfo {pages} {052326} (\bibinfo {year}
  {2018})}\BibitemShut {NoStop}%
\bibitem [{\citenamefont {Helsen}\ \emph {et~al.}(2019)\citenamefont {Helsen},
  \citenamefont {Battistel},\ and\ \citenamefont {Terhal}}]{Helsen2019-fi}%
  \BibitemOpen
  \bibfield  {author} {\bibinfo {author} {\bibfnamefont {J.}~\bibnamefont
  {Helsen}}, \bibinfo {author} {\bibfnamefont {F.}~\bibnamefont {Battistel}},\
  and\ \bibinfo {author} {\bibfnamefont {B.~M.}\ \bibnamefont {Terhal}},\
  }\bibfield  {title} {\bibinfo {title} {Spectral quantum tomography},\ }\href
  {https://doi.org/10.1038/s41534-019-0189-0} {\bibfield  {journal} {\bibinfo
  {journal} {npj Quantum Information}\ }\textbf {\bibinfo {volume} {5}},\
  \bibinfo {pages} {74} (\bibinfo {year} {2019})}\BibitemShut {NoStop}%
\bibitem [{\citenamefont {Huang}\ \emph {et~al.}(2020)\citenamefont {Huang},
  \citenamefont {Kueng},\ and\ \citenamefont {Preskill}}]{Huang2020-ll}%
  \BibitemOpen
  \bibfield  {author} {\bibinfo {author} {\bibfnamefont {H.-Y.}\ \bibnamefont
  {Huang}}, \bibinfo {author} {\bibfnamefont {R.}~\bibnamefont {Kueng}},\ and\
  \bibinfo {author} {\bibfnamefont {J.}~\bibnamefont {Preskill}},\ }\bibfield
  {title} {\bibinfo {title} {Predicting many properties of a quantum system
  from very few measurements},\ }\href
  {https://doi.org/10.1038/s41567-020-0932-7} {\bibfield  {journal} {\bibinfo
  {journal} {Nature Physics}\ }\textbf {\bibinfo {volume} {16}},\ \bibinfo
  {pages} {1050} (\bibinfo {year} {2020})}\BibitemShut {NoStop}%
\bibitem [{\citenamefont {Blume-Kohout}\ \emph {et~al.}()\citenamefont
  {Blume-Kohout}, \citenamefont {Gamble}, \citenamefont {Nielsen},
  \citenamefont {Rudinger}, \citenamefont {Maunz}, \citenamefont {Lobser},
  \citenamefont {Fortier}, \citenamefont {Silva}, \citenamefont {Riste},
  \citenamefont {Ware},\ and\ \citenamefont {Ryan}}]{robin2016bbn}%
  \BibitemOpen
  \bibfield  {author} {\bibinfo {author} {\bibfnamefont {R.~J.}\ \bibnamefont
  {Blume-Kohout}}, \bibinfo {author} {\bibfnamefont {J.~K.}\ \bibnamefont
  {Gamble}}, \bibinfo {author} {\bibfnamefont {E.}~\bibnamefont {Nielsen}},
  \bibinfo {author} {\bibfnamefont {K.~M.}\ \bibnamefont {Rudinger}}, \bibinfo
  {author} {\bibfnamefont {P.~L.~W.}\ \bibnamefont {Maunz}}, \bibinfo {author}
  {\bibfnamefont {D.}~\bibnamefont {Lobser}}, \bibinfo {author} {\bibfnamefont
  {K.~M.}\ \bibnamefont {Fortier}}, \bibinfo {author} {\bibfnamefont {M.~d.}\
  \bibnamefont {Silva}}, \bibinfo {author} {\bibfnamefont {D.}~\bibnamefont
  {Riste}}, \bibinfo {author} {\bibfnamefont {M.}~\bibnamefont {Ware}},\ and\
  \bibinfo {author} {\bibfnamefont {C.}~\bibnamefont {Ryan}},\ }\bibfield
  {title} {\bibinfo {title} {Gate set tomography and beyond.}\ }\href
  {https://www.osti.gov/biblio/1345878} {}\BibitemShut {NoStop}%
\bibitem [{\citenamefont {Zhang}\ \emph {et~al.}(2022)\citenamefont {Zhang},
  \citenamefont {Majumder}, \citenamefont {Leung}, \citenamefont {Crain},
  \citenamefont {Wang}, \citenamefont {Fang}, \citenamefont {Debroy},
  \citenamefont {Kim},\ and\ \citenamefont {Brown}}]{zhang2022hidden}%
  \BibitemOpen
  \bibfield  {author} {\bibinfo {author} {\bibfnamefont {B.}~\bibnamefont
  {Zhang}}, \bibinfo {author} {\bibfnamefont {S.}~\bibnamefont {Majumder}},
  \bibinfo {author} {\bibfnamefont {P.~H.}\ \bibnamefont {Leung}}, \bibinfo
  {author} {\bibfnamefont {S.}~\bibnamefont {Crain}}, \bibinfo {author}
  {\bibfnamefont {Y.}~\bibnamefont {Wang}}, \bibinfo {author} {\bibfnamefont
  {C.}~\bibnamefont {Fang}}, \bibinfo {author} {\bibfnamefont {D.~M.}\
  \bibnamefont {Debroy}}, \bibinfo {author} {\bibfnamefont {J.}~\bibnamefont
  {Kim}},\ and\ \bibinfo {author} {\bibfnamefont {K.~R.}\ \bibnamefont
  {Brown}},\ }\bibfield  {title} {\bibinfo {title} {Hidden inverses: Coherent
  error cancellation at the circuit level},\ }\href@noop {} {\bibfield
  {journal} {\bibinfo  {journal} {Physical Review Applied}\ }\textbf {\bibinfo
  {volume} {17}},\ \bibinfo {pages} {034074} (\bibinfo {year}
  {2022})}\BibitemShut {NoStop}%
\bibitem [{\citenamefont {Kim}\ \emph {et~al.}(2022)\citenamefont {Kim},
  \citenamefont {Yun}, \citenamefont {Jang}, \citenamefont {Jang},
  \citenamefont {Park}, \citenamefont {Song}, \citenamefont {Cho},
  \citenamefont {Sim}, \citenamefont {Sohn}, \citenamefont {Jung} \emph
  {et~al.}}]{kim2022approaching}%
  \BibitemOpen
  \bibfield  {author} {\bibinfo {author} {\bibfnamefont {J.}~\bibnamefont
  {Kim}}, \bibinfo {author} {\bibfnamefont {J.}~\bibnamefont {Yun}}, \bibinfo
  {author} {\bibfnamefont {W.}~\bibnamefont {Jang}}, \bibinfo {author}
  {\bibfnamefont {H.}~\bibnamefont {Jang}}, \bibinfo {author} {\bibfnamefont
  {J.}~\bibnamefont {Park}}, \bibinfo {author} {\bibfnamefont {Y.}~\bibnamefont
  {Song}}, \bibinfo {author} {\bibfnamefont {M.-K.}\ \bibnamefont {Cho}},
  \bibinfo {author} {\bibfnamefont {S.}~\bibnamefont {Sim}}, \bibinfo {author}
  {\bibfnamefont {H.}~\bibnamefont {Sohn}}, \bibinfo {author} {\bibfnamefont
  {H.}~\bibnamefont {Jung}}, \emph {et~al.},\ }\bibfield  {title} {\bibinfo
  {title} {Approaching ideal visibility in singlet-triplet qubit operations
  using energy-selective tunneling-based hamiltonian estimation},\ }\href@noop
  {} {\bibfield  {journal} {\bibinfo  {journal} {Physical Review Letters}\
  }\textbf {\bibinfo {volume} {129}},\ \bibinfo {pages} {040501} (\bibinfo
  {year} {2022})}\BibitemShut {NoStop}%
\bibitem [{\citenamefont {Geller}(2021)}]{geller2021conditionally}%
  \BibitemOpen
  \bibfield  {author} {\bibinfo {author} {\bibfnamefont {M.~R.}\ \bibnamefont
  {Geller}},\ }\bibfield  {title} {\bibinfo {title} {Conditionally rigorous
  mitigation of multiqubit measurement errors},\ }\href@noop {} {\bibfield
  {journal} {\bibinfo  {journal} {Physical Review Letters}\ }\textbf {\bibinfo
  {volume} {127}},\ \bibinfo {pages} {090502} (\bibinfo {year}
  {2021})}\BibitemShut {NoStop}%
\bibitem [{\citenamefont {Mooney}\ \emph {et~al.}(2021)\citenamefont {Mooney},
  \citenamefont {White}, \citenamefont {Hill},\ and\ \citenamefont
  {Hollenberg}}]{mooney2021generation}%
  \BibitemOpen
  \bibfield  {author} {\bibinfo {author} {\bibfnamefont {G.~J.}\ \bibnamefont
  {Mooney}}, \bibinfo {author} {\bibfnamefont {G.~A.}\ \bibnamefont {White}},
  \bibinfo {author} {\bibfnamefont {C.~D.}\ \bibnamefont {Hill}},\ and\
  \bibinfo {author} {\bibfnamefont {L.~C.}\ \bibnamefont {Hollenberg}},\
  }\bibfield  {title} {\bibinfo {title} {Generation and verification of
  27-qubit greenberger-horne-zeilinger states in a superconducting quantum
  computer},\ }\href@noop {} {\bibfield  {journal} {\bibinfo  {journal}
  {Journal of Physics Communications}\ }\textbf {\bibinfo {volume} {5}},\
  \bibinfo {pages} {095004} (\bibinfo {year} {2021})}\BibitemShut {NoStop}%
\bibitem [{\citenamefont {White}\ \emph
  {et~al.}(2021{\natexlab{b}})\citenamefont {White}, \citenamefont {Pollock},
  \citenamefont {Hollenberg}, \citenamefont {Hill},\ and\ \citenamefont
  {Modi}}]{white2021many}%
  \BibitemOpen
  \bibfield  {author} {\bibinfo {author} {\bibfnamefont {G.~A.}\ \bibnamefont
  {White}}, \bibinfo {author} {\bibfnamefont {F.~A.}\ \bibnamefont {Pollock}},
  \bibinfo {author} {\bibfnamefont {L.~C.}\ \bibnamefont {Hollenberg}},
  \bibinfo {author} {\bibfnamefont {C.~D.}\ \bibnamefont {Hill}},\ and\
  \bibinfo {author} {\bibfnamefont {K.}~\bibnamefont {Modi}},\ }\bibfield
  {title} {\bibinfo {title} {From many-body to many-time physics},\ }\href@noop
  {} {\bibfield  {journal} {\bibinfo  {journal} {arXiv preprint
  arXiv:2107.13934}\ } (\bibinfo {year} {2021}{\natexlab{b}})}\BibitemShut
  {NoStop}%
\bibitem [{\citenamefont {Zhang}(2021)}]{zhang2021improving}%
  \BibitemOpen
  \bibfield  {author} {\bibinfo {author} {\bibfnamefont {B.}~\bibnamefont
  {Zhang}},\ }\emph {\bibinfo {title} {Improving Circuit Performance in a
  Trapped-Ion Quantum Computer}},\ \href@noop {} {Ph.D. thesis},\ \bibinfo
  {school} {Duke University} (\bibinfo {year} {2021})\BibitemShut {NoStop}%
\bibitem [{\citenamefont {Moueddene}\ \emph {et~al.}(2021)\citenamefont
  {Moueddene}, \citenamefont {Khammassi}, \citenamefont {Feld},\ and\
  \citenamefont {Hamdioui}}]{moueddene2021context}%
  \BibitemOpen
  \bibfield  {author} {\bibinfo {author} {\bibfnamefont {A.~A.}\ \bibnamefont
  {Moueddene}}, \bibinfo {author} {\bibfnamefont {N.}~\bibnamefont
  {Khammassi}}, \bibinfo {author} {\bibfnamefont {S.}~\bibnamefont {Feld}},\
  and\ \bibinfo {author} {\bibfnamefont {S.}~\bibnamefont {Hamdioui}},\
  }\bibfield  {title} {\bibinfo {title} {A context-aware gate set tomography
  characterization of superconducting qubits},\ }\href@noop {} {\bibfield
  {journal} {\bibinfo  {journal} {arXiv preprint arXiv:2103.09922}\ } (\bibinfo
  {year} {2021})}\BibitemShut {NoStop}%
\bibitem [{\citenamefont {Chen}(2021)}]{chen2021quantum}%
  \BibitemOpen
  \bibfield  {author} {\bibinfo {author} {\bibfnamefont {Y.}~\bibnamefont
  {Chen}},\ }\emph {\bibinfo {title} {Quantum Computing: Characterization of
  Resource States with Symmetry and of Near-Term Quantum Circuits}},\
  \href@noop {} {Ph.D. thesis},\ \bibinfo  {school} {State University of New
  York at Stony Brook} (\bibinfo {year} {2021})\BibitemShut {NoStop}%
\bibitem [{\citenamefont {Luhman}(2021)}]{luhman2021control}%
  \BibitemOpen
  \bibfield  {author} {\bibinfo {author} {\bibfnamefont {D.}~\bibnamefont
  {Luhman}},\ }\href@noop {} {\emph {\bibinfo {title} {Control and
  Characterization of a Spin-Orbit-Driven Singlet-Triplet Qubit in
  Silicon.}}},\ \bibinfo {type} {Tech. Rep.}\ (\bibinfo  {institution} {Sandia
  National Lab.(SNL-NM), Albuquerque, NM (United States)},\ \bibinfo {year}
  {2021})\BibitemShut {NoStop}%
\bibitem [{\citenamefont {White}\ \emph {et~al.}(2020)\citenamefont {White},
  \citenamefont {Hill}, \citenamefont {Pollock}, \citenamefont {Hollenberg},\
  and\ \citenamefont {Modi}}]{white2020demonstration}%
  \BibitemOpen
  \bibfield  {author} {\bibinfo {author} {\bibfnamefont {G.~A.}\ \bibnamefont
  {White}}, \bibinfo {author} {\bibfnamefont {C.~D.}\ \bibnamefont {Hill}},
  \bibinfo {author} {\bibfnamefont {F.~A.}\ \bibnamefont {Pollock}}, \bibinfo
  {author} {\bibfnamefont {L.~C.}\ \bibnamefont {Hollenberg}},\ and\ \bibinfo
  {author} {\bibfnamefont {K.}~\bibnamefont {Modi}},\ }\bibfield  {title}
  {\bibinfo {title} {Demonstration of non-markovian process characterisation
  and control on a quantum processor},\ }\href@noop {} {\bibfield  {journal}
  {\bibinfo  {journal} {Nature Communications}\ }\textbf {\bibinfo {volume}
  {11}},\ \bibinfo {pages} {6301} (\bibinfo {year} {2020})}\BibitemShut
  {NoStop}%
\bibitem [{\citenamefont {Geller}(2020)}]{geller2020rigorous}%
  \BibitemOpen
  \bibfield  {author} {\bibinfo {author} {\bibfnamefont {M.~R.}\ \bibnamefont
  {Geller}},\ }\bibfield  {title} {\bibinfo {title} {Rigorous measurement error
  correction},\ }\href@noop {} {\bibfield  {journal} {\bibinfo  {journal}
  {Quantum Science and Technology}\ }\textbf {\bibinfo {volume} {5}},\ \bibinfo
  {pages} {03LT01} (\bibinfo {year} {2020})}\BibitemShut {NoStop}%
\bibitem [{\citenamefont {Wang}\ \emph {et~al.}(2020)\citenamefont {Wang},
  \citenamefont {Crain}, \citenamefont {Fang}, \citenamefont {Zhang},
  \citenamefont {Huang}, \citenamefont {Liang}, \citenamefont {Leung},
  \citenamefont {Brown},\ and\ \citenamefont {Kim}}]{wang2020high}%
  \BibitemOpen
  \bibfield  {author} {\bibinfo {author} {\bibfnamefont {Y.}~\bibnamefont
  {Wang}}, \bibinfo {author} {\bibfnamefont {S.}~\bibnamefont {Crain}},
  \bibinfo {author} {\bibfnamefont {C.}~\bibnamefont {Fang}}, \bibinfo {author}
  {\bibfnamefont {B.}~\bibnamefont {Zhang}}, \bibinfo {author} {\bibfnamefont
  {S.}~\bibnamefont {Huang}}, \bibinfo {author} {\bibfnamefont
  {Q.}~\bibnamefont {Liang}}, \bibinfo {author} {\bibfnamefont {P.~H.}\
  \bibnamefont {Leung}}, \bibinfo {author} {\bibfnamefont {K.~R.}\ \bibnamefont
  {Brown}},\ and\ \bibinfo {author} {\bibfnamefont {J.}~\bibnamefont {Kim}},\
  }\bibfield  {title} {\bibinfo {title} {High-fidelity two-qubit gates using a
  microelectromechanical-system-based beam steering system for individual qubit
  addressing},\ }\href@noop {} {\bibfield  {journal} {\bibinfo  {journal}
  {Physical Review Letters}\ }\textbf {\bibinfo {volume} {125}},\ \bibinfo
  {pages} {150505} (\bibinfo {year} {2020})}\BibitemShut {NoStop}%
\bibitem [{\citenamefont {Rudinger}\ \emph {et~al.}(2019)\citenamefont
  {Rudinger}, \citenamefont {Proctor}, \citenamefont {Langharst}, \citenamefont
  {Sarovar}, \citenamefont {Young},\ and\ \citenamefont
  {Blume-Kohout}}]{Rudinger-PRX2019}%
  \BibitemOpen
  \bibfield  {author} {\bibinfo {author} {\bibfnamefont {K.}~\bibnamefont
  {Rudinger}}, \bibinfo {author} {\bibfnamefont {T.}~\bibnamefont {Proctor}},
  \bibinfo {author} {\bibfnamefont {D.}~\bibnamefont {Langharst}}, \bibinfo
  {author} {\bibfnamefont {M.}~\bibnamefont {Sarovar}}, \bibinfo {author}
  {\bibfnamefont {K.}~\bibnamefont {Young}},\ and\ \bibinfo {author}
  {\bibfnamefont {R.}~\bibnamefont {Blume-Kohout}},\ }\bibfield  {title}
  {\bibinfo {title} {Probing context-dependent errors in quantum processors},\
  }\href {https://doi.org/10.1103/PhysRevX.9.021045} {\bibfield  {journal}
  {\bibinfo  {journal} {Phys. Rev. X}\ }\textbf {\bibinfo {volume} {9}},\
  \bibinfo {pages} {021045} (\bibinfo {year} {2019})}\BibitemShut {NoStop}%
\bibitem [{\citenamefont {Crain}\ \emph {et~al.}(2019)\citenamefont {Crain},
  \citenamefont {Cahall}, \citenamefont {Vrijsen}, \citenamefont {Wollman},
  \citenamefont {Shaw}, \citenamefont {Verma}, \citenamefont {Nam},\ and\
  \citenamefont {Kim}}]{crain2019high}%
  \BibitemOpen
  \bibfield  {author} {\bibinfo {author} {\bibfnamefont {S.}~\bibnamefont
  {Crain}}, \bibinfo {author} {\bibfnamefont {C.}~\bibnamefont {Cahall}},
  \bibinfo {author} {\bibfnamefont {G.}~\bibnamefont {Vrijsen}}, \bibinfo
  {author} {\bibfnamefont {E.~E.}\ \bibnamefont {Wollman}}, \bibinfo {author}
  {\bibfnamefont {M.~D.}\ \bibnamefont {Shaw}}, \bibinfo {author}
  {\bibfnamefont {V.~B.}\ \bibnamefont {Verma}}, \bibinfo {author}
  {\bibfnamefont {S.~W.}\ \bibnamefont {Nam}},\ and\ \bibinfo {author}
  {\bibfnamefont {J.}~\bibnamefont {Kim}},\ }\bibfield  {title} {\bibinfo
  {title} {High-speed low-crosstalk detection of a 171yb+ qubit using
  superconducting nanowire single photon detectors},\ }\href@noop {} {\bibfield
   {journal} {\bibinfo  {journal} {Communications Physics}\ }\textbf {\bibinfo
  {volume} {2}},\ \bibinfo {pages} {97} (\bibinfo {year} {2019})}\BibitemShut
  {NoStop}%
\bibitem [{\citenamefont {Hu}\ \emph {et~al.}(2018)\citenamefont {Hu},
  \citenamefont {Cui}, \citenamefont {Santos}, \citenamefont {Huang},
  \citenamefont {Sarandy}, \citenamefont {Li},\ and\ \citenamefont
  {Guo}}]{hu2018experimental}%
  \BibitemOpen
  \bibfield  {author} {\bibinfo {author} {\bibfnamefont {C.-K.}\ \bibnamefont
  {Hu}}, \bibinfo {author} {\bibfnamefont {J.-M.}\ \bibnamefont {Cui}},
  \bibinfo {author} {\bibfnamefont {A.~C.}\ \bibnamefont {Santos}}, \bibinfo
  {author} {\bibfnamefont {Y.-F.}\ \bibnamefont {Huang}}, \bibinfo {author}
  {\bibfnamefont {M.~S.}\ \bibnamefont {Sarandy}}, \bibinfo {author}
  {\bibfnamefont {C.-F.}\ \bibnamefont {Li}},\ and\ \bibinfo {author}
  {\bibfnamefont {G.-C.}\ \bibnamefont {Guo}},\ }\bibfield  {title} {\bibinfo
  {title} {Experimental implementation of generalized transitionless quantum
  driving},\ }\href@noop {} {\bibfield  {journal} {\bibinfo  {journal} {Optics
  Letters}\ }\textbf {\bibinfo {volume} {43}},\ \bibinfo {pages} {3136}
  (\bibinfo {year} {2018})}\BibitemShut {NoStop}%
\bibitem [{\citenamefont {O’Brien}\ \emph {et~al.}(2017)\citenamefont
  {O’Brien}, \citenamefont {Tarasinski},\ and\ \citenamefont
  {DiCarlo}}]{o2017density}%
  \BibitemOpen
  \bibfield  {author} {\bibinfo {author} {\bibfnamefont {T.~E.}\ \bibnamefont
  {O’Brien}}, \bibinfo {author} {\bibfnamefont {B.}~\bibnamefont
  {Tarasinski}},\ and\ \bibinfo {author} {\bibfnamefont {L.}~\bibnamefont
  {DiCarlo}},\ }\bibfield  {title} {\bibinfo {title} {Density-matrix simulation
  of small surface codes under current and projected experimental noise},\
  }\href@noop {} {\bibfield  {journal} {\bibinfo  {journal} {npj Quantum
  Information}\ }\textbf {\bibinfo {volume} {3}},\ \bibinfo {pages} {39}
  (\bibinfo {year} {2017})}\BibitemShut {NoStop}%
\bibitem [{\citenamefont {Rudinger}\ \emph
  {et~al.}(2017{\natexlab{b}})\citenamefont {Rudinger}, \citenamefont {Kimmel},
  \citenamefont {Lobser},\ and\ \citenamefont {Maunz}}]{RudingerRPE_PRL2017}%
  \BibitemOpen
  \bibfield  {author} {\bibinfo {author} {\bibfnamefont {K.}~\bibnamefont
  {Rudinger}}, \bibinfo {author} {\bibfnamefont {S.}~\bibnamefont {Kimmel}},
  \bibinfo {author} {\bibfnamefont {D.}~\bibnamefont {Lobser}},\ and\ \bibinfo
  {author} {\bibfnamefont {P.}~\bibnamefont {Maunz}},\ }\bibfield  {title}
  {\bibinfo {title} {Experimental demonstration of a cheap and accurate phase
  estimation},\ }\href {https://doi.org/10.1103/PhysRevLett.118.190502}
  {\bibfield  {journal} {\bibinfo  {journal} {Phys. Rev. Lett.}\ }\textbf
  {\bibinfo {volume} {118}},\ \bibinfo {pages} {190502} (\bibinfo {year}
  {2017}{\natexlab{b}})}\BibitemShut {NoStop}%
\bibitem [{\citenamefont {Dehollain}\ \emph
  {et~al.}(2016{\natexlab{b}})\citenamefont {Dehollain}, \citenamefont
  {Muhonen}, \citenamefont {Blume-Kohout}, \citenamefont {Rudinger},
  \citenamefont {Gamble}, \citenamefont {Nielsen}, \citenamefont {Laucht},
  \citenamefont {Simmons}, \citenamefont {Kalra}, \citenamefont {Dzurak},\ and\
  \citenamefont {Morello}}]{GST2016}%
  \BibitemOpen
  \bibfield  {author} {\bibinfo {author} {\bibfnamefont {J.~P.}\ \bibnamefont
  {Dehollain}}, \bibinfo {author} {\bibfnamefont {J.~T.}\ \bibnamefont
  {Muhonen}}, \bibinfo {author} {\bibfnamefont {R.}~\bibnamefont
  {Blume-Kohout}}, \bibinfo {author} {\bibfnamefont {K.~M.}\ \bibnamefont
  {Rudinger}}, \bibinfo {author} {\bibfnamefont {J.~K.}\ \bibnamefont
  {Gamble}}, \bibinfo {author} {\bibfnamefont {E.}~\bibnamefont {Nielsen}},
  \bibinfo {author} {\bibfnamefont {A.}~\bibnamefont {Laucht}}, \bibinfo
  {author} {\bibfnamefont {S.}~\bibnamefont {Simmons}}, \bibinfo {author}
  {\bibfnamefont {R.}~\bibnamefont {Kalra}}, \bibinfo {author} {\bibfnamefont
  {A.~S.}\ \bibnamefont {Dzurak}},\ and\ \bibinfo {author} {\bibfnamefont
  {A.}~\bibnamefont {Morello}},\ }\bibfield  {title} {\bibinfo {title}
  {Optimization of a solid-state electron spin qubit using gate set
  tomography},\ }\href {https://doi.org/10.1088/1367-2630/18/10/103018}
  {\bibfield  {journal} {\bibinfo  {journal} {New Journal of Physics}\ }\textbf
  {\bibinfo {volume} {18}},\ \bibinfo {pages} {103018} (\bibinfo {year}
  {2016}{\natexlab{b}})}\BibitemShut {NoStop}%
\bibitem [{\citenamefont {Xue}\ \emph {et~al.}(2022)\citenamefont {Xue},
  \citenamefont {Russ}, \citenamefont {Samkharadze}, \citenamefont {Undseth},
  \citenamefont {Sammak}, \citenamefont {Scappucci},\ and\ \citenamefont
  {Vandersypen}}]{xue2022quantum}%
  \BibitemOpen
  \bibfield  {author} {\bibinfo {author} {\bibfnamefont {X.}~\bibnamefont
  {Xue}}, \bibinfo {author} {\bibfnamefont {M.}~\bibnamefont {Russ}}, \bibinfo
  {author} {\bibfnamefont {N.}~\bibnamefont {Samkharadze}}, \bibinfo {author}
  {\bibfnamefont {B.}~\bibnamefont {Undseth}}, \bibinfo {author} {\bibfnamefont
  {A.}~\bibnamefont {Sammak}}, \bibinfo {author} {\bibfnamefont
  {G.}~\bibnamefont {Scappucci}},\ and\ \bibinfo {author} {\bibfnamefont
  {L.~M.}\ \bibnamefont {Vandersypen}},\ }\bibfield  {title} {\bibinfo {title}
  {Quantum logic with spin qubits crossing the surface code threshold},\
  }\href@noop {} {\bibfield  {journal} {\bibinfo  {journal} {Nature}\ }\textbf
  {\bibinfo {volume} {601}},\ \bibinfo {pages} {343} (\bibinfo {year}
  {2022})}\BibitemShut {NoStop}%
\bibitem [{\citenamefont {Dahlhauser}\ and\ \citenamefont
  {Humble}(2022)}]{dahlhauser2022benchmarking}%
  \BibitemOpen
  \bibfield  {author} {\bibinfo {author} {\bibfnamefont {M.~L.}\ \bibnamefont
  {Dahlhauser}}\ and\ \bibinfo {author} {\bibfnamefont {T.~S.}\ \bibnamefont
  {Humble}},\ }\bibfield  {title} {\bibinfo {title} {Benchmarking
  characterization methods for noisy quantum circuits},\ }\href@noop {}
  {\bibfield  {journal} {\bibinfo  {journal} {arXiv preprint arXiv:2201.02243}\
  } (\bibinfo {year} {2022})}\BibitemShut {NoStop}%
\bibitem [{\citenamefont {Dahlhauser}(2021)}]{dahlhauser2021characterization}%
  \BibitemOpen
  \bibfield  {author} {\bibinfo {author} {\bibfnamefont {M.~L.}\ \bibnamefont
  {Dahlhauser}},\ }\bibfield  {title} {\bibinfo {title} {Characterization and
  benchmarking of quantum computers},\ }\href@noop {} {\  (\bibinfo {year}
  {2021})}\BibitemShut {NoStop}%
\bibitem [{\citenamefont {Hong}\ \emph
  {et~al.}(2020{\natexlab{b}})\citenamefont {Hong}, \citenamefont {Papageorge},
  \citenamefont {Sivarajah}, \citenamefont {Crossman}, \citenamefont {Didier},
  \citenamefont {Polloreno}, \citenamefont {Sete}, \citenamefont {Turkowski},
  \citenamefont {da~Silva},\ and\ \citenamefont
  {Johnson}}]{hong2020demonstration}%
  \BibitemOpen
  \bibfield  {author} {\bibinfo {author} {\bibfnamefont {S.~S.}\ \bibnamefont
  {Hong}}, \bibinfo {author} {\bibfnamefont {A.~T.}\ \bibnamefont
  {Papageorge}}, \bibinfo {author} {\bibfnamefont {P.}~\bibnamefont
  {Sivarajah}}, \bibinfo {author} {\bibfnamefont {G.}~\bibnamefont {Crossman}},
  \bibinfo {author} {\bibfnamefont {N.}~\bibnamefont {Didier}}, \bibinfo
  {author} {\bibfnamefont {A.~M.}\ \bibnamefont {Polloreno}}, \bibinfo {author}
  {\bibfnamefont {E.~A.}\ \bibnamefont {Sete}}, \bibinfo {author}
  {\bibfnamefont {S.~W.}\ \bibnamefont {Turkowski}}, \bibinfo {author}
  {\bibfnamefont {M.~P.}\ \bibnamefont {da~Silva}},\ and\ \bibinfo {author}
  {\bibfnamefont {B.~R.}\ \bibnamefont {Johnson}},\ }\bibfield  {title}
  {\bibinfo {title} {Demonstration of a parametrically activated entangling
  gate protected from flux noise},\ }\href@noop {} {\bibfield  {journal}
  {\bibinfo  {journal} {Physical Review A}\ }\textbf {\bibinfo {volume}
  {101}},\ \bibinfo {pages} {012302} (\bibinfo {year}
  {2020}{\natexlab{b}})}\BibitemShut {NoStop}%
\bibitem [{\citenamefont {Hughes}\ \emph {et~al.}(2020)\citenamefont {Hughes},
  \citenamefont {Sch{\"a}fer}, \citenamefont {Thirumalai}, \citenamefont
  {Nadlinger}, \citenamefont {Woodrow}, \citenamefont {Lucas},\ and\
  \citenamefont {Ballance}}]{hughes2020benchmarking}%
  \BibitemOpen
  \bibfield  {author} {\bibinfo {author} {\bibfnamefont {A.}~\bibnamefont
  {Hughes}}, \bibinfo {author} {\bibfnamefont {V.}~\bibnamefont {Sch{\"a}fer}},
  \bibinfo {author} {\bibfnamefont {K.}~\bibnamefont {Thirumalai}}, \bibinfo
  {author} {\bibfnamefont {D.}~\bibnamefont {Nadlinger}}, \bibinfo {author}
  {\bibfnamefont {S.}~\bibnamefont {Woodrow}}, \bibinfo {author} {\bibfnamefont
  {D.}~\bibnamefont {Lucas}},\ and\ \bibinfo {author} {\bibfnamefont
  {C.}~\bibnamefont {Ballance}},\ }\bibfield  {title} {\bibinfo {title}
  {Benchmarking a high-fidelity mixed-species entangling gate},\ }\href@noop {}
  {\bibfield  {journal} {\bibinfo  {journal} {Physical Review Letters}\
  }\textbf {\bibinfo {volume} {125}},\ \bibinfo {pages} {080504} (\bibinfo
  {year} {2020})}\BibitemShut {NoStop}%
\bibitem [{\citenamefont {Govia}\ \emph {et~al.}(2020)\citenamefont {Govia},
  \citenamefont {Ribeill}, \citenamefont {Rist{\`e}}, \citenamefont {Ware},\
  and\ \citenamefont {Krovi}}]{govia2020bootstrapping}%
  \BibitemOpen
  \bibfield  {author} {\bibinfo {author} {\bibfnamefont {L.}~\bibnamefont
  {Govia}}, \bibinfo {author} {\bibfnamefont {G.}~\bibnamefont {Ribeill}},
  \bibinfo {author} {\bibfnamefont {D.}~\bibnamefont {Rist{\`e}}}, \bibinfo
  {author} {\bibfnamefont {M.}~\bibnamefont {Ware}},\ and\ \bibinfo {author}
  {\bibfnamefont {H.}~\bibnamefont {Krovi}},\ }\bibfield  {title} {\bibinfo
  {title} {Bootstrapping quantum process tomography via a perturbative
  ansatz},\ }\href@noop {} {\bibfield  {journal} {\bibinfo  {journal} {Nature
  communications}\ }\textbf {\bibinfo {volume} {11}},\ \bibinfo {pages} {1084}
  (\bibinfo {year} {2020})}\BibitemShut {NoStop}%
\bibitem [{\citenamefont {Moueddene}\ \emph {et~al.}(2020)\citenamefont
  {Moueddene}, \citenamefont {Khammassi}, \citenamefont {Bertels},\ and\
  \citenamefont {Almudever}}]{moueddene2020realistic}%
  \BibitemOpen
  \bibfield  {author} {\bibinfo {author} {\bibfnamefont {A.~A.}\ \bibnamefont
  {Moueddene}}, \bibinfo {author} {\bibfnamefont {N.}~\bibnamefont
  {Khammassi}}, \bibinfo {author} {\bibfnamefont {K.}~\bibnamefont {Bertels}},\
  and\ \bibinfo {author} {\bibfnamefont {C.~G.}\ \bibnamefont {Almudever}},\
  }\bibfield  {title} {\bibinfo {title} {Realistic simulation of quantum
  computation using unitary and measurement channels},\ }\href@noop {}
  {\bibfield  {journal} {\bibinfo  {journal} {Physical Review A}\ }\textbf
  {\bibinfo {volume} {102}},\ \bibinfo {pages} {052608} (\bibinfo {year}
  {2020})}\BibitemShut {NoStop}%
\bibitem [{\citenamefont {Hashim}\ \emph {et~al.}(2023)\citenamefont {Hashim},
  \citenamefont {Seritan}, \citenamefont {Proctor}, \citenamefont {Rudinger},
  \citenamefont {Goss}, \citenamefont {Naik}, \citenamefont {Kreikebaum},
  \citenamefont {Santiago},\ and\ \citenamefont
  {Siddiqi}}]{hashim2023benchmarking}%
  \BibitemOpen
  \bibfield  {author} {\bibinfo {author} {\bibfnamefont {A.}~\bibnamefont
  {Hashim}}, \bibinfo {author} {\bibfnamefont {S.}~\bibnamefont {Seritan}},
  \bibinfo {author} {\bibfnamefont {T.}~\bibnamefont {Proctor}}, \bibinfo
  {author} {\bibfnamefont {K.}~\bibnamefont {Rudinger}}, \bibinfo {author}
  {\bibfnamefont {N.}~\bibnamefont {Goss}}, \bibinfo {author} {\bibfnamefont
  {R.}~\bibnamefont {Naik}}, \bibinfo {author} {\bibfnamefont {J.~M.}\
  \bibnamefont {Kreikebaum}}, \bibinfo {author} {\bibfnamefont
  {D.}~\bibnamefont {Santiago}},\ and\ \bibinfo {author} {\bibfnamefont
  {I.}~\bibnamefont {Siddiqi}},\ }\bibfield  {title} {\bibinfo {title}
  {Benchmarking quantum logic operations for achieving fault tolerance},\
  }\href@noop {} {\bibfield  {journal} {\bibinfo  {journal} {Bulletin of the
  American Physical Society}\ } (\bibinfo {year} {2023})}\BibitemShut {NoStop}%
\bibitem [{\citenamefont {Wang}\ \emph {et~al.}(2021)\citenamefont {Wang},
  \citenamefont {Fontana}, \citenamefont {Cerezo}, \citenamefont {Sharma},
  \citenamefont {Sone}, \citenamefont {Cincio},\ and\ \citenamefont
  {Coles}}]{wang2021noise}%
  \BibitemOpen
  \bibfield  {author} {\bibinfo {author} {\bibfnamefont {S.}~\bibnamefont
  {Wang}}, \bibinfo {author} {\bibfnamefont {E.}~\bibnamefont {Fontana}},
  \bibinfo {author} {\bibfnamefont {M.}~\bibnamefont {Cerezo}}, \bibinfo
  {author} {\bibfnamefont {K.}~\bibnamefont {Sharma}}, \bibinfo {author}
  {\bibfnamefont {A.}~\bibnamefont {Sone}}, \bibinfo {author} {\bibfnamefont
  {L.}~\bibnamefont {Cincio}},\ and\ \bibinfo {author} {\bibfnamefont {P.~J.}\
  \bibnamefont {Coles}},\ }\bibfield  {title} {\bibinfo {title} {Noise-induced
  barren plateaus in variational quantum algorithms},\ }\href@noop {}
  {\bibfield  {journal} {\bibinfo  {journal} {Nature communications}\ }\textbf
  {\bibinfo {volume} {12}},\ \bibinfo {pages} {6961} (\bibinfo {year}
  {2021})}\BibitemShut {NoStop}%
\bibitem [{\citenamefont {Cincio}\ \emph {et~al.}(2021)\citenamefont {Cincio},
  \citenamefont {Rudinger}, \citenamefont {Sarovar},\ and\ \citenamefont
  {Coles}}]{cincio2020machine}%
  \BibitemOpen
  \bibfield  {author} {\bibinfo {author} {\bibfnamefont {L.}~\bibnamefont
  {Cincio}}, \bibinfo {author} {\bibfnamefont {K.}~\bibnamefont {Rudinger}},
  \bibinfo {author} {\bibfnamefont {M.}~\bibnamefont {Sarovar}},\ and\ \bibinfo
  {author} {\bibfnamefont {P.~J.}\ \bibnamefont {Coles}},\ }\bibfield  {title}
  {\bibinfo {title} {Machine learning of noise-resilient quantum circuits},\
  }\href {https://doi.org/10.1103/PRXQuantum.2.010324} {\bibfield  {journal}
  {\bibinfo  {journal} {PRX Quantum}\ }\textbf {\bibinfo {volume} {2}},\
  \bibinfo {pages} {010324} (\bibinfo {year} {2021})}\BibitemShut {NoStop}%
\bibitem [{\citenamefont {Xue}\ \emph {et~al.}(2021)\citenamefont {Xue},
  \citenamefont {Russ}, \citenamefont {Samkharadze}, \citenamefont {Undseth},
  \citenamefont {Sammak}, \citenamefont {Scappucci},\ and\ \citenamefont
  {Vandersypen}}]{xue2021computing}%
  \BibitemOpen
  \bibfield  {author} {\bibinfo {author} {\bibfnamefont {X.}~\bibnamefont
  {Xue}}, \bibinfo {author} {\bibfnamefont {M.}~\bibnamefont {Russ}}, \bibinfo
  {author} {\bibfnamefont {N.}~\bibnamefont {Samkharadze}}, \bibinfo {author}
  {\bibfnamefont {B.}~\bibnamefont {Undseth}}, \bibinfo {author} {\bibfnamefont
  {A.}~\bibnamefont {Sammak}}, \bibinfo {author} {\bibfnamefont
  {G.}~\bibnamefont {Scappucci}},\ and\ \bibinfo {author} {\bibfnamefont
  {L.~M.}\ \bibnamefont {Vandersypen}},\ }\bibfield  {title} {\bibinfo {title}
  {Computing with spin qubits at the surface code error threshold},\
  }\href@noop {} {\bibfield  {journal} {\bibinfo  {journal} {arXiv preprint
  arXiv:2107.00628}\ } (\bibinfo {year} {2021})}\BibitemShut {NoStop}%
\bibitem [{\citenamefont {Zhang}\ \emph
  {et~al.}(2020{\natexlab{b}})\citenamefont {Zhang}, \citenamefont {Lu},
  \citenamefont {Zhang}, \citenamefont {Chen}, \citenamefont {Li},
  \citenamefont {Zhang},\ and\ \citenamefont {Kim}}]{zhang2020error}%
  \BibitemOpen
  \bibfield  {author} {\bibinfo {author} {\bibfnamefont {S.}~\bibnamefont
  {Zhang}}, \bibinfo {author} {\bibfnamefont {Y.}~\bibnamefont {Lu}}, \bibinfo
  {author} {\bibfnamefont {K.}~\bibnamefont {Zhang}}, \bibinfo {author}
  {\bibfnamefont {W.}~\bibnamefont {Chen}}, \bibinfo {author} {\bibfnamefont
  {Y.}~\bibnamefont {Li}}, \bibinfo {author} {\bibfnamefont {J.-N.}\
  \bibnamefont {Zhang}},\ and\ \bibinfo {author} {\bibfnamefont
  {K.}~\bibnamefont {Kim}},\ }\bibfield  {title} {\bibinfo {title}
  {Error-mitigated quantum gates exceeding physical fidelities in a trapped-ion
  system},\ }\href@noop {} {\bibfield  {journal} {\bibinfo  {journal} {Nature
  communications}\ }\textbf {\bibinfo {volume} {11}},\ \bibinfo {pages} {587}
  (\bibinfo {year} {2020}{\natexlab{b}})}\BibitemShut {NoStop}%
\bibitem [{\citenamefont {Blume-Kohout}\ \emph {et~al.}(2022)\citenamefont
  {Blume-Kohout}, \citenamefont {da~Silva}, \citenamefont {Nielsen},
  \citenamefont {Proctor}, \citenamefont {Rudinger}, \citenamefont {Sarovar},\
  and\ \citenamefont {Young}}]{rbk2022taxonomy}%
  \BibitemOpen
  \bibfield  {author} {\bibinfo {author} {\bibfnamefont {R.}~\bibnamefont
  {Blume-Kohout}}, \bibinfo {author} {\bibfnamefont {M.~P.}\ \bibnamefont
  {da~Silva}}, \bibinfo {author} {\bibfnamefont {E.}~\bibnamefont {Nielsen}},
  \bibinfo {author} {\bibfnamefont {T.}~\bibnamefont {Proctor}}, \bibinfo
  {author} {\bibfnamefont {K.}~\bibnamefont {Rudinger}}, \bibinfo {author}
  {\bibfnamefont {M.}~\bibnamefont {Sarovar}},\ and\ \bibinfo {author}
  {\bibfnamefont {K.}~\bibnamefont {Young}},\ }\bibfield  {title} {\bibinfo
  {title} {A taxonomy of small markovian errors},\ }\href
  {https://doi.org/10.1103/PRXQuantum.3.020335} {\bibfield  {journal} {\bibinfo
   {journal} {PRX Quantum}\ }\textbf {\bibinfo {volume} {3}},\ \bibinfo {pages}
  {020335} (\bibinfo {year} {2022})}\BibitemShut {NoStop}%
\bibitem [{\citenamefont {Ly}\ \emph {et~al.}(2017)\citenamefont {Ly},
  \citenamefont {Marsman}, \citenamefont {Verhagen}, \citenamefont {Grasman},\
  and\ \citenamefont {Wagenmakers}}]{ly2017tutorial}%
  \BibitemOpen
  \bibfield  {author} {\bibinfo {author} {\bibfnamefont {A.}~\bibnamefont
  {Ly}}, \bibinfo {author} {\bibfnamefont {M.}~\bibnamefont {Marsman}},
  \bibinfo {author} {\bibfnamefont {J.}~\bibnamefont {Verhagen}}, \bibinfo
  {author} {\bibfnamefont {R.~P.}\ \bibnamefont {Grasman}},\ and\ \bibinfo
  {author} {\bibfnamefont {E.-J.}\ \bibnamefont {Wagenmakers}},\ }\bibfield
  {title} {\bibinfo {title} {A tutorial on fisher information},\ }\href@noop {}
  {\bibfield  {journal} {\bibinfo  {journal} {Journal of Mathematical
  Psychology}\ }\textbf {\bibinfo {volume} {80}},\ \bibinfo {pages} {40}
  (\bibinfo {year} {2017})}\BibitemShut {NoStop}%
\bibitem [{\citenamefont {Nielsen}\ \emph {et~al.}(2019)\citenamefont
  {Nielsen}, \citenamefont {Blume-Kohout}, \citenamefont {Rudinger},
  \citenamefont {Proctor}, \citenamefont {Saldyt} \emph
  {et~al.}}]{nielsen2019python}%
  \BibitemOpen
  \bibfield  {author} {\bibinfo {author} {\bibfnamefont {E.}~\bibnamefont
  {Nielsen}}, \bibinfo {author} {\bibfnamefont {R.~J.}\ \bibnamefont
  {Blume-Kohout}}, \bibinfo {author} {\bibfnamefont {K.~M.}\ \bibnamefont
  {Rudinger}}, \bibinfo {author} {\bibfnamefont {T.~J.}\ \bibnamefont
  {Proctor}}, \bibinfo {author} {\bibfnamefont {L.}~\bibnamefont {Saldyt}},
  \emph {et~al.},\ }\href@noop {} {\emph {\bibinfo {title} {Python GST
  Implementation (PyGSTi) v. 0.9}}},\ \bibinfo {type} {Tech. Rep.}\ (\bibinfo
  {institution} {Sandia National Lab.(SNL-NM), Albuquerque, NM (United
  States)},\ \bibinfo {year} {2019})\BibitemShut {NoStop}%
\bibitem [{\citenamefont {Gambetta}\ \emph
  {et~al.}(2012{\natexlab{b}})\citenamefont {Gambetta}, \citenamefont
  {C\'orcoles}, \citenamefont {Merkel}, \citenamefont {Johnson}, \citenamefont
  {Smolin}, \citenamefont {Chow}, \citenamefont {Ryan}, \citenamefont
  {Rigetti}, \citenamefont {Poletto}, \citenamefont {Ohki}, \citenamefont
  {Ketchen},\ and\ \citenamefont {Steffen}}]{gambetta2012simultaneousRB}%
  \BibitemOpen
  \bibfield  {author} {\bibinfo {author} {\bibfnamefont {J.~M.}\ \bibnamefont
  {Gambetta}}, \bibinfo {author} {\bibfnamefont {A.~D.}\ \bibnamefont
  {C\'orcoles}}, \bibinfo {author} {\bibfnamefont {S.~T.}\ \bibnamefont
  {Merkel}}, \bibinfo {author} {\bibfnamefont {B.~R.}\ \bibnamefont {Johnson}},
  \bibinfo {author} {\bibfnamefont {J.~A.}\ \bibnamefont {Smolin}}, \bibinfo
  {author} {\bibfnamefont {J.~M.}\ \bibnamefont {Chow}}, \bibinfo {author}
  {\bibfnamefont {C.~A.}\ \bibnamefont {Ryan}}, \bibinfo {author}
  {\bibfnamefont {C.}~\bibnamefont {Rigetti}}, \bibinfo {author} {\bibfnamefont
  {S.}~\bibnamefont {Poletto}}, \bibinfo {author} {\bibfnamefont {T.~A.}\
  \bibnamefont {Ohki}}, \bibinfo {author} {\bibfnamefont {M.~B.}\ \bibnamefont
  {Ketchen}},\ and\ \bibinfo {author} {\bibfnamefont {M.}~\bibnamefont
  {Steffen}},\ }\bibfield  {title} {\bibinfo {title} {Characterization of
  addressability by simultaneous randomized benchmarking [supplementary
  material]},\ }\href {https://doi.org/10.1103/PhysRevLett.109.240504}
  {\bibfield  {journal} {\bibinfo  {journal} {Phys. Rev. Lett.}\ }\textbf
  {\bibinfo {volume} {109}},\ \bibinfo {pages} {240504} (\bibinfo {year}
  {2012}{\natexlab{b}})}\BibitemShut {NoStop}%
\bibitem [{\citenamefont {Hashim}()}]{akel}%
  \BibitemOpen
  \bibfield  {author} {\bibinfo {author} {\bibfnamefont {A.}~\bibnamefont
  {Hashim}},\ }\href@noop {} {}\bibinfo {howpublished} {personal
  communication}\BibitemShut {NoStop}%
\bibitem [{\citenamefont {{QSCOUT Collaboration}}()}]{qscout}%
  \BibitemOpen
  \bibfield  {author} {\bibinfo {author} {\bibnamefont {{QSCOUT
  Collaboration}}},\ }\href@noop {} {}\bibinfo {howpublished} {personal
  communication}\BibitemShut {NoStop}%
\bibitem [{\citenamefont {Bruzewicz}\ \emph {et~al.}(2019)\citenamefont
  {Bruzewicz}, \citenamefont {Chiaverini}, \citenamefont {McConnell},\ and\
  \citenamefont {Sage}}]{bruzewicz2019trapped}%
  \BibitemOpen
  \bibfield  {author} {\bibinfo {author} {\bibfnamefont {C.~D.}\ \bibnamefont
  {Bruzewicz}}, \bibinfo {author} {\bibfnamefont {J.}~\bibnamefont
  {Chiaverini}}, \bibinfo {author} {\bibfnamefont {R.}~\bibnamefont
  {McConnell}},\ and\ \bibinfo {author} {\bibfnamefont {J.~M.}\ \bibnamefont
  {Sage}},\ }\bibfield  {title} {\bibinfo {title} {Trapped-ion quantum
  computing: Progress and challenges},\ }\href@noop {} {\bibfield  {journal}
  {\bibinfo  {journal} {Applied Physics Reviews}\ }\textbf {\bibinfo {volume}
  {6}} (\bibinfo {year} {2019})}\BibitemShut {NoStop}%
\bibitem [{\citenamefont {Stemp}()}]{holly}%
  \BibitemOpen
  \bibfield  {author} {\bibinfo {author} {\bibfnamefont {H.}~\bibnamefont
  {Stemp}},\ }\href@noop {} {}\bibinfo {howpublished} {personal
  communication}\BibitemShut {NoStop}%
\bibitem [{\citenamefont {{Resende, Mauricio G.C. and Ribeiro, Celso
  C.}}(2006)}]{resende2006grasp}%
  \BibitemOpen
  \bibfield  {author} {\bibinfo {author} {\bibnamefont {{Resende, Mauricio G.C.
  and Ribeiro, Celso C.}}},\ }\bibfield  {title} {\bibinfo {title} {{Handbook
  of Metaheuristics}}\ }(\bibinfo  {publisher} {Springer US},\ \bibinfo {year}
  {2006})\ Chap.\ \bibinfo {chapter} {{Greedy Randomized Adaptive Search
  Procedures}}, pp.\ \bibinfo {pages} {219--249}\BibitemShut {NoStop}%
\end{thebibliography}%


\appendix

\section{Gauge Invariance}
\label{app:gauge_invariance}
Self-consistent descriptions of entire gate sets introduce an important new phenomenon in the form of gauge degrees-of-freedom. While a detailed treatment of the gauge is beyond the scope of this paper (see \cite{Nielsen2020_GST} for one), the basic idea is that in GST (and any QCVV experiment, really), all we ever have access to are the observed outcome probabilities, which are all of the form

\begin{equation}
    \bbra{E'_i} G_{k_n} \cdots G_{k_1} \kket{\rho'_j}.
\end{equation}

All of these experimentally accessible probabilities are invariant under a simultaneous linear transformation of the form

\begin{equation}
    \begin{aligned}
        G_k & \rightarrow M G_k M^{-1}\\
        \kket{\rho'_j} & \rightarrow M \kket{\rho'_j}\\
        \bbra{E'_i} & \rightarrow \bbra{E'_i} M^{-1},
    \end{aligned}
    \label{eqn:gauge_transformation}
\end{equation}

\noindent where $M$ is an invertible linear transformation applied to every state preparation, gate and measurement. To see this invariance explicitly, consider the effect of applying this linear transformation to all of the elements of the expression for some generic probability:

\begin{equation}
    \bbra{E'_i}M^{-1} M G_{k_n} M^{-1} \cdots M G_{k_1} M^{-1} M\kket{\rho'_j}= \bbra{E'_i} G_{k_n} \cdots G_{k_1} \kket{\rho'_j}.
\end{equation}

The above gauge invariance is analogous to the unobservability of global phases for quantum states. For the purposes of this work the importance of the gauge is two-fold. First, the gauge freedom results in directions in parameter space we can move in which formally change the description of our gate set model, but are physically unobservable. Many of the most commonly used distance measures in the quantum information literature are gauge-variant quantities and, as such, their use must be treated with care. This includes quantities such as fidelity, trace distance, diamond distance as well as many others. In Section \ref{sec:results} we will present results for the average gate set diamond distance scaling for simulated GST experiments, for which we rely on a gauge-fixing procedure called gauge-optimization \cite{Nielsen2020_GST}. The presence of gauge degrees-of-freedom is relevant to our interpretation of the Fisher information for GST experiment designs in Section \ref{sec:fisher_information}.

\section{Estimated Wall Clock Times}
\label{app:wall_clock}

While the reduction in experiment design size is most easily quantified in total number of circuits that need to be run, it can still be difficult to determine the real world impact of the circuit reduction techniques presented here.
Therefore, we estimate wall clock times on several quantum devices for a traditional two-qubit GST experiment for the XYCPHASE gate set against various reduced experiments in Table \ref{tab:estimated-wall-times}.
The estimated running times are given by a simple model where the total time executing circuits is given as
\begin{equation}
    T_{c} = N_{\rm circ} N_{\rm shots} \left(T_{\rm measure/reset} + \sum_{l}^{N_{\rm layer}} \max_i T_{l,i} \right)
    \label{eqn:circ_time}
\end{equation}
where $N_{\rm circ}$ is the number of circuits, $N_{\rm shots}$ is the number of shots per circuit, and $N_{\rm layer}$ is the number of layers in the circuit. Component $i$ of layer $l$ takes $T_{l,i}$ time such that the total time for layer $l$ is given by $\max_i T_{l,i}$, and there is a measure/reset time of $T_{\rm measure/reset}$ between each circuit.

Typically, there is also some overhead in processing and uploading circuits for execution.
The total upload time is given by
\begin{equation}
    T_{u} = T_{\rm latency} \left\lceil \frac{N_{\rm circ}}{N_{\rm circ/batch}} \right\rceil \left\lceil \frac{N_{\rm shots}}{N_{\rm shots/circ/batch}} \right\rceil
    \label{eqn:upload_time}
\end{equation}
where $N_{\rm circ/batch}$ is the number of circuits per batch, $N_{\rm shots/circ/batch}$ is the number of shots per circuit per batch, and $T_{\rm latency}$ is the time needed to upload new circuits between batches.
The second term is the number of batches needed, while the third term is the number of rounds needed, i.e. how many times does each batch need to be run in order to achieve $N_{\rm shots}$. The total wall time is then simply $T_c + T_u$.

The parameters of Eqs. \ref{eqn:circ_time}-\ref{eqn:upload_time} are listed in Table \ref{tab:estimated-device-params}.  The parameter values used are taken from the literature and private communications.  They should be considered representative but not \textit{authoritative}.  That is to say that while these values typify extant hardware properties, there may exist hardware with different corresponding parameter values today, and we should expect hardware advances in the future to non-trivially modify these values.  In particular, advances in classical hardware may significantly improve interbatch latency and number of circuits per batch.  Additionally, 100 shots per circuit-batch is used as representative of what is often use in GST experiments (and is sufficient for good statistics), but that number can often be chosen to be non-trivially larger.

\begin{table}[htbp]
    \centering
    \begin{tabularx}{\textwidth}{>{\centering\arraybackslash}X|c|cc|cc|cc}
        \toprule
        \multirow{2}{*}{\begin{tabular}{c}Device\\Architecture\end{tabular}}&
        \begin{tabular}{c}Robust Germs\\Full Fids.\\(104,002 Circs.)\end{tabular} &
        \multicolumn{2}{c|}{\begin{tabular}{c}Standard Germs\\Full Fids.\\(24,042 Circs.)\end{tabular}} &
        \multicolumn{2}{c|}{\begin{tabular}{c}Standard Germs\\Per-Germ Fids.\\(10,725 Circs.)\end{tabular}} &
        \multicolumn{2}{c}{\begin{tabular}{c}Standard Germs\\12.5\% Random Fids.\\(3,683 Circs.)\end{tabular}} \\
        \cline{2-8}
        & Time & Time & Speedup & Time & Speedup & Time & Speedup \\
        \colrule
        Transmons & 19 min & 4.4 min & 4.3x & 2.4 min & 7.9x & 40 s & 28.9x \\ 
        Trapped Ions & 34 hr & 7.5 hr & 4.3x & 3.5 hr & 9.4x & 1 hr & 33.5x \\ 
        SiMOS & 9.6 hr & 2.3 hr & 4.3x & 1.0 hr & 9.2x & 23 min & 25.3x \\ 
        \colrule
        
        \botrule
    \end{tabularx}
    \caption{Estimated wall clock times for various quantum device architectures for two-qubit with X, Y, and CPHASE gates with germ power up to $L=1024$ with 100 total shots per circuit. Speedups are reported compared to the robust germs/full fiducial design.}
    \label{tab:estimated-wall-times}
\end{table}

\begin{table}[htbp]
    \centering
    \begin{tabularx}{\textwidth}{>{\centering\arraybackslash}X|cccccc}
        \toprule
        Device  & One-Qubit  & Two-Qubit & Measure/Reset & Interbatch & Circuits & Shots per  \\
        Architecture & Gate Time & Gate Time & Time & Latency & per Batch & Circuit-Batch\\
        \colrule
        Transmons \cite{akel} & 20 ns & 200 ns & 1 $\mu$s & 1 s & 100 & 100  \\
        Trapped Ions \cite{qscout,bruzewicz2019trapped} & 10 $\mu$s & 200 $\mu$s & 3.5 ms & 1 s & 200 & 100 \\
        SiMOS \cite{holly} & 0.5 $\mu$s & 1 $\mu$s & 200 $\mu$s & 5 min & 2500 & 100  \\
        \botrule
    \end{tabularx}
    \caption{Representative device parameters used to estimated wall clock times in Table \ref{tab:estimated-wall-times}.}
    \label{tab:estimated-device-params}
\end{table}

\section{Circuit Selection \& Reduction Algorithms}
\label{app:algorithms}
Here we provide more detailed algorithms for circuit selection and reduction as implemented in the \texttt{pyGSTi} package.
The algorithm for fiducial selection is presented in Section~\ref{sec:fiducial_selection}, and consists of constructing the Gram matrix for the state preparation or measurement fiducials and maximizing over the smallest non-trivial eigenvalue. Algorithms for performing germ selection and fiducial pair reduction are presented below.

\subsection{Germ Selection}

The germ selection procedure is given in Algorithm~\ref{alg:germ_selection}, which takes as input some set of models $\mathcal{M}$ and a set of candidate germs $\germs_{\rm cand}$, and returns a subset of germs $\germs\subset\germs_{\rm cand}$ that amplifies the observable parameters of all models.
The computational bottleneck is the computation of the twirled derivatives $J_{ij}^\infty$ in line~\ref{alg:germ_selection_J} and the eigendecomposition required to compute the number of amplified parameters $N_{ij}$ in line~\ref{alg:germ_selection_N}.
Although $N_{ij}$ is the target quantity of interest, it is not necessarily a good choice as the metric for optimization;
instead, we use a score function that can differentiate between germ sets that amplify the same number of parameters.
In \texttt{pyGSTi}, the score function can either be the sum or the minimum of the inverse eigenvalues of $J_{ij}^\infty$.
Regularizations can be applied to either penalize the total number of operations in each germ, or the number of germs in the germ set.
The best germ set for each iteration is the one with the best score over its worst-performing model.

\begin{algorithm}
\caption{Germ Selection}\label{alg:germ_selection}
\SetKwInOut{Input}{Input}\SetKwInOut{Output}{Output}
\Input{Models $\mathcal{M}$ with $N$ non-gauge parameters, and candidate germs $\germs_{\rm cand}$}
\Output{Germ set $\germs \subset \germs_{\rm cand}$ which amplifies all $N$ parameters}
$\germs \gets \set{}$\;
$N_{\rm amp} \gets 0$\;
\While{$N_{\rm amp} < N$ {\bf and} $\germs\neq\germs_{\rm cand}$}{
    $\set{\germs_j} \gets $\texttt{ gen\_test\_sets}$(\germs, \germs_{\rm cand})$ \Comment*[r]{User-controlled}
    \For{$\germs_j \in \set{\germs_j}$}{
        \For{$M_i \in \mathcal{M}$}{
            $J_{ij}^\infty \gets $\texttt{compute\_twirled\_deriv}$( M_i,\germs_j)$ 
            \Comment*[r]{Eqn.~\ref{eqn:infl_germset_jacobian}}
            \label{alg:germ_selection_J}
            $N_{ij} \gets $\texttt{rank}$(J_{ij}^\infty)$ 
            \Comment*[r]{Params amplified by $\mathcal{A}_j$ for $M_i$}
            \label{alg:germ_selection_N}
            $S_{ij} \gets $\texttt{compute\_score}$(J_{ij}^\infty)$ 
            \Comment*[r]{Score (not just $N_{ij}$, user-controlled)}
        }
        $k \gets \argmin_i S_{ij}$ \Comment*[r]{Take worst score over models}
        $S_j \gets S_{kj}$\;
    }
    $l \gets \argmax_j S_j$ \Comment*[r]{Take best score over test germ sets}
    $\germs \gets \germs_l$\;
    $N_{\rm amp} \gets N_{kl}$\;
}
\If{$N_{\rm amp} < N$}{
    $\germs_{\rm cand}$ is not amplificationally complete, try again with more candidate germs\;
}
\end{algorithm}

Another point of flexibility in the germ selection algorithm is the generation of test germ sets at each iteration.
One simple choice is a greedy algorithm: given the current germ set $\germs$, generate a new test set $\germs \cup \set{g_j}$ for each $g_j \in \germs_{\rm cand} \setminus \germs$.
This will select the germ that increases the number of amplified parameters the most in each iteration.
One additional option available in \texttt{pyGSTi} is a greedy randomized adaptive search procedure (GRASP)~\cite{resende2006grasp}, where initial solutions are found via a greedy approach and then improved upon with randomized local optimizations for some fixed number of iterations.
Test sets for the randomized local optimization stage are generated by swapping entries from the current feasible germ set with those in the candidate germ set that have not been selected.

Since the candidate set of germs $\germs_{\rm cand}$ is user-specified, it is possible for Algorithm~\ref{alg:germ_selection} to complete without amplifying all the parameters if $\germs_{\rm cand}$ itself does not amplify all model parameters.
One reasonable choice for $\germs_{\rm cand}$ is to use all possible combinations of gates up to some length excluding cycles and repeats. Taking all such combinations up to length 6 is typically more than sufficient for most one- and two-qubit gate sets. Nonetheless, it can be wise to pre-test the candidate germ set to ensure that $\germs_{\rm cand}$ is AC. While this adds an additional initial cost to germ selection, it does prevent the worst-case run time of a greedy optimization that checks all possible germ sets. It is recommended to only skip the pre-test in cases where the user is confident that there is an AC subset in $\germs_{\rm cand}$.

\subsection{Relaxing Amplificational Completeness}
Relaxing the AC constraint actually requires no modifications to Algorithm \ref{alg:germ_selection};
rather, this corresponds to a choice for the initial set of test models $\mathcal{M}$.
When performing germ selection to generate the robust germ set, we use as input into Algorithm \ref{alg:germ_selection} a set of some number of randomized models, each perturbed by small unitary errors (empirically we have found 5 typically suffices).
The goal of adding unitary perturbations is to break degeneracies in the eigenspectrum of the germs, as discussed in Sections \ref{sec:germ_reduction} and \ref{sec:connection_to_fisher_info}.
By contrast, when constructing the standard germ set we simply use the target model as the input to the algorithm. As a result, germ selection will typically conclude that we are able to amplify parameters within degenerate blocks of a germ's kite structure which the true physical gate set cannot amplify. This effect is most pronounced for gates with high degeneracy, such as idle gates as shown in Fig. \ref{fig:projected_incremental_fisher_information_spectra_full_lite}.
In addition to resulting in a smaller germ set, using the target model only as input into Algorithm \ref{alg:germ_selection} is also faster as it reduces the number of twirled derivative and eigendecomposition calls required.

\subsection{Per-Germ Fiducial Pair Reduction}
The per-germ fiducial pair reduction (FPR) procedure is presented in Algorithm~\ref{alg:per_germ_fpr}, and takes as input the state preparation and measurement fiducials, the germ set.
The algorithm is roughly split into two stages. In the first stage between lines~\ref{alg:per_germ_fpr_reparam}--\ref{alg:per_germ_fpr_for_ij_end}, the germ is reparameterized by its generalized eigendecomposition (i.e. kite structure), where $\theta_k$ are elements within each of the blocks of this decomposition. Then, the Jacobian for each circuit outcome is computed with respect to the $\theta_k$ parameters and cached for each fiducial pair.

\begin{algorithm}
\caption{Per-Germ Fiducial Pair Reduction}\label{alg:per_germ_fpr}
\SetKwInOut{Input}{Input}\SetKwInOut{Output}{Output}
\Input{Prep fiducials $\pfids$ with $N_F$ elements, measure fiducials $\mfids$ with $N_H$ entries, germs $\germs$, and relative minimum eigenvalue threshold $\epsilon_\lambda$}
\Output{Fiducial pairs $\mathcal{P}_k\in\pfids\times\mfids$ for each germ $g_k\in\germs$}
$\mathcal{P}_{\rm all} \gets \set{(f_j,h_i)}\quad\forall f_j\in\pfids,h_i\in\mfids$\;
\For{$g_k \in \germs$}{
    $\mathcal{P}_k \gets \set{}$\;
    $g(\theta_k) \gets $\texttt{reparameterize\_by\_kite}$(g_k)$
    \Comment*[r]{$\theta_k$ are entries in Fig.~\ref{fig:kite_structure}}
    \label{alg:per_germ_fpr_reparam}
    \For{$i=0$ \KwTo $N_H$}{
        \For{$j=0$ \KwTo $N_F$}{
            $f_{ij}(\theta_k) \gets \bbra{E_i^\prime}g(\theta_k)\kket{\rho_j^\prime}$
            \Comment*[r]{Probabilities from Eqn.~\ref{eqn:frequencies}}
            $J_{ij} \gets \frac{\partial f_{ij}(\theta_k)}{\partial \theta_k}$
            \Comment*[r]{Jacobian from Eqn.~\ref{eqn:per_germ_fpr_jacobian}}
        }
    } \label{alg:per_germ_fpr_for_ij_end}
    $\lambda_0^{\rm all} \gets $\texttt{get\_min\_eigval}$(J_{ij}, \mathcal{P}_{\rm all})$\;
    \label{alg:per_germ_fpr_lambda_all}
    $\set{\mathcal{P}_{\rm test}} \gets $\texttt{gen\_fidpair\_sets}$(\mathcal{P}_{\rm all})$
    \Comment*[r]{User-controlled}
    \For{$\mathcal{P}_{\rm test} \in \set{\mathcal{P}_{\rm test}}$}{
        $\lambda_0 \gets $\texttt{get\_min\_eigval}$(J_{ij}, \mathcal{P}_{\rm test})$\;
        \If{$\lambda_0 \geq \epsilon_\lambda\lambda_0^{\rm all}$}{
            $\mathcal{P}_k \gets \mathcal{P}_{\rm test}$
            \Comment*[r]{Accept if still sensitive to min eigval}
            Break from $g_k$ loop\;
        }
    } \label{alg:per_germ_fpr_for_Ptest_end}
}
\end{algorithm}

With all the Jacobians computed, the algorithm now enters the second stage which occurs between lines~\ref{alg:per_germ_fpr_lambda_all}--\ref{alg:per_germ_fpr_for_Ptest_end}.
The goal of FPR is to identify a subset of the complete set of fiducial pairs which are sensitive to changes in all of the amplified parameters. A test set of fiducial pairs, $\mathcal{P}_{\rm test}$,  is a valid candidate solution if, when concatenating the Jacobians for each of the elements in $\mathcal{P}_{\rm test}$, the resulting Jacobian for that set has rank equal to the number of amplified parameters.

Just as with germ selection, the rank itself is not sufficient; for example, a particular set of fiducial pairs may have a concatenated Jacobian with the correct rank but a poor condition number, indicating that we are only weakly sensitive to some of the amplified parameters. Instead, we impose a threshold, measured relative to the complete fiducial set, which the smallest non-gauge eigenvalue of the test set must exceed, and denote the relative factor by $\epsilon_\lambda$. Since the largest eigenvalue of the test set is upper bounded by the largest eigenvalue of the full set, checking the smallest eigenvalue implicitly tests both the rank and condition number.

In selecting a magnitude for $\epsilon_\lambda$, the goal is selecting a value which maintains good sensitivity to all of the amplified parameters while yielding significant reductions in the number of circuits needed. Of course, there is invariably a tradeoff between the magnitude of $\epsilon_\lambda=$ and the reduction in experiment design size which needs to be balanced in practice. For the one-qubit experiment designs utilizing per-germ FPR presented in this work we used a value of $\epsilon_\lambda=\frac{1}{30}$, while the two-qubit experiment designs used $\epsilon_\lambda=\frac{1}{2}$.

As with germ selection, there is some flexibility in the generation of test sets of fiducial pairs. The most common strategy used in \texttt{pyGSTi} is to simply randomly sample fiducial pairs. The number of fiducial pairs starts at the number required to observe the correct number of non-gauge parameters. A user-specified number of sets of fiducial pairs of a given size are tested at each iteration of the random search, with each subsequent iteration incrementing upward the size of the candidate sets until an acceptable solution is found

\subsection{Per-Germ Power-Random Fiducial Pair Reduction}

\begin{algorithm}
\caption{Per-Germ Power-Random Fiducial Pair Reduction}\label{alg:random_fpr}
\SetKwInOut{Input}{Input}\SetKwInOut{Output}{Output}
\Input{Prep fiducials $\pfids$ with $N_F$ elements, measure fiducials $\mfids$ with $N_H$ entries, germs $\germs$, germ powers $\set{L}$, and keep fraction $\gamma$}
\Output{Fiducial pairs $\mathcal{P}_{k,l}\in\pfids\times\mfids$ for each germ $g_k\in\germs$ and power $l\in\set{L}$}
$\mathcal{P}_{\rm all} \gets \set{(f_j,h_i)}\quad\forall f_j\in\pfids,h_i\in\mfids$\;
\For{$g_k \in \germs$}{
    \For{$l \in {L}$}{
        $\mathcal{P}_{k,l} \gets $\texttt{choose\_without\_replacement}$(\mathcal{P}_{\rm all}, \lceil\gamma N_F N_H \rceil)$\;
    }
}
\end{algorithm}

The per-germ power-random FPR method is the most straightforward of the FPR schemes.
As shown in Algorithm~\ref{alg:random_fpr}, this simply consists of selecting a random fraction of fiducial pairs to keep for each germ-power pair.
While there is no guarantee that the random selection is sensitive to all parameters in a germ, selecting a different set of fiducial pairs for each germ \textit{and} power provides some robustness to the random nature of this procedure. Moreover, while it doesn't come with the same performance guarantees as the more structured per-germ FPR method, we've found empirically that this works well in practice for sufficiently large fractions of fiducial pairs kept (see Figure \ref{fig:one_and_two_qubit_ddists_vs_num_ckts}).

\end{document}